\documentclass[prd,twocolumn,lengthcheck,superscriptaddress,showpacs,letterpaper,nofootinbib,aps]{revtex4-1}

\usepackage{color}
\usepackage{graphicx}  
\usepackage{amsmath}
\usepackage{times}
\usepackage{bm}
\usepackage{acronym}
\usepackage{subfig}
\usepackage[bookmarksnumbered, bookmarksopen, breaklinks, colorlinks]{hyperref}

\def\be{\begin{equation}}
\def\ee{\end{equation}}

\def\bea{\begin{eqnarray}}
\def\eea{\end{eqnarray}}

\begin{document}

\title{Investigating the effect of precession on searches for
neutron-star--black-hole binaries with Advanced LIGO}

\author{Ian W.\ Harry}
\affiliation{Department of Physics, Syracuse University, Syracuse, New 
York 13244, USA}
\affiliation{Kavli Institute of Theoretical Physics, University of California, 
Santa Barbara, California 93106, USA}

\author{Alexander H.\ Nitz}
\affiliation{Department of Physics, Syracuse University, Syracuse, New 
York 13244, USA}

\author{Duncan A.\ Brown}
\affiliation{Department of Physics, Syracuse University, Syracuse, New 
York 13244, USA}
\affiliation{Kavli Institute of Theoretical Physics, University of California, 
Santa Barbara, California 93106, USA}
\affiliation{LIGO Laboratory, California Institute of Technology, Pasadena, 
California 91125, USA}

\author{Andrew P.\ Lundgren}
\affiliation{Albert-Einstein-Institut, Max-Planck-Institut f\"ur
Gravitationsphysik, D-30167 Hannover, Germany}
\affiliation{Leibniz Universit\"at Hannover, D-30167 Hannover, Germany}
\affiliation{Kavli Institute of Theoretical Physics, University of California, 
Santa Barbara, California 93106, USA}

\author{Evan Ochsner}
\affiliation{Center for Gravitation and Cosmology, University of
Wisconsin-Milwaukee,
Milwaukee, Wisconsin 53211, USA}
\affiliation{Kavli Institute of Theoretical Physics, University of California, 
Santa Barbara, California 93106, USA}

\author{Drew Keppel}
\affiliation{Albert-Einstein-Institut, Max-Planck-Institut f\"ur
Gravitationsphysik, D-30167 Hannover, Germany}
\affiliation{Leibniz Universit\"at Hannover, D-30167 Hannover, Germany}

\begin{abstract}
The first direct detection of neutron-star--black-hole binaries will
likely be made with gravitational-wave observatories. Advanced LIGO and
Advanced Virgo will be able to observe neutron-star--black-hole mergers at a
maximum distance of 900 Mpc. To achieve this sensitivity, gravitational-wave
searches will rely on using a bank of filter waveforms that
accurately model the expected gravitational-wave signal. The emitted signal
will depend on the masses of the black hole and the neutron star and also the
angular momentum of both components. The angular momentum
of the black hole is expected to be comparable to the orbital angular momentum 
when the system is emitting gravitational waves in Advanced LIGO's and Advanced 
Virgo's sensitive band.
This angular momentum will affect the dynamics of the inspiralling system and
alter the phase evolution of the emitted gravitational-wave signal. In
addition, if the black hole's angular momentum is not aligned with the orbital
angular momentum it will cause the orbital plane of the system to precess.
In this work we demonstrate that if the effect of the black
hole's angular momentum is neglected in the waveform models used in
gravitational-wave searches, the detection rate of $(10+1.4)M_{\odot}$ 
neutron-star--black-hole systems with isotropic spin distributions would be 
reduced by $33\%--37\%$ in comparison to a hypothetical perfect search at a 
fixed signal-to-noise ratio threshold. The error in this measurement is due
to uncertainty in the post-Newtonian approximations that are used to model the
gravitational-wave signal of neutron-star--black-hole inspiralling binaries. We
describe a new method for creating a bank of filter waveforms where the black
hole has nonzero angular momentum that is aligned with the orbital angular
momentum. With this bank we find that the detection rate of $(10+1.4)M_{\odot}$
neutron-star--black-hole systems would be reduced by $26\%--33\%$. Systems that
will not be detected are ones where the precession of the orbital plane causes
the gravitational-wave signal to match poorly with nonprecessing filter
waveforms. We identify the regions of parameter space where such systems occur
and suggest methods for searching for highly precessing
neutron-star--black-hole binaries.
\end{abstract}

\maketitle

\acrodef{aLIGO}[aLIGO]
{The Advanced Laser Interferometer Gravitational Wave Observatory}
\acrodef{AdV}[AdV]{Advanced Virgo}
\acrodef{CBC}[CBC]{compact binary coalescence}
\acrodef{S6}[S6]{LIGO's sixth science run}
\acrodef{VSR23}[VSR2 and VSR3]{Virgo's second and third science runs}
\acrodef{EM}[EM]{electromagnetic}
\acrodef{GW}[GW]{gravitational wave}
\acrodef{NS}[NS]{neutron star}
\acrodef{BH}[BH]{black hole}
\acrodef{BNS}[BNS]{binary neutron star}
\acrodef{NSBH}[NSBH]{neutron-star--black-hole}
\acrodef{GRB}[GRB]{gamma-ray burst}
\acrodef{S5}[S5]{LIGO's fifth science run}
\acrodef{S4}[S4]{LIGO's fourth science run}
\acrodef{VSR1}[VSR1]{Virgo's first science run}

\acrodef{PSD}[PSD]{power spectral density}
\acrodef{VSR3}[VSR3]{Virgo's third science run}
\acrodef{BBH}[BBH]{binary black holes}
\acrodef{SNR}[SNR]{signal-to-noise ratio}
\acrodef{SPA}[SPA]{stationary-phase approximation}
\acrodef{LHO}[LHO]{LIGO Hanford Observatory}
\acrodef{LLO}[LLO]{LIGO Livingston Observatory}
\acrodef{LSC}[LSC]{LIGO Scientific Collaboration}
\acrodef{PN}[PN]{Post-Newtonian}
\acrodef{DQ}[DQ]{data quality}
\acrodef{IFO}[IFO]{interferometer}
\acrodef{DTF}[DTF]{detection template families}
\acrodef{FAR}[FAR]{false alarm rate}
\acrodef{FAP}[FAP]{false alarm probability}
\acrodef{PTF}[PTF]{physical template family}
\acrodef{ADE}[ADE]{advanced detector era}
\acrodef{FFT}[FFT]{Fast Fourier Transformation}
\acrodef{GPU}[GPU]{graphical processing unit}
\acrodef{ISCO}[ISCO]{inner-most stable circular orbit}
\acrodef{MECO}[MECO]{minimum energy condition}

\section{Introduction}
\label{sec:intro}

\ac{aLIGO} will
begin observing the gravitational-wave sky in 2015~\cite{Aasi:2013wya}. When
\ac{aLIGO} reaches design sensitivity, it will be sensitive to a volume of the
universe 1000 times greater than the first-generation LIGO
detectors~\cite{Harry:2010zz}. The French-Italian \ac{AdV} detector
will begin observations shortly after \ac{aLIGO}, forming a worldwide network
of gravitational-wave
observatories~\cite{Aasi:2013wya,aVirgo,AdV2}.
One of the most interesting sources for \ac{aLIGO} and \ac{AdV} is the inspiral
and merger of \ac{NSBH} binaries. It has been argued that
Cyg X-3 is a possible \ac{NSBH} \emph{progenitor}~\cite{Belczynski:2012jc};
however, \ac{NSBH} binaries have not been observed by radio or other
electromagnetic observations. The first direct detection of a \ac{NSBH} binary
will likely be made with \ac{aLIGO} and \ac{AdV}. Population-synthesis models 
of binary evolution predict
that \ac{aLIGO} should see 0.2--300 \ac{NSBH} binaries per
year~\cite{Abadie:2010cf}. Direct detection of the gravitational waves from NSBH
binaries would confirm their existence and allow us to explore the astrophysics
behind the formation and evolution of these systems.

The gravitational waves radiated by \ac{NSBH} binaries are expected to be
significantly affected by the black hole's angular momentum
(spin), which is expected to be comparable to the orbital angular momentum of
the binary~\cite{Cutler:1992tc,Apostolatos:1994mx,Kidder:1992fr,Kidder:1995zr}.
Spin-orbit coupling changes the gravitational waveform of the binary's inspiral
and merger and can cause the orbital plane of the binary to
precess~\cite{Apostolatos:1994mx}. Coupling between the black-hole spin and the
neutron-star spin~\cite{Kidder:1995zr}, the quadrupole-monopole interaction due
to the spheroidal deformation of spinning black holes and neutron
stars~\cite{Poisson:1997ha} and the ``self-spin''
interaction~\cite{Mikoczi:2005dn} will also affect the gravitational waveform
emitted during a \ac{NSBH} binary inspiral.
The resulting changes in the waveform observed by \ac{aLIGO}
carry a great deal of information about the dynamics of the binary. However,
optimal searches of \ac{aLIGO} data must incorporate this dynamics into their
waveform models to avoid a reduction in sensitivity and hence the rate of
detected events. Variation between the available waveform models, and with 
nature's waveforms, will also cause a reduction in sensitivity, we investigate 
this issue in a companion work~\cite{Nitz:2013mxa}.

Gravitational-wave searches for the merger of two compact objects rely on
matched filtering against compact binary merger gravitational waveform
models~\cite{Wainstein,Helstrom,Allen:2005fk}.
Compact binary mergers in quasicircular orbit are described by 15
parameters: the masses, spin magnitude, spin orientations, source orientation,
sky location, distance and time and phase of
coalescence~\cite{Peters:1963ux,Th300}. Matched-filter
searches must be capable of detecting binary mergers regardless of the
parameters of the system. For nonprecessing systems and restricting to the
dominant gravitational-wave mode, the extrinsic parameters--source orientation,
sky location, distance and coalescence phase--only affect the overall phase and
amplitude of the observed gravitational-wave system. Therefore, it is possible
to analytically maximize over these extrinsic parameters \cite{Allen:2005fk}.

Changing the masses and spin magnitudes of a nonprecessing system will change
the intrinsic phase evolution of the system. To be able to detect \ac{NSBH}
systems within the desired parameter range a set of waveforms or ``template
bank'' must be
constructed~\cite{Sathyaprakash:1991mt,Poisson:1995ef,Owen:1995tm,Owen:1998dk,
Babak:2006ty,Balasubramanian:1995bm,Cokelaer:2007kx}. These waveforms should
span the desired range of mass and spins.
The standard practice is to construct a bank of waveforms such that any
waveform within the parameter space of interest would be recovered with at
least 97\% of the optimal \ac{SNR} by at least one waveform in the template
bank~\cite{Babak:2006ty,Babak:2012zx} . However, the geometrical
placement algorithms employed in the most recent searches for compact binary 
coalescences in 
LIGO and Virgo data are only applicable for compact binary systems whose 
components have no angular momentum--non-spinning 
systems~\cite{Abbott:2009tt,Abbott:2009qj,Abadie:2010yba,Abadie:2011nz}. 
Stochastic
placement algorithms~\cite{Babak:2008rb,Harry:2009ea,Manca:2009xw,Ajith:2012mn}
are capable of placing banks of waveforms where the spin of the
black hole is aligned with the orbital angular momentum (aligned-spin
\ac{NSBH})~\cite{Ajith:2012mn}. However,
these algorithms are known to need more templates to cover a parameter space
when compared to geometric algorithms~\cite{Harry:2009ea}. In 
\cite{Brown:2012qf} we developed a new geometrical placement algorithm that 
could place template banks of aligned-spin \ac{BNS} signals. In this work we 
expand that method to be able to place template banks of aligned-spin \ac{NSBH} 
signals.

When precessing systems are considered as template waveforms, the matched-filter
search becomes more complex. In this case the extrinsic parameters no longer
enter as overall phase and amplitude shifts in the
waveform~\cite{Apostolatos:1994mx}. Previous work has been conducted to explore
the affect of precession on gravitational-wave searches and to develop methods
to detect precessing
systems~\cite{Apostolatos:1996rf,Buonanno:2002fy,Grandclement:2002dv,
Grandclement:2002vx,Grandclement:2003ck,Pan:2003qt,Buonanno:2004yd,
Buonanno:2005pt, Abbott:2007ai,VanDenBroeck:2009gd,Fazi:2009,Harry:2011qh,
Ajith:2012mn, Brown:2012gs,Lundgren:2013jla}. However, these searches, when
applied to initial LIGO and Virgo data, have not shown an increase in efficiency
with respect to nonprecessing searches~\cite{VanDenBroeck:2009gd}. This is
because the filtering codes allow for increased, and unphysical, freedom when
maximizing over extrinsic parameters and because no suitable method to
distinguish gravitational-wave signals from non-Gaussian instrumental noise has
been developed for these searches. Therefore, searches for \ac{NSBH} binaries
in data from LIGO and Virgo's most recent science runs ignored spin effects and
used quasicircular templates to search for \ac{NSBH}
binaries~\cite{Abbott:2009tt,Abbott:2009qj,Abadie:2010yba,Abadie:2011nz}.

The majority of previous work considered the initial LIGO detectors. \ac{aLIGO}
will have a substantially different noise curve than initial
LIGO~\cite{Aasi:2013wya}. Conclusions drawn using the initial LIGO sensitivity
curve may not hold when considering \ac{aLIGO}. A previous study considering
\ac{aLIGO} sensitivity curves has suggested that it may be possible to detect
generic, precessing \ac{NSBH} binaries using aligned-spin 
waveforms~\cite{Ajith:2012mn}. However, other studies have suggested that
precession may significantly change the gravitational waveform seen by
\ac{aLIGO}, requiring templates that explicitly capture this
effect~\cite{Brown:2012gs}.

In this paper, we first investigate the effect of ignoring spin on optimal
(matched-filter) searches for \ac{NSBH} binaries with \ac{aLIGO}. We demonstrate
that the quasicircular templates used in initial LIGO will reduce the detection
rate by $33\% - 37\%$ for \ac{NSBH} systems with masses uniformly distributed 
between
$(10\pm0.5,1.4\pm0.05)M_{\odot}$, an isotropic black-hole spin distribution and 
spin magnitude uniformly distributed between 0 and 1. Over a wider range of 
uniformly distributed masses, $(3-15,1-3)M_{\odot}$, we find that the detection 
rate would be reduced
by $31\% - 36\%$. In both cases this loss in detection rate is 
compared against a template bank where every signal is matched
exactly by the bank of filters. The loss in event rate is greatest for
\ac{NSBH} binaries with large black-hole spins and large mass ratios. The range
quoted in both measurements is due to uncertainty in the waveform models used to
simulate \ac{NSBH} gravitational-wave signals. These values also strongly 
depend on the signal distributions that we selected. If nature does not provide 
a uniform distribution of masses and an isotropic distribution of masses then 
these averaged values will change. To account for this, we explore the ability 
to recover \ac{NSBH} signals as a function of their spins and masses in Sec. 
\ref{sec:resultsII_generic}.

We expand upon the method we
introduced in \cite{Brown:2012qf} and
construct a bank of templates for aligned-spin \ac{NSBH} binaries. We
demonstrate that this template bank is effectual for recovering the population
of aligned-spin \ac{NSBH} systems that it is designed to detect. We
assess the ability of an aligned-spin template bank to detect a population of
generic
\ac{NSBH} binaries where the black-hole spin is not constrained to be parallel
to the orbital angular momentum. We find using the aligned-spin
bank will reduce the detection rate by $17\%-23\%$ compared to using a bank 
where
every signal matches exactly with one of the filter waveforms when searching 
for \ac{NSBH} waveforms with masses $(3-15,1-3)M_{\odot}$. When restricting the 
mass range to $(10\pm0.5,1.4\pm0.05)M_{\odot}$ we find that the detection rate 
is reduced by $26\%-33\%$. We find that
there are regions of the \ac{NSBH} signal parameter space where precession
effects cause a significant reduction in signal-to-noise ratio. These regions 
are those where the black hole's angular momentum is large in comparison to 
the orbital angular momentum. We suggest possible methods for constructing 
searches that recover these systems. By considering several \ac{NSBH} waveform 
models, we demonstrate that our results are robust against possible errors in 
the post-Newtonian phasing for \ac{NSBH} binaries.

There has been a great deal of recent work focused on numerically 
modeling the merger of a black hole and a neutron 
star~\cite{Duez:2009yz,Shibata:2011jka,Pannarale:2012ux,Lackey:2013axa,
Foucart:2013psa}. However, there is not currently any widely available waveform 
model 
that includes both the full evolution of a \ac{NSBH} coalescence \emph{and} 
includes precessional effects over the full parameter space that we consider.
Therefore, in this work we have restricted ourselves to considering 
post-Newtonian, inspiral-only signal waveforms and consider only the case of two 
point particles. If a full 
inspiral-merger-ringdown, precessing \ac{NSBH} waveform model 
becomes available, it would be informative to compare results with that model 
against those presented here. However, in this work the black-hole mass is 
restricted to be less than $15M_{\odot}$. It has been demonstrated that 
inspiral-only template banks recover $> 95\%$ of the signal power of 
numerically modeled $(3+15)M_{\odot}$ binary black-hole 
waveforms~\cite{Brown:2012nn,Smith:2013mfa}. It has also been demonstrated that 
nonspinning \ac{NSBH} mergers with total mass $\sim 10M_{\odot}$ are 
indistinguishable from binary black-hole mergers with the same 
masses~\cite{Foucart:2013psa}. With these observations we expect that 
our results are qualitatively valid in the parameter space we study.

The layout of this work is as follows. In Sec. \ref{sec:nsbhpop} we describe
the set of \ac{NSBH} systems that we use to assess
the performance of our template banks. In Sec. \ref{sec:waveform_model} we
discuss the waveform models that we use in our simulations. In Sec.
\ref{sec:bank_testing} we discuss the methods we use to test the template
banks. In Sec. \ref{sec:bank_method} we describe our new method to
create banks of aligned-spin filter waveforms and use these methods in Sec.
\ref{sec:bank_construction} to create our template banks. In Sec.
\ref{sec:bank_validation} we validate our template banks against the
aligned-spin signal models they are constructed to detect. In Sec.
\ref{sec:resultsII_generic} we assess the performance of non-spinning template 
banks to search for generic \ac{NSBH} signals and assess the performance of 
aligned-spin template banks to detect the same signals. We conclude in Sec. 
\ref{sec:conclusion}. Throughout this work we will use $G = c = 1$.

\section{A population of NSBH binaries}
\label{sec:nsbhpop}

In this section, we describe our large simulated set of \ac{NSBH} binaries. 
This is used to assess the loss in detection rate when using
nonspinning and
aligned-spin template banks to search for generic \ac{NSBH} binaries. To
construct this set we incorporate current astrophysical knowledge to choose the
distribution of masses and spins. However, this astrophysical knowledge is 
limited due to the fact that no \ac{NSBH} binaries have been directly 
observed. Nevertheless, both
\acp{NS} and \acp{BH} have been observed in other binary systems, and these
observations can be used to make inferences about the mass and spin
distributions that might be expected in \ac{NSBH} binaries.
We begin by giving the distributions that we use in this work, before 
describing the astrophysical knowledge that motivated these choices.

We simulate 100000 \ac{NSBH}
binaries with parameters drawn from the following distribution.
The black-hole mass is chosen uniformly between 3 and 15 solar masses;
the neutron-star mass is chosen uniformly between 1 and 3 solar masses; the
black-hole dimensionless spin magnitude is chosen uniformly between 0 and 1 and
the neutron-star dimensionless spin magnitude is chosen uniformly between 0 and
0.05. The initial spin orientation for both bodies, the source orientation and
the sky location are all chosen from an isotropic distribution.

Black holes observed in x-ray binaries can be used to estimate the \ac{BH}
mass distribution, though it is difficult to disentangle the individual masses
and inclination angle with only electromagnetic observations~\cite{Ozel:2010su}.
Using a population of $\sim20$ low-mass x-ray binary systems with estimated 
masses, two separate works found that a \ac{BH} mass distribution of $7.8 \pm 
1.2 M_{\odot}$ fits the observed data well~\cite{Ozel:2010su,Farr:2010tu}. 
There is evidence that there is a ``mass gap'' between $3M_{\odot}$ and 
$5M_{\odot}$ where \acp{BH} will not form~\cite{Ozel:2010su,Farr:2010tu}, 
although this may be due to observational bias~\cite{Kreidberg:2012ud}. When 
high-mass x-ray binary systems are considered the mass distribution increases 
to $9.2012 \pm 3 M_{\odot}$, although a Gaussian model is a poor fit for these 
systems~\cite{Farr:2010tu}. Evidence exists for a stellar mass black hole with 
mass $> 20 M_{\odot}$ in the IC 10 X-1 x-ray 
binary~\cite{Prestwich:2007mj,Silverman:2008ss}. We choose to use a uniform 
range of 3-15 solar masses for the black holes in our \ac{NSBH} signal 
population. This is partly motivated by the considerations above, and partly by 
our concern of the validity of inspiral-only, point particle waveform models 
for high-mass \ac{NSBH} systems.
Observations of black-hole spin have found spin values that span the minimum and
maximum possible values for Kerr black holes~\cite{Miller:2009cw}; therefore, 
we use a uniform black-hole spin distribution between 0 and 1.

Observations of \acp{NS} in binary systems other than \ac{NSBH} binaries can be
used to estimate the \ac{NS} mass distribution. Using a
population of six \ac{BNS} systems with well constrained masses,
Ozel \textit{et al.}~\cite{Ozel:2012ax}
found that the \ac{NS} mass distribution was well fitted by $1.33 \pm 0.05
M_{\odot}$, in agreement with Kiziltan, Kottas and Thorsett's result
of $1.35 \pm 0.13 M_{\odot}$~\cite{Kiziltan:2010ct}. However, nonrecycled
\acp{NS} in eclipsing high-mass binaries, as well as slow
pulsars, are found to have a much wider mass distribution of $1.28 \pm
0.24 M_{\odot}$~\cite{Ozel:2012ax}. Recycled \acp{NS} are found to have a
higher range of masses, $1.48 \pm 0.2 M_{\odot}$, due to
accretion~\cite{Ozel:2012ax}. However, it is expected that the black hole would
form first in the vast majority of cases, which would remove the
possibility of recycling.
There is also evidence for a \ac{NS}
with a mass as high as $\sim 3 M_{\odot}$ \cite{Freire:2007jd}, which is very
close to the theoretical upper limit on a \ac{NS}
mass of $\sim3.2 M_{\odot}$ \cite{Rhoades:1974fn}. While a conservative choice, 
we choose to use a uniform mass distribution between 1 and 3 solar masses for 
the \acp{NS} in our \ac{NSBH} signal population.

The magnitude of the dimensionless spin, $\bm{\chi} = {\bm{S}/ 
{m}^2}$, of
a neutron star cannot be larger than $\sim$ 0.7 \cite{Lo:2010bj} as the
neutron star would break apart under the rotational force.
However, it is rather
unlikely that \ac{NS} spins will have values as large as this in \ac{NSBH}
systems. At birth, neutron star spins are believed to be in the range
10-140 ms, corresponding to $\chi < 0.04$~\cite{Lorimer:2008se,Mandel:2009nx}.
Recycled neutron stars can have larger spin values~\cite{Bildsten:1997vw};
however, they are unlikely to have periods less than
1~ms~\cite{Chakrabarty:2008gz}, corresponding to a
dimensionless spin of $\chi \sim 0.4$. The fastest spinning recycled neutron
star observed in a \ac{BNS} binary has a spin period of only 
23~ms~\cite{Burgay:2003jj}. As astrophysical observations seem to suggest that 
large neutron spins will be unlikely in \ac{NSBH} binaries we choose a uniform 
\ac{NS} spin distribution between 0 and 0.05.

\section{Waveform models}
\label{sec:waveform_model}

Matched-filter searches require an accurate model of compact binary mergers.
In a companion work we investigate the agreement of different waveform
families in the \ac{NSBH} region of parameter space and find a considerable
disagreement between waveforms produced by different waveform models, which
will reduce detection efficiency~\cite{Nitz:2013mxa}.

In this work we wish to investigate the effects of spin, especially spin-induced
precession, while understanding and mitigating any bias in our results due to 
the choice of waveform approximant. We therefore run all our simulations using 
two waveform approximants: TaylorT2~\cite{Damour:2000zb} and 
TaylorT4~\cite{Buonanno:2002fy}.

\ac{PN} waveforms, such as TaylorT2 and TaylorT4, are constructed by solving the
\ac{PN} equations of motion to obtain the binary orbits. It is assumed that the
binary evolves adiabatically through a series of quasicircular orbits. This is
a reasonable assumption as it is expected that the emission of gravitational
radiation will circularize the orbits of isolated binaries~\cite{Peters:1964zz}. 
The
equations of motion then reduce to series expansions of the center-of-mass
energy $E(v)$ and the gravitational-wave flux $\mathcal{F}(v)$, which are
expanded as a power series in the orbital velocity $v$:
\begin{align}
\label{ref:PNeqs}
E(v) &= E_{\mathrm{N}} v^2 \left(1+\sum_{n=2}^{6}E_i v^i\right), \\
\mathcal{F}(v) &= F_{\mathrm{N}} v^{10} \left(1+\sum_{n=2}^{7}
\sum_{j = 0}^{1}F_{i,j} v^i \log^j v\right).
\end{align}
The various coefficients ($E_N$,$E_i$,$\mathcal{F}_N$,$\mathcal{F}_i$) are
reviewed in \cite{Arun:2008kb,Buonanno:2009zt}. For terms involving the orbital
contribution, the center-of-mass energy and gravitational-wave flux are known 
to 3.5\ac{PN} order [$n=7$ in the parentheses of (\ref{ref:PNeqs})]
\cite{Wiseman:1993aj,Blanchet:1995fg,Blanchet:1995ez,Blanchet:1996pi,
Blanchet:2001ax, Blanchet:2004ek}. For terms
involving the spin of the objects, the expansions of the energy and
flux are complete to 2.5\ac{PN} order [$n=5$ in the parenthesis of 
(\ref{ref:PNeqs})]
\cite{Kidder:1992fr,Kidder:1995zr,Arun:2008kb}. In recent work,
terms relating to the coupling between the component spins and the orbit have
also been computed to 3.5\ac{PN} order \cite{Blanchet:2012sm,Bohe:2013cla}. We
choose not to use these terms in this work because terms relating to the
spin(1)-spin(2), quadrupole-monopole and self-spin contributions are not yet
known at 3 \ac{PN} order, so we restrict the spin-related terms to 2.5 \ac{PN}
where these terms are fully known. We do not expect these terms to change the 
main conclusions of the work as these additional phase evolution terms will 
have little effect on the precessional evolution of a system.

The orbital phase, $\phi$ is then obtained via the energy balance equation
\begin{equation}
 \frac{dE}{dt} = - \mathcal{F}
\end{equation}
and by
\begin{equation}
  \frac{d\phi}{dt} = \pi f.
\end{equation}
Here the gravitational-wave frequency $f$ is given by twice the orbital phasing
frequency and is related to the orbital velocity by $v = (\pi M f)^{1/3}$, 
where $M$ denotes the total mass of the binary.

The various approximants are constructed via \emph{different} ways of
obtaining the gravitational-wave phase from the equations above.

\subsection{TaylorT2 and TaylorF2}
\label{ssec:taylort2}

The TaylorT2 approximant is constructed by first calculating
\begin{equation}\label{eq:t2}
 B(v) = \left[ \frac{E'(v)}{-\mathcal{F}(v)} \right].
\end{equation}
Here $[X]$ is used to indicate that $X$ is calculated by first expanding it
as a Taylor series. Then orbital terms larger
than 3.5\ac{PN} and spin terms larger than 2.5\ac{PN} are discarded. This is 
because terms of this order would also depend on unknown terms in the 
expansion of the center-of-mass energy and the gravitational-wave phase.
As $B(v) = dt / dv$ the gravitational-wave phase is therefore obtained 
according to
\begin{equation}\label{eq:phaset2}
\phi(v) = \int \frac{v^3}{M} B(v) dv,
\end{equation}
which can be integrated analytically. In the same manner $t(v)$ can be 
calculated according to
\begin{equation}
 t(v) = \int B(v) dv.
\end{equation}
$\phi(v)$ and $t(v)$ can then be numerically inverted to obtain $\phi(t)$ and 
$v(t)$, which are used to construct the waveform.

When constructing a TaylorT2 waveform, one begins at a fiducial 
starting frequency, chosen to be smaller than the lowest frequency over which to 
perform the matched filter. In this work, we use $14$ Hz as the starting 
frequency. The waveform is terminated when the frequency reaches the 
\ac{MECO}, which is the point where
\begin{equation}
\frac{dE(v)}{dv} = 0.
\end{equation}

The TaylorF2 approximant is a frequency-domain equivalent of the TaylorT2
approximant and is constructed using the stationary phase approximation
\cite{Sathyaprakash:1991mt,Cutler:1994ys,Droz:1999qx,Blanchet:2006zz}.
The TaylorF2 waveforms can be expressed as an analytic expression of the form
\begin{equation}
 \tilde h(f) = A(f;\mathcal{M},D_L\theta_x) e^{ - i \Psi(f;\lambda_i)},
\end{equation}
where $\tilde h(f)$ denotes the Fourier transform of $h(t)$, the time-domain
gravitational-wave strain, $\mathcal{M}$ denotes the chirp mass, $D_L$ the 
luminosity distance to the source and $\theta_x$ describes the various 
orientation angles that only affect the
amplitude and overall phase of the observed gravitational
waveform~\cite{Allen:2005fk}. The phase $\Psi$ is given by
\begin{equation} \label{eq:phase_exp}
\Psi = 2 \pi f t_c - \phi_c(\theta_x) + 
\sum_{i = 0}^{7} \sum_{j = 0}^{1} \lambda_{i, j} f^{(i-5)/3} \log^j f,
\end{equation}
where $t_c$ is the coalescence time and $\phi_c$ is a constant phase offset. The
$\lambda$ terms give the various coefficients of the orbital phase,
which are summarized in \cite{Arun:2008kb,Buonanno:2009zt}. TaylorF2 waveforms
are usually terminated at the frequency corresponding to the \ac{ISCO} of a
nonspinning system with the given masses \cite{Allen:2005fk}.

\subsection{TaylorT4 and TaylorR2F4}
\label{ssec:taylort4}

In contrast to the TaylorT2 approximant, the TaylorT4 approximant
introduced in~\cite{Buonanno:2002fy} is formed by calculating
\begin{equation}\label{eq:t4}
\frac{dv}{dt} = \left[ \frac{-\mathcal{F}(v)}{E'(v)} \right] = 
A(v).
\end{equation}
Similar to the TaylorT2 approximant, orbital terms larger than 3.5\ac{PN} and
spin terms larger than 2.5\ac{PN} are discarded from $A(v)$. This is
numerically solved to obtain $v(t)$ which can then be used to obtain the
gravitational-wave phase. The TaylorT4 approximant uses the same start and
termination conditions as the TaylorT2 approximant.

The TaylorR2F4 approximant, introduced in~\cite{Nitz:2013mxa}, is a
frequency-domain analytical approximation of the TaylorT4 waveform model. It is
constructed in the same manner as TaylorF2; however, it uses
\begin{equation}
\frac{dt}{dv} = \left[ \frac{1}{A(v)} \right]
\end{equation}
instead of Eq.~(\ref{eq:t2}). In this case, while $A(v)$ is restricted as
described above, $1 / A(v)$ is truncated to a
higher order in $v$. The additional ``partial'' terms that are obtained in the
resulting \ac{PN} expansion describe the difference between the TaylorT2 and
TaylorT4 models. It has empirically been found that TaylorR2F4 matches best 
with TaylorT4
when $1 / A(v)$ is expanded to 4.5\ac{PN} order or 6\ac{PN} order
\cite{Nitz:2013mxa}. We only consider these two expansions of TaylorR2F4 in this
work.

\section{Method for assessing the performance of NSBH searches}
\label{sec:bank_testing}

In this section we describe the methods we use to assess the efficiency of
template banks and the terminology that we will use in the rest of this work.
The ``overlap'' between two waveforms $h_1$ and $h_2$ is defined as
\begin{equation}
 \mathcal{O}(h_1,h_2) = (\hat{h}_1|\hat{h}_2) =
\dfrac{(h_1|h_2)}{\sqrt{(h_1|h_1)(h_2|h_2)}},
\end{equation}
where $(h_1,h_2)$ denotes the noise-weighted inner product
\begin{equation}
(h_1|h_2) = 4 \, \mathrm{Re}
\int^{\infty}_{f_{\mathrm{min}}}\dfrac{\tilde{h}_1(f)\tilde{h}_2^*(f)}{S_n(f)} 
df.
\end{equation}
Here, $S_n(f)$ denotes the one-sided power spectral density of the noise in the
interferometer. In this work, we model $S_n(f)$ with the \ac{aLIGO}
zero-detuned, high-power design sensitivity curve~\cite{Harry:2010zz} and use a
lower frequency cutoff, $f_{\mathrm{min}}$, of 15 Hz.

As gravitational-wave searches for binary mergers analytically maximize over an
overall phase and time shift, we define the ``match'' between two waveforms
to be the overlap maximized over a phase and time shift
\begin{equation}
\mathcal{M}(h_1,h_2) =
\underset{\phi_c,t_c}{\max}(\hat{h}_1|\hat{h}_2(\phi_c,t_c)).
\end{equation}
One can understand this match as the fraction of the optimal \ac{SNR} that would
be recovered if a template $h_1$ was used to search for a signal $h_2$.

We define the ``fitting factor'' between a waveform $h_s$ with unknown
parameters
and a bank of templates $h_b$ to be the maximum match between $h_s$ and all the
waveforms in the template bank~\cite{Apostolatos:1995pj}:
\begin{equation}
\mathrm{FF}(h_s) = \max_{h \in \{h_b\}} \mathcal{M}(h_s,h).
\end{equation}
The mismatch
\begin{equation}
\mathrm{MM} = 1 - \mathrm{FF}(h_s)
\end{equation}
describes the fraction of \ac{SNR} that is lost due to the fact that the
template in the bank that best matches $h_s$ will not match it exactly due to
the discreteness of the bank and due to any disagreement between the waveform 
families used to model the templates and the signals. In previous searches of 
LIGO and Virgo data using
nonspinning template banks, the banks of signals were constructed so that the
fitting factor would be greater than 0.97 for any nonspinning signal within
the parameter space \cite{Babak:2012zx}. This was chosen as a balance between
detection efficiency and computational cost. We also construct our 
aligned-spinning banks with this criterion.

For the precessing \ac{NSBH} signals that we consider in this work, the fitting
factor will depend on the masses, spin magnitudes, spin orientations, sky
locations and orientation of the \ac{NSBH} system used to produce the waveform
$h$. We only sample the fitting factor at discrete points corresponding to
the distribution of systems described in Sec. \ref{sec:nsbhpop}. When showing
results from this set of fitting factors we often do so as a function of only 
two
of the various parameters that the fitting factor depends on. When doing this
we split the set of fitting factors into a series of bins corresponding to 
ranges in both of the parameters we are interested in. For each bin we then 
calculate an ``average fitting factor'' within that bin. This is done by taking
the mean value of the fitting factor from all points within each bin
\begin{equation}
 \textrm{FF}_{\textrm{av}} =  \left\langle \textrm{FF}
\right\rangle,
\end{equation}
where $\left\langle X \right\rangle$ denotes the mean average of $X$.
However, this measure can often be misleading.
The \ac{aLIGO} detectors have a direction-dependent and
orientation-dependent sensitivity. Systems that are poorly aligned with respect
to the detector may not have sufficient \ac{SNR} to be detected, regardless of
the fitting factor. A number of systems for which precessional effects are most
prominent are ones in which the precessing orbital plane moves through points
where the detector has very little sensitivity~\cite{Brown:2012gs}.
To account for this we make use of the ``effective fitting factor,''
first defined in \cite{Buonanno:2002fy} as
\begin{equation}
 \textrm{FF}_{\textrm{eff}} = \left( 
 \frac{ \left\langle \textrm{FF}^3 \sigma_i^3 \right\rangle }{\left\langle
\sigma_i^3 \right\rangle} \right)^{1/3}.
\end{equation}
Here $\sigma_i = \sqrt{(h_i|h_i)}$, which describes the optimal \ac{SNR}
of $h_i$. For each bin, the cube of the effective fitting factor gives, above an
arbitrary \ac{SNR} threshold, the ratio between the fraction of \ac{NSBH}
signals that would be recovered with the discrete template bank that was used
and a theoretical continuous template bank that would recover 100\% of signal
power for \emph{any} \ac{NSBH} waveform. We therefore define the ``signal
recovery fraction'' as $\textrm{FF}_{\textrm{eff}}^3$.

\section{A new algorithm for constructing template banks of aligned-spin NSBH 
waveforms}
\label{sec:bank_method}

In \cite{Brown:2012qf} we proposed a method for generating a
geometrically placed bank of aligned-spin
systems that can be used to search for \ac{BNS} systems in the advanced detector
era. In this section we adapt the methods presented in that work to the case of
\ac{NSBH} systems and describe how to generate template banks that can recover
aligned-spin \ac{NSBH} waveforms. These banks are applicable for waveforms
modeled using either the TaylorT2 approximant or the TaylorT4 approximant.

A bank of templates should be placed such that any putative signal
within the parameter space of interest would be recovered with a loss in SNR
that is always
less than some predefined value, usually taken to be 3\%
\cite{Poisson:1995ef,Owen:1995tm,Owen:1998dk,Babak:2006ty,
Cokelaer:2007kx,Babak:2012zx}.
To determine the maximum spacing between templates that meets this criterion,
the parameter space metric is used. This approximates the distance between any
two points that are close in the parameter space \cite{Owen:1995tm}
\begin{equation}
\label{eq:cbc_metric2}
\mathcal{O}(h(\bm{\theta}),h(\bm{\theta}+\delta\bm{\theta})) =
  1 - \sum_{ij} g_{ij}(\bm{\theta}) \,\delta\theta^i \,\delta\theta^j,
\end{equation}
with the metric given by
\begin{equation}
\label{eq:cbc_metric}
g_{ij}(\boldsymbol{\theta}) = - \frac{1}{2} \frac{\partial^2
\mathcal{O}}{\partial \delta\theta^i \partial
\delta\theta^j} = \left(\frac{\partial h(\boldsymbol{\theta})}{\partial
\theta^i} \bigg|
\frac{\partial h(\boldsymbol{\theta})}{\partial \theta^j}\right).
\end{equation}
Here $\bm{\theta}$ describes the parameters of the signal, in this case the
masses and the spins. This is also commonly referred to as the Fisher 
information matrix. 

Obtaining an analytic solution for Eq. (\ref{eq:cbc_metric}) is much simpler in 
the frequency domain and therefore frequency-domain waveform models are 
commonly used when placing a bank of 
templates~\cite{Owen:1995tm,Owen:1998dk,Babak:2012zx}. We follow that approach 
and consider only frequency-domain metrics here. It is important to carefully 
consider which
coordinates to use as parameters when using this metric as an approximation to
the parameter space distance. If one were to naively use the masses and spins
directly as coordinates it
would result in a parameter space metric with a large amount of extrinsic
curvature, and Eq.~(\ref{eq:cbc_metric2})
would only be valid for small ranges of $\delta\theta^i$. In previous searches
for nonspinning systems,
the ``chirp times'' were used \cite{Owen:1998dk}, defined as
\begin{subequations}
\begin{align}
 \tau_0 &= \frac{5}{128} \left(\pi M \right)^{-5/3} \eta^{-1} \\
 \tau_3 &= \frac{\pi}{4} \left(\pi M \right)^{-2/3} \eta^{-1},
\end{align}
\end{subequations}
as these are the two combinations of the masses that minimize extrinsic
curvature.

When the template waveforms include spin it is difficult to identify a
parameterization of the waveform for which the metric is locally flat.
Instead, in~\cite{Brown:2012qf} we constructed a metric
that uses the various coefficients of the expansion of the orbital
phase, given by the various $\lambda_i$ terms in Eq.~(\ref{eq:phase_exp}),
directly as coordinates. Using these coordinates, the parameter space is
globally flat. However, for the TaylorF2 metric including terms up to
3.5\ac{PN} order, the parameter space is eight-dimensional. The physical 
subspace forms a four-dimensional manifold within this parameter 
space.

To deal with the increased dimensionality of the space we perform two
coordinate transformations~\cite{Brown:2012qf,Ohme:2013nsa}. These two 
coordinate
transformations map points from the $\lambda_i$ coordinates into a
Cartesian coordinate system where the principal directions are mapped using
coordinates denoted by $\xi_i$. Specifically, the first coordinate 
transformation uses the eigenvectors and
eigenvalues of the $\lambda_i$ metric to transform to a Cartesian
coordinate system. A principal component analysis is then performed to rotate
into the frame given by the principal directions of the manifold describing 
the physical range of masses and spins we 
consider within the eight-dimensional parameter space. In 
this Cartesian coordinate system of principal directions we can
assess the effective dimension of the parameter space, i.e., the number of
directions in which templates actually need to be placed in order to achieve the
desired coverage. For the case of the \ac{BNS} parameter space with the 
\ac{aLIGO} power spectral density we found that many of the directions had an 
extremely small 
extent and
could be neglected entirely. We found that a two-dimensional lattice could
efficiently cover the entire space of aligned-spin \ac{BNS}
waveforms~\cite{Brown:2012qf}.

Our geometrical placement method is not specific to the \ac{BNS} area of the
parameter space. However, some modifications to the method were necessary when
placing a template bank of \ac{NSBH} waveforms. Our \ac{BNS} aligned-spin
template bank, as described in~\cite{Brown:2012qf}, was given in terms of the
positions of the points in the eight-dimensional Euclidean parameter space 
$\xi_i$.
These points do not correspond directly to physical masses and spins. For this
study we want to use time-domain template families and therefore we
must translate the bank into physical parameters. However, if a set of $\xi_i$ 
values is given it will, in general, not be possible to find a set of masses 
and spins that give the exact $\xi_i$ values. As 
templates are normally placed in a two-dimensional lattice, we need only to 
find a physical point that has the corresponding values of $\xi_1$ and $\xi_2$ 
and \emph{any} value of the other $\xi_i$ values that correspond to a waveform 
within the physically allowed manifold. For some cases where a two-dimensional 
lattice is not sufficient to cover the space we will also specify values of 
$\xi_3$ and $\xi_4$. We attempt to find a 
physical solution that is sufficiently close to the desired point using a
numerical solution. We generate a large set of points in
the mass and spin space and map these points to the $\xi_i$ parameters. For 
each template we then find the closest point from our large 
set of physical points.
We then proceed to iteratively test physical points in the vicinity to find a
match of at least 0.9999 with the intended position.
If the template is within the physically allowed parameter space we can 
generally find a physical point that has the desired match with the intended 
$\xi_i$ point.
Templates on the boundaries of the space, might have an
overlap as low as $\sim 0.97$ with the edge of the physical parameter space.
Our method pushes such points back into
the desired physical space thereby providing a slight \emph{improvement} in the
bank coverage. This method also provides an easy method to determine the extent 
of the physical space: if \emph{no} physical point is found with 0.97 or higher 
match with the $\xi_i$ position then that point is not within the physical 
extent of the parameter space and no template needs to be placed there.

The downside to our brute-force numerical method is that it is currently not
computationally efficient; generating a bank with this numerical technique can 
take $O(10)$ hours when running on $\sim 500$
CPU cores. The cost of placing a bank using this method, however, is negligible
when compared to the cost of filtering data against a bank of templates if a 
single bank is used to filter \textit{O}(days) of data. If the bank is 
regenerated every 
hour, as in previous searches of LIGO and Virgo data~\cite{Babak:2012zx}, this 
cost would not be negligible.
We note that it should be possible to optimize our implementation to obtain a 
significant speed increase over what we quote above.

The TaylorF2 metric can be used to place a bank of waveforms modeled with the
TaylorT2 approximant. However, we also require that our template placement
algorithm place a bank of waveforms that can detect aligned-spin signals
modeled using TaylorT4 with no more loss in \ac{SNR} than that specified by 
the minimal match of the bank. This will allow us to investigate the efficiency 
of aligned-spin banks to search for precessing \ac{NSBH} signals using two 
waveform models. Using two models will help to mitigate any bias in our 
results that arises due to the choice of waveform approximant.
We investigate the distribution of fitting factors when using a template bank
constructed using the TaylorF2 metric to search for aligned-spin TaylorT4
\ac{NSBH} signals in Sec. \ref{sec:bank_validation} and find that this would
result in a reduction of sensitivity. We therefore make use of
a metric that models the TaylorT4 waveform well. To do this we use the
TaylorR2F4 waveform model. We have found that restricting the TaylorR2F4 model
to terms no larger than 4.5\ac{PN} and placing a bank of templates using the
ensuing metric is sufficient to cover the TaylorT4 parameter space. This is
a 12-dimensional metric. We then perform the same rotations as for the TaylorF2
metric to identify the $\xi_i$ directions for our TaylorR2F4 parameter
space and proceed in the same manner as described above.

In contrast to \ac{BNS} mergers, \ac{NSBH} systems can merge in
the sensitive band of the advanced detectors. Existing nonspinning
template placement algorithms
\cite{Poisson:1995ef,Owen:1995tm,Owen:1998dk,Babak:2006ty,
Cokelaer:2007kx} as well as our aligned-spin algorithm must use the same 
termination frequency when modeling waveforms across the parameter space. 
The standard approach is to assume that the waveforms will follow the TaylorF2, 
or TaylorR2F4, evolution up to the Nyquist frequency, usually 2048 Hz. For 
\ac{BNS} systems, the merger generally occurs above 1000 Hz where the 
sensitivity of
gravitational wave interferometers falls off and therefore little power is
incurred between 1000 Hz and Nyquist. Even a $(3 + 3)M_{\odot}$ \ac{BNS}
has an \ac{ISCO} with a frequency of 730 Hz. In contrast, a $(15 + 3)M_{\odot}$
\ac{NSBH} system has an \ac{ISCO} frequency at 240 Hz. We must therefore
consider what frequency cutoff is most appropriate to use when placing a bank
of \ac{NSBH} waveforms.

We found that using an upper frequency cutoff that is higher than the waveform's
termination frequency results in overcoverage in the parameter space. This 
result is expected as the subdominant \ac{PN} terms can have a
significant effect in the late part of the evolution, causing
systems with the same chirp masses but different spins and mass ratio to diverge
faster. Therefore we use an upper frequency cutoff of 1000 Hz for all waveforms
within the \ac{NSBH} parameter space to generate a template bank that will cover
to the desired minimal match. However, as this template bank will overcover
at least the high-mass end of the parameter space we also investigate the
efficiency of banks placed with smaller upper frequency cutoffs in Sec.
\ref{ssec:stoch_fup_compare}. This choice will be an important consideration
in the advanced detector era given limits on computational power for conducting
\ac{NSBH} searches.

\section{Constructing template banks of aligned-spin NSBH waveforms with our 
new algorithm}
\label{sec:bank_construction}

\begin{table*}
    \centering
    \begin{minipage}[l]{2.0\columnwidth}
    \centering
\begin{tabular}{c | c | c | c}
 Template bank & Approximant & Waveform cutoff frequency & Number of templates
in bank \\ \hline \hline
 Geometric nonspinning bank & TaylorF2 & 1000 Hz & 117,632 \\
 Geometric nonspinning bank & TaylorR2F4 (up to 4.5PN) & 1000 Hz & 99,309 \\
 Geometric aligned-spin bank & TaylorF2 & 1000 Hz & 817,460 \\
 Geometric aligned-spin bank & TaylorF2 & 400 Hz & 432,537 \\
 Geometric aligned-spin bank & TaylorF2 & 240 Hz & 282,090 \\
 Stochastic aligned-spin bank & TaylorF2 & Dynamic & 971,105 \\
 Geometric aligned-spin bank & TaylorR2F4 (up to 4.5PN) & 1000 Hz & 1,100,277 \\
 Geometric aligned-spin bank & TaylorR2F4 (up to 4.5PN) & 400 Hz & 504,132 \\
 Geometric aligned-spin bank & TaylorR2F4 (up to 4.5PN) & 240 Hz & 260,325 \\
 Stochastic aligned-spin bank & TaylorR2F4 (up to 4.5PN) & Dynamic & 1,327,175\\
\end{tabular}
\caption{\label{tab:banksizes}
The sizes of the various template banks that are used in this work. All of these
banks are valid for
aligned-spin \acp{NSBH} with BH mass $\in [3,15) M_{\odot}$; NS mass $\in 
[1,3)M_{\odot}$; BH
dimensionless spin $\in [-1,1]$;
NS dimensionless spin $\in [-0.05,0.05]$. For all banks the \ac{aLIGO}
zero-detuned, high-power noise curve is used with a lower frequency cutoff of
15 Hz.
}
\end{minipage}
\end{table*}

\begin{figure*}
    \centering
    \begin{minipage}[l]{2.0\columnwidth}
    \centering
\includegraphics[width=0.45\textwidth]
{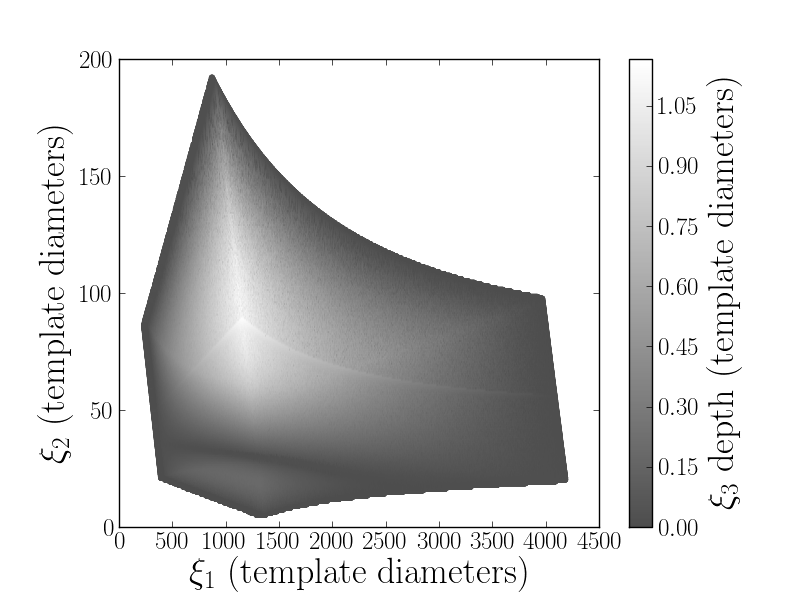}
\includegraphics[width=0.45\textwidth]
{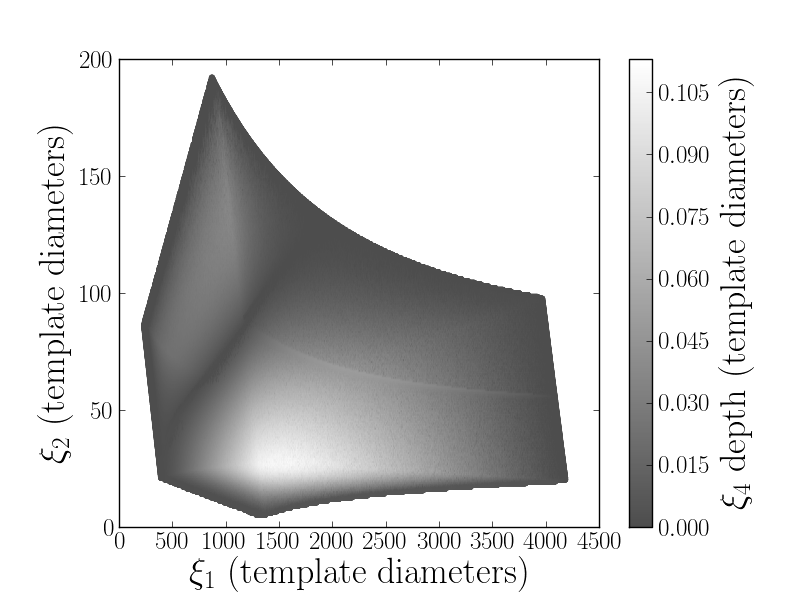}
\caption{\label{fig:bankF2depths}
The depth of the physically possible range of $\xi_3$ (left) and $\xi_4$ (right)
values as a function of
$\xi_1$ and $\xi_2$ shown for the TaylorF2 \ac{NSBH} parameter space.
The $\xi_i$ coordinates have been scaled
such that one unit corresponds to the coverage diameter of a template
at 0.97 mismatch. Shown using the zero-detuned, high-power advanced LIGO
sensitivity curve with a 15 Hz lower frequency cutoff
and a 1000 Hz upper frequency cutoff.}
\end{minipage}
\end{figure*}

We begin by creating a template bank using the TaylorF2 parameter space metric.
We first explore the space to assess the effective dimensionality and to
determine whether the two-dimensional placement used to cover the \ac{BNS}
space in~\cite{Brown:2012qf} is applicable to the \ac{NSBH}
space. We do this by creating a set of $10^7$ points drawn uniformly from the
chosen range of \ac{NSBH} masses and spins. We then transform these points
into the $\xi_i$ coordinates. In Fig.~\ref{fig:bankF2depths} we show the
extent of the dominant two directions ($\xi_1$ and $\xi_2$). The color shows,
respectively, the depth of the third direction ($\xi_3$) and the fourth
direction ($\xi_4$). The fifth and subsequent directions are, as in the \ac{BNS}
space, small enough to be ignored completely.

From these plots we can see that the extent of the space in all but the $\xi_1$
and $\xi_2$ directions is small in most regions. In these areas a
two-dimensional lattice of template points would suffice to cover the parameter
space. However, there is a
small region in the center of the parameter space where the depth of the third
direction is not negligible. Therefore, to cover this space we
follow~\cite{Brown:2012qf} and initially place a two-dimensional lattice in the
$\xi_1$, $\xi_2$ coordinates. Then, where necessary, templates are stacked in
the $\xi_3$ direction. The density of this stacking is chosen such that the loss
in match due to the depth of the third direction can never be larger than 0.01.
As the two-dimensional lattice is placed to ensure that matches will not be less
than 0.97 in a two-dimensional plane, and as each direction in our Euclidean 
parameter space is orthogonal, there are therefore regions of the parameter
space where the fitting factor can be as low as 0.96. However, these regions are
small and the mean fitting factor, as we will show, is still much larger than
0.97. This bank, constructed using the TaylorF2 parameter space metric, contains
801,183 templates, of which 134,807 were added by the stacking process. For
ease of comparison Table \ref{tab:banksizes} gives the sizes and properties of
all the banks that are used in this work.

\begin{figure*}
    \centering
    \begin{minipage}[l]{2.0\columnwidth}
    \centering
\includegraphics[width=0.45\textwidth]
{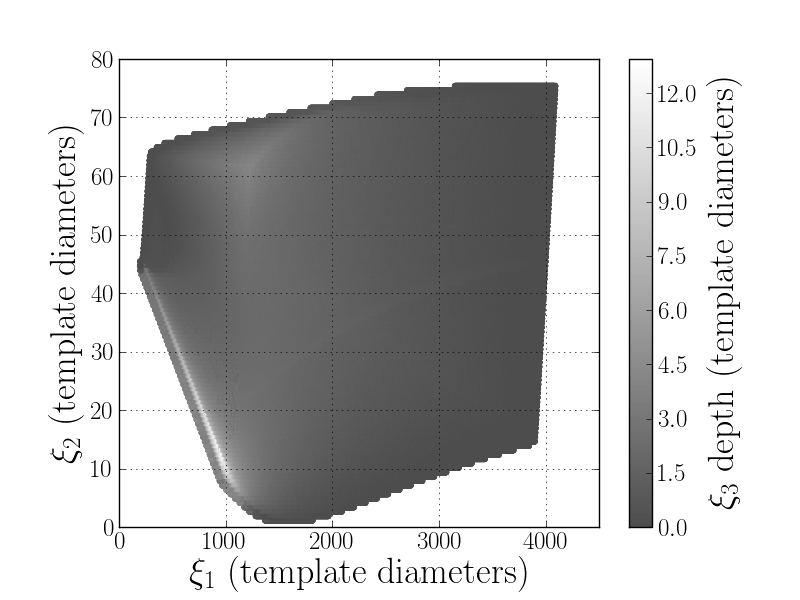}
\includegraphics[width=0.45\textwidth]
{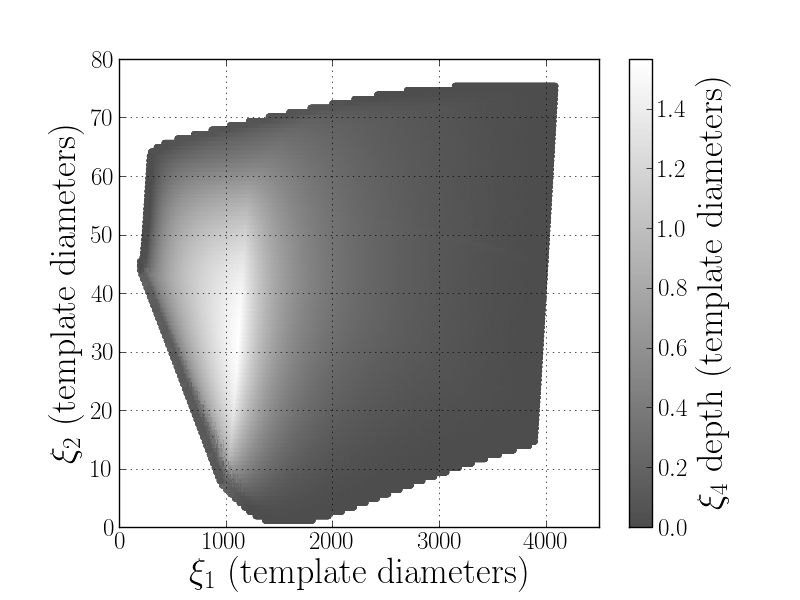}
\caption{\label{fig:bankF4depths}
The depth of the
physically possible range of $\xi_3$ (left) and $\xi_4$ (right) values as a
function of
$\xi_1$ and $\xi_2$ shown for the TaylorR2F4 \ac{NSBH} parameter
space. The $\xi_i$ coordinates have been scaled
such that one unit corresponds to the coverage diameter of a template
at 0.97 mismatch. Shown using the zero-detuned, high-power advanced LIGO
sensitivity curve with a 15 Hz
lower frequency cutoff and a 1000 Hz upper frequency cutoff.}
\end{minipage}
\end{figure*}

We next construct a bank of template waveforms using the TaylorR2F4 parameter
space metric.
We begin by exploring the parameter space to assess the effective
dimensionality.
In Fig.~\ref{fig:bankF4depths} we show the depths of the $\xi_3$ and $\xi_4$
directions as a function of $\xi_1$ and $\xi_2$ for the TaylorR2F4 parameter
space. We immediately notice that the degeneracies present in the TaylorF2
space, which allow us to use a two-dimensional placement,
are much weaker in the TaylorR2F4 parameter space. For this space
there is substantial depth in the third direction. In one small region it is
wider than ten template diameters. The median depth in this direction, however,
is only one template diameter.

If the depth in the third direction was larger in all regions, the most
efficient placement scheme would be to place a template bank in a 
three-dimensional
$A_n^{\star}$ lattice \cite{Conway:1993}. However, in regions where the depth of
the third direction is small, the three-dimensional lattice, when flattened 
into the 
two-dimensional space, would cause an overcoverage. We therefore tried both a
three-dimensional lattice placement and a two-dimensional placement, followed
by stacking in the third direction as we used for the TaylorF2 bank. 
Additionally,
unlike in the TaylorF2 space, the depth of the fourth dimension is not
negligible. However, as in most places the width in that direction is small, the
stacking technique can also be used to cover the depth of the fourth dimension 
when needed.

When we choose to employ a three-dimensional lattice we find that 1,805,036
templates are needed to cover the space, 90,463 of which were added due to
stacking in the fourth direction. In contrast, when we use a hexagonal lattice
followed by stacking in both the third and fourth directions we find that 
1,100,277
templates are needed, of which 741,626
were added by the stacking process. It may seem surprising that the 2D
hexagonal lattice requires less templates
than the 3D $A_n^{\star}$ lattice. In fact, it would still require less
templates even if the depth of the third
direction was large in all regions of the space. The reason for this is that the
$A_n^{\star}$ placement \emph{guarantees}
that all points within the three-dimensional space will have a fitting factor 
of at
least 0.97. With the
hexagonal placement followed by stacking, there are points in the space where
the fitting factor can be as low as 0.96
(when the depth of the fourth dimension is significant this can be as low as 
0.95).
If we were to require that all points
within the space \emph{must} have a fitting factor of at least 0.97, our
hexagonal lattice would need to be placed
to a minimal match of 0.98.
For comparison, we generated a three-dimensional lattice with a minimal match 
of 
0.96; this bank contained 1,175,523 templates. The three-dimensional lattice is 
still less efficient than the two-dimensional lattice. This can be attributed, 
as 
described above, to the fact that the depth of the third direction is not large 
in all areas of the parameter space. In some areas a two-dimensional lattice, 
without any stacking, is
sufficient to cover the parameter space. An alternative approach might be to use 
a three-dimensional lattice of points only in regions where it is needed and a 
two-dimensional lattice elsewhere; we did not investigate that here. For the 
simulations in the following sections, we use the hexagonal lattice with
stacking as the method for placing banks of templates for the TaylorT4
approximant.

\section{Results I: Validating the new template bank placement for 
aligned-spin systems}
\label{sec:bank_validation}

In this section we demonstrate that our aligned-spin template banks achieve the
level of coverage they are constructed for when used to search for aligned-spin
signals. We also compare our banks to banks generated using a stochastic
placement algorithm~\cite{Harry:2009ea,Babak:2008rb,Manca:2009xw,Ajith:2012mn}
and show that our method achieves the same level of coverage with fewer
templates. 

\begin{figure}
\includegraphics[width=0.45\textwidth]
{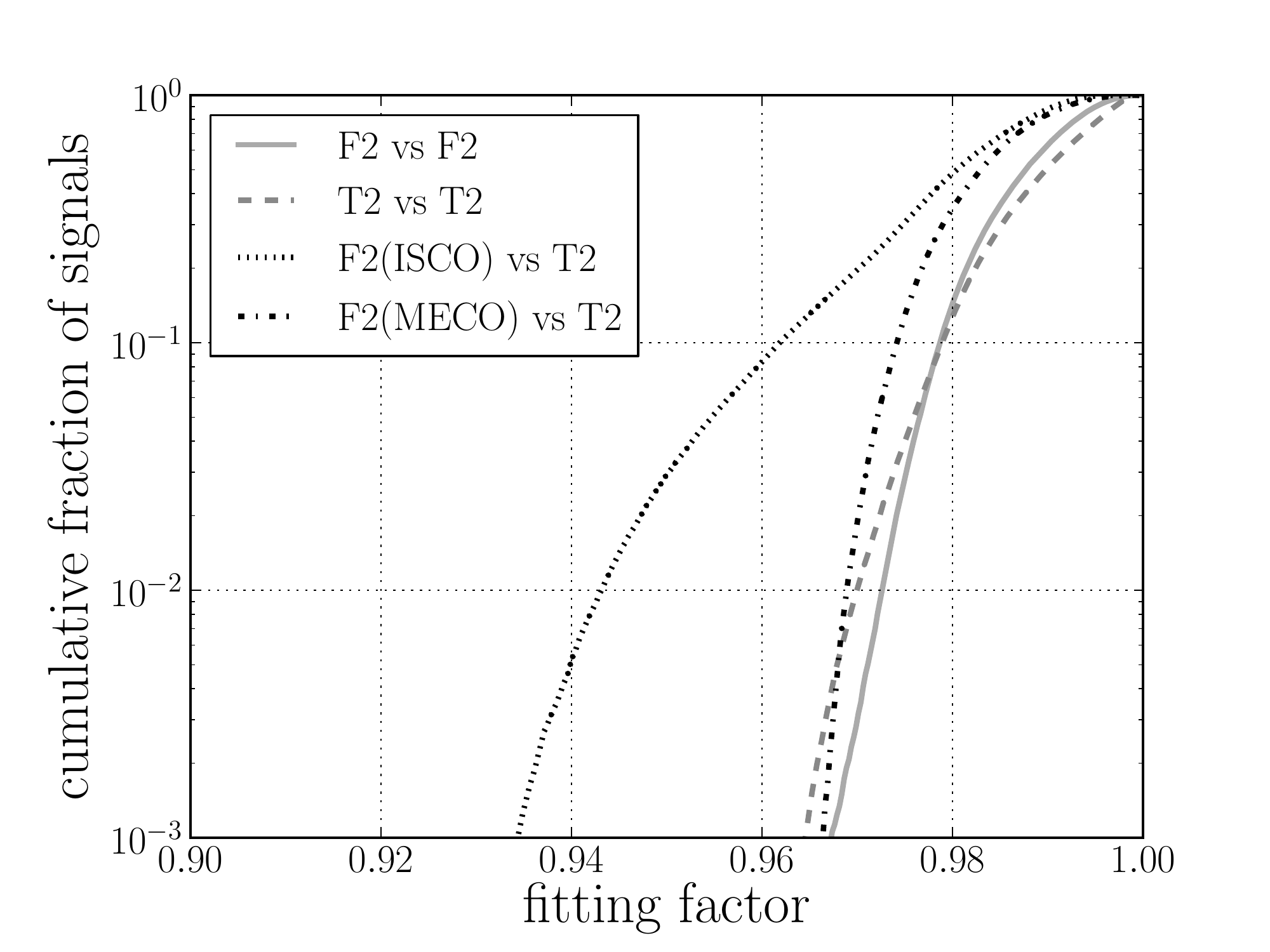}
\caption{\label{fig:bankF2verification}
Fitting factor between a set of aligned-spin \ac{NSBH} signals and our
geometrically placed aligned-spin template bank placed using the TaylorF2
metric. Shown when both templates and signals are generated using
the TaylorF2 approximant (gray solid line) and when both are modeled with
TaylorT2 (gray
dashed line). Also shown when the signals are modeled with TaylorT2 and the
templates modeled
with TaylorF2 waveforms terminated at ISCO (black dotted line) and TaylorF2
waveforms terminated at MECO (black dot-dashed line). Results obtained
using the zero-detuned, high-power advanced LIGO sensitivity curve with a 15 Hz
lower frequency cutoff.
}
\end{figure}

\begin{figure}
\includegraphics[width=0.45\textwidth]
{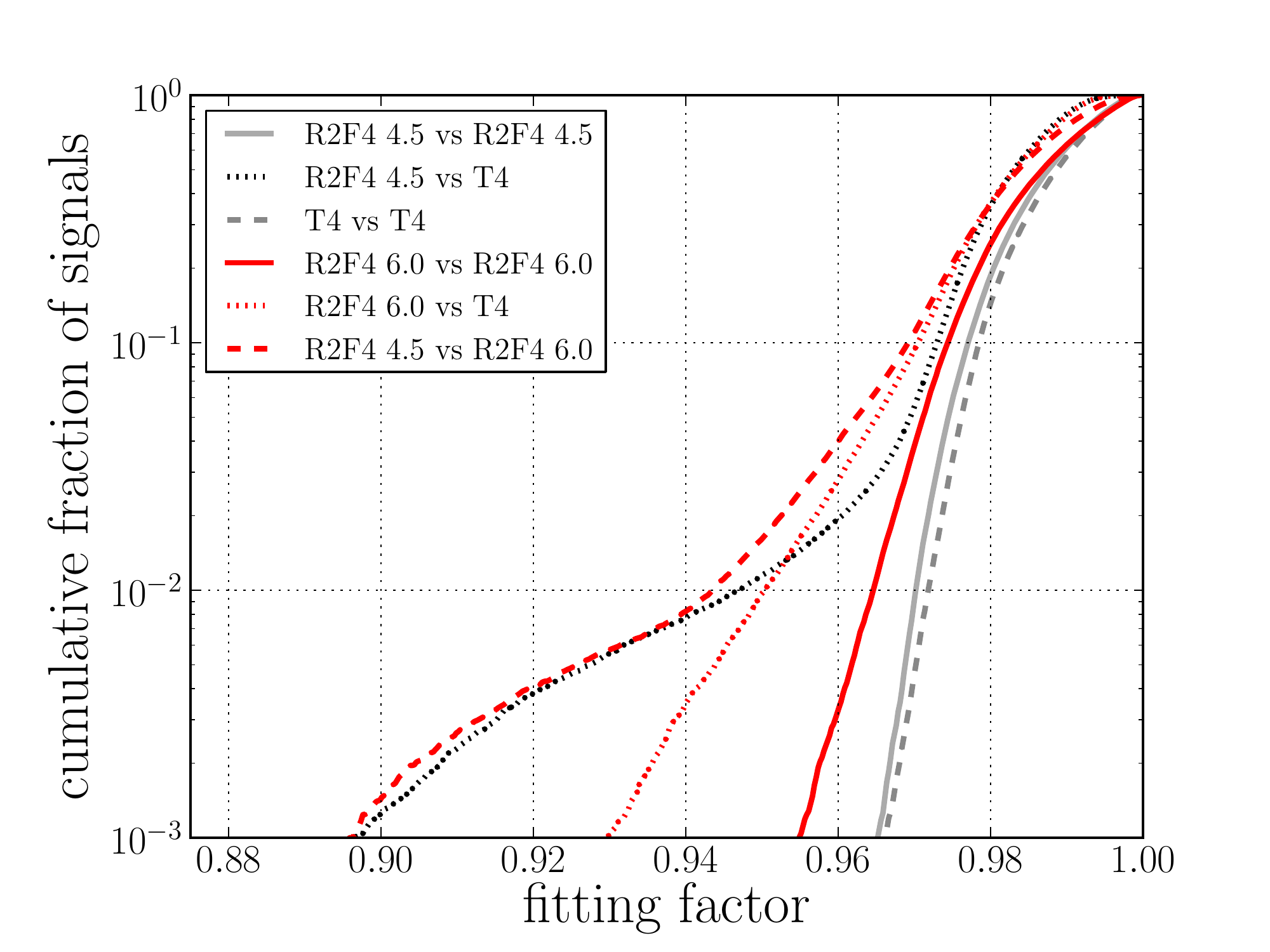}
\caption{\label{fig:bankF4verification}
Fitting factor between a set of aligned-spin \ac{NSBH} signals and our
geometrically placed aligned-spin template bank placed using the TaylorR2F4
metric. Shown are comparisons between
TaylorT4 waveforms, TaylorR2F4 waveforms
including terms to 4.5PN order and TaylorR2F4 waveforms including terms to 6PN
order. Results obtained
using the zero-detuned, high-power advanced LIGO sensitivity curve with a 15 Hz
lower frequency cutoff.
}
\end{figure}

To verify the performance of our aligned-spin template banks we compute the
fitting factors between the banks and a set of 100,000 aligned-spin \ac{NSBH}
waveforms. These waveforms are drawn from the
distribution that we describe in Sec. \ref{sec:nsbhpop}, except that the
spins are all aligned (or antialigned) with the orbital angular momentum.

In Fig.~\ref{fig:bankF2verification} we show the results of this test using
the template bank constructed with the TaylorF2 metric. We show results when
both template waveforms and signals are modeled using the TaylorF2
approximant, when both are modeled using the TaylorT2 approximant and when we
model the template waveforms with TaylorF2 and the signals with TaylorT2.
In both cases where the same waveform model was used almost all of the fitting
factors were greater than 0.97. The bank generation was successful.

The lowest matches in the TaylorF2 vs TaylorF2 results were in cases
where a system with low mass ratio was recovered with a template with a high
mass ratio, or vice versa. These are systems where the degeneracy between the
spins and the mass ratio \cite{Baird:2012cu} causes
the phase evolution of the two systems to be very similar and therefore the
match predicted by the metric is higher than 0.97.
However, the system with the larger black-hole mass will terminate at a 
significantly lower frequency than the system with the smaller black-hole mass
and some power is lost due to the difference in termination frequencies, which
is not predicted by the metric.

The difference in termination conditions is also the reason why we see
comparatively poorer performance when using TaylorF2 waveforms, terminated
at the \ac{ISCO} frequency, to search for TaylorT2 signals. The TaylorT2
signals terminate when the evolution becomes unphysical,
either at the \ac{MECO} or where the frequency spuriously begins to drop. In
some cases, especially when the spins
are large, these can correspond to rather different termination frequencies. To
demonstrate this we also show the performance
of searching for TaylorT2 signals with TaylorF2 waveforms,
but where we terminate the TaylorF2 waveforms using the same cutoff frequency
that TaylorT2 waveforms would have at the given masses and spins.
This gives a much more comparable performance to the TaylorF2 vs TaylorF2
and TaylorT2 vs TaylorT2 cases.

In Fig.~\ref{fig:bankF4verification} we repeat this test using
the template bank constructed with the TaylorR2F4 metric, with terms restricted
to 4.5\ac{PN} order. We show results when the template waveforms and signals
are modeled with varying approximants. We use TaylorR2F4 with terms up to
4.5\ac{PN} order, TaylorR2F4 with terms up to 6\ac{PN} order and TaylorT4.
We can see from this figure that using TaylorR2F4 template waveforms
with terms only to 4.5\ac{PN} order would not be satisfactory when conducting
searches for signals modeled with the TaylorT4 approximant. However, we note
that when this bank is used with either TaylorT4 templates or TaylorR2F4
templates including terms up to 6\ac{PN} order the coverage is much better.
When TaylorT4 is used to model both the signals and the template waveforms we
find that $>99\%$ of the fitting factors are greater than
0.97. In this plot the TaylorR2F4 waveforms are terminated at
the same frequency (the \ac{MECO} frequency) as the TaylorT4 waveforms.

The TaylorR2F4 metric, with terms up to
4.5\ac{PN}, is sufficient to place
a bank of templates to cover waveforms modeled by the TaylorT4 approximant.
However, when performing the
matched filtering the templates must be modeled with either TaylorT4 or
TaylorR2F4 with terms up to 6\ac{PN} order.

\begin{figure}
\includegraphics[width=0.45\textwidth]
{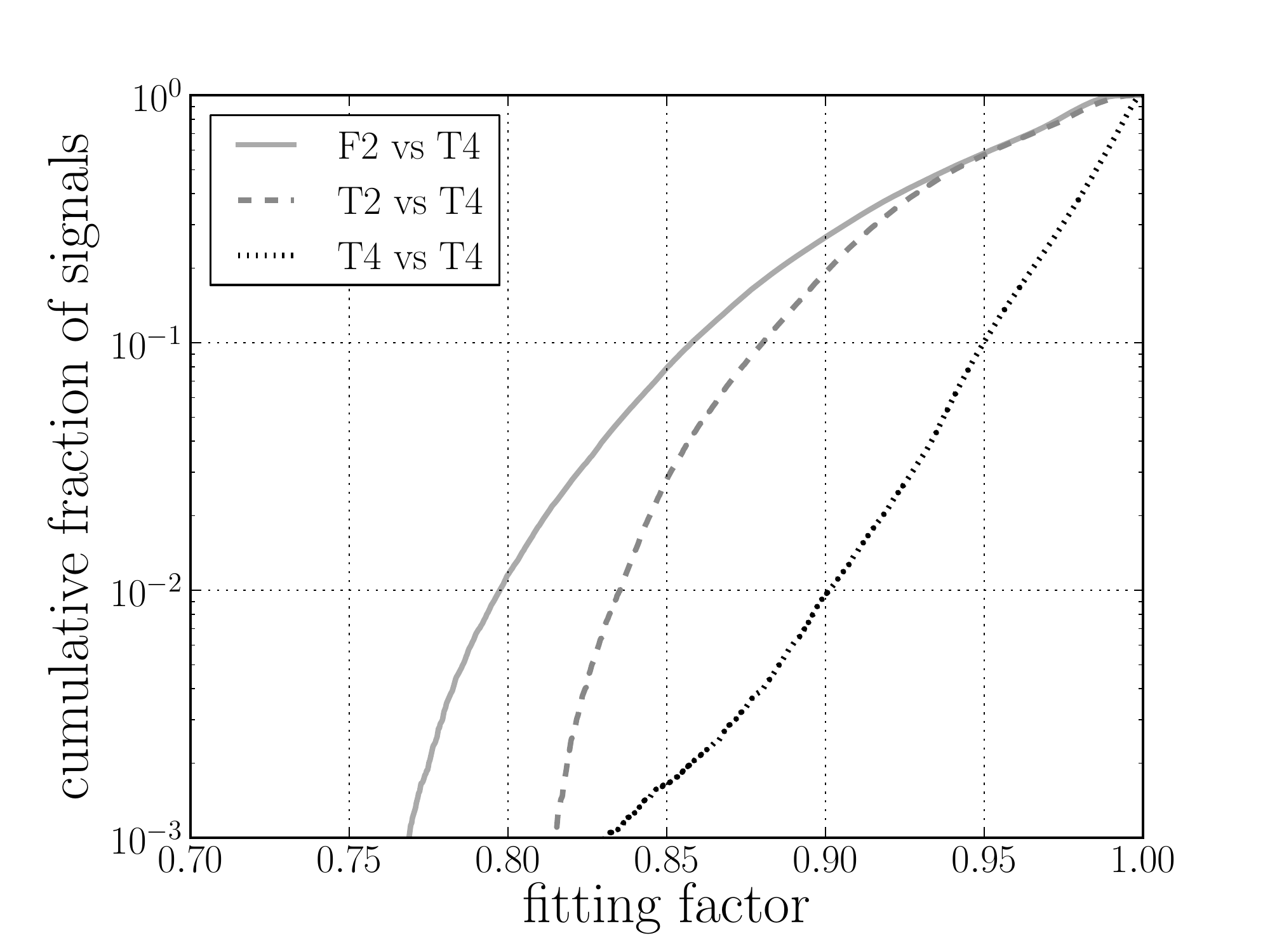}
\caption{\label{fig:bankF2T4testing}
Fitting factor between a set of aligned-spin \ac{NSBH} signals modeled with the
TaylorT4 approximant and our template bank of aligned-spin signals placed using
the TaylorF2 parameter space metric. Shown are the fitting factors when the
templates used are modeled using the TaylorF2 approximant (gray solid line),
TaylorT2 (gray dashed line) and TaylorT4 (black dotted line). Results obtained
using the zero-detuned, high-power advanced LIGO sensitivity curve with a 15 Hz
lower frequency cutoff.
}
\end{figure}

In Fig.~\ref{fig:bankF2T4testing} we also show the performance of a bank
placed using the TaylorF2 metric to search for TaylorT4 aligned-spin signals.
We assess the performance when the templates are modeled using TaylorF2,
TaylorT2 and TaylorT4 approximants. Even when TaylorT4 is used to model both
template waveforms and signals, $10\%$ of signals are recovered with fitting
factors smaller than 0.95. The TaylorF2 metric does not achieve the desired
coverage for TaylorT4 waveforms. In a companion work we investigate how the 
disagreement of different waveform families in the \ac{NSBH} region of 
parameter space will reduce detection efficiency~\cite{Nitz:2013mxa}.

\subsection{Varying the upper frequency cutoff and comparison with stochastic
placement algorithms}
\label{ssec:stoch_fup_compare}

\begin{figure*}
    \centering
    \begin{minipage}[l]{2.0\columnwidth}
    \centering
\includegraphics[width=0.45\textwidth]
{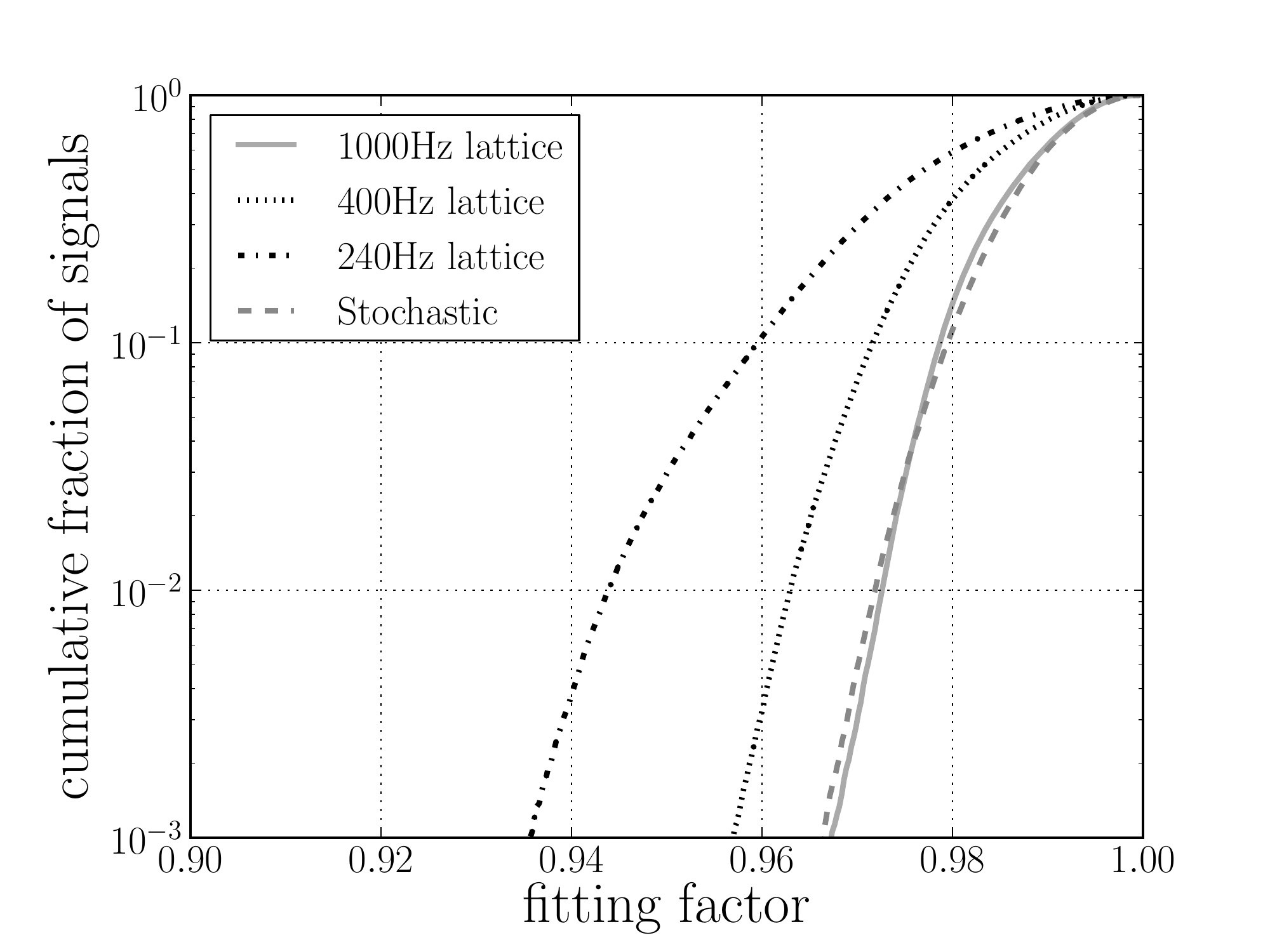}
\includegraphics[width=0.45\textwidth]
{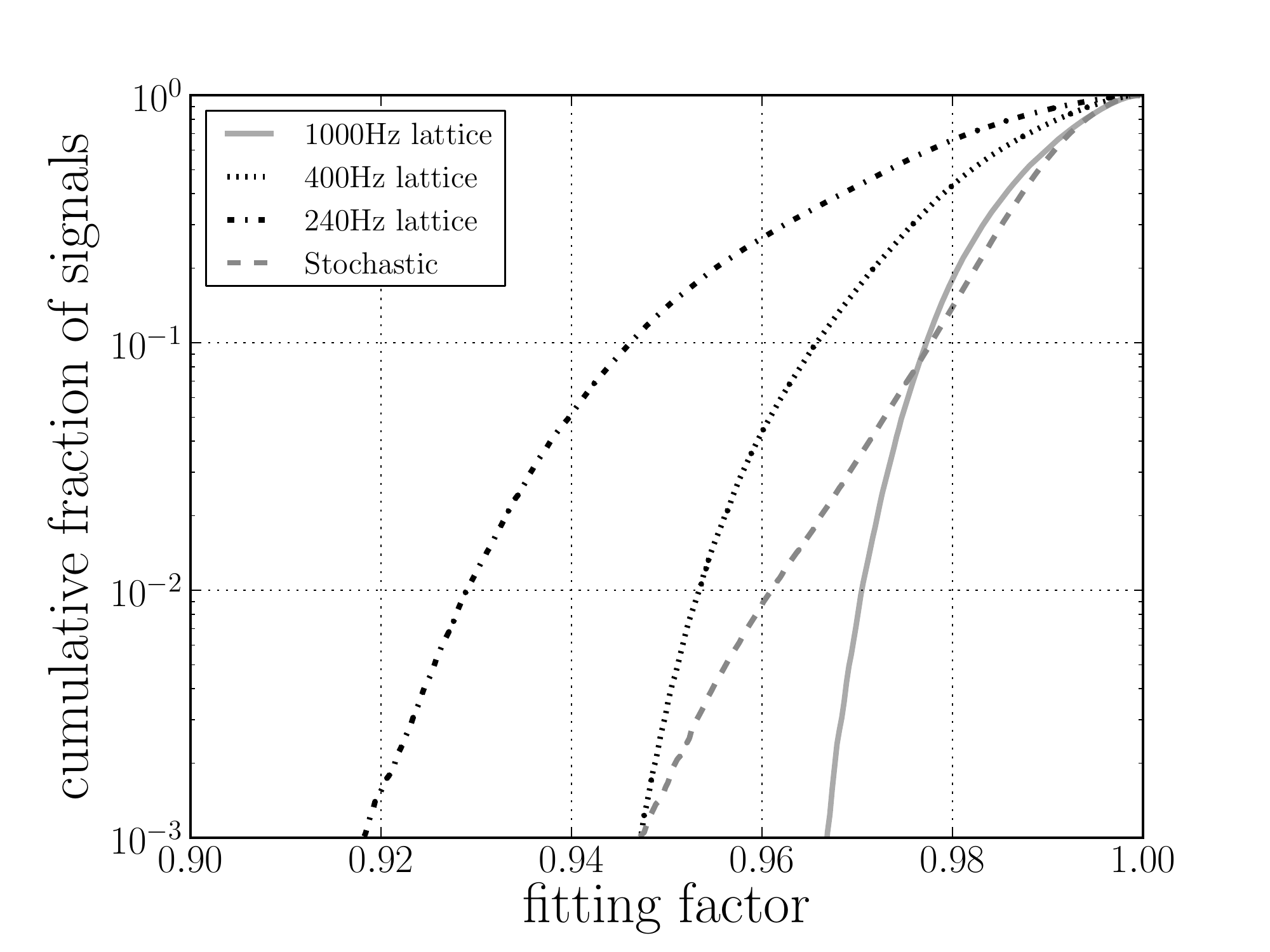}
\caption{\label{fig:bankfcutoff}
Fitting factor between a set of aligned-spin \ac{NSBH} signals
and a template bank of aligned-spin waveforms for varying values of the upper
frequency cutoff used in the construction metric. Shown for template banks
placed using the TaylorF2 metric and with both templates and signals 
modeled using the TaylorF2 approximant (left). Also shown for template banks
placed using the TaylorR2F4 metric and with both templates and signals 
modeled using the TaylorR2F4 approximant (right). The
performance of using a stochastically placed template bank with varying upper
frequency cutoff is also plotted. Results obtained
using the zero-detuned, high-power advanced LIGO sensitivity curve with a 15 Hz
lower frequency cutoff.
}
\end{minipage}
\end{figure*}

Filtering ${\sim}10^6$ templates against data from advanced
gravitational-wave detectors will require a large amount of computing power. It
would therefore be desirable if we could reduce the overcoverage that
is incurred in the high-mass region of the parameter space when using an upper
frequency cutoff of 1000 Hz. An alternative ``stochastic'' placement scheme,
based on randomly picking points in the space and only retaining points which
are not close to points already in the bank
\cite{Harry:2009ea,Babak:2008rb,Manca:2009xw}, is capable of using an upper
frequency cutoff that varies with mass \cite{Ajith:2012mn}. However, this
method is known to pack templates more densely than a geometrical
lattice~\cite{Harry:2009ea}.
We found that using a stochastic method to cover this \ac{NSBH} space with the
same covering criterion required 971,105 (1,327,175) templates when using
the TaylorF2 (TaylorR2F4) metric to place the bank. In both cases this is $\sim
20\%$ larger than our geometric algorithm using a constant upper frequency
cutoff of 1000 Hz. It is also possible to generate the geometric bank with a
lower upper frequency cutoff. This will require less
templates but will not reach the desired coverage in the lower-mass regions of
the parameter space. In Fig.~\ref{fig:bankfcutoff} we compare the efficiency
of geometric banks placed using a 240, 1000 and 400 Hz upper frequency
cutoff. These correspond to roughly the lowest possible \ac{ISCO}
frequency, the highest and an ``average'' system. The sizes of these banks are
shown in Table \ref{tab:banksizes}. As expected we notice a number of systems
recovered with fitting factors less than 0.97 when the upper frequency cutoff is
reduced. We also compare with the performance of a
stochastic placement algorithm, which uses a varying upper frequency cutoff.
The performance of the
stochastic bank is very comparable to the 1000 Hz bank when using the TaylorF2
metric. When using the TaylorR2F4 metric  the stochastic bank, which was placed
using $10^9$ seed points, seems to be struggling to achieve the necessary
coverage in certain regions of the space. As the stochastic placement algorithm
only uses a finite number of sample points, it is known that it can leave holes
in the parameter space, resulting in undercoverage~\cite{Harry:2009ea}.

We plan to adapt the geometric placement algorithm to allow the upper frequency
cutoff to vary over the space; however, we leave this investigation for future
work. We note that the minimal match and \emph{lower} frequency cutoff of the
bank can also be modified to reduce the number of templates and balance the
computational cost~\cite{Keppel:2013yia}.

\section{Results II: Template bank performance when searching for generic NSBH 
signals}
\label{sec:resultsII_generic}

In this section we evaluate the efficiency of searching for generic
\ac{NSBH} systems using template banks of nonspinning waveforms. Template 
banks of nonspinning waveforms were used to search for \ac{NSBH} 
signals in data from LIGO and Virgo's most recent science runs 
\cite{Abbott:2009tt,Abbott:2009qj,Abadie:2010yba,Abadie:2011nz}.
We demonstrate that ignoring the effects of spin when conducting searches
for \ac{NSBH} systems in the advanced detector era will significantly decrease
the rate of \ac{NSBH} observations and impose a selection bias against systems
with large spins and large $m_{BH} / m_{NS}$. We then evaluate the efficiency 
of 
searching for generic \ac{NSBH} systems using our new template bank of 
aligned-spin waveforms. We calculate the improvement gained by using our new 
bank when compared to a nonspinning bank.

\subsection{Performance of nonspinning template banks when searching for 
generic
NSBH signals}
\label{sec:non_spinning}

We compute fitting factors between a set of 100,000 generic, precessing
\ac{NSBH} signals and a bank of nonspinning template waveforms. The
precessing signals are drawn from the distribution
that we describe in Sec. \ref{sec:nsbhpop}.
To mitigate any bias that arises due to the choice of waveform approximant
we run the simulation twice. First we use the TaylorT2 approximant for both
signal and template waveforms and a template bank designed to obtain
a fitting factor of at least 0.97 for any TaylorT2 non-spinning signal.
The simulation was then repeated using the TaylorT4 approximant for both signal
and template waveforms and a bank designed with the same fitting factor
criterion for TaylorT4 signals. These banks were constructed using the methods
described to create aligned-spin banks in section \ref{sec:bank_construction}
but with the spins set to 0.

\begin{figure*}
    \centering
    \begin{minipage}[l]{2.0\columnwidth}
    \centering
\includegraphics[width=0.45\textwidth]
{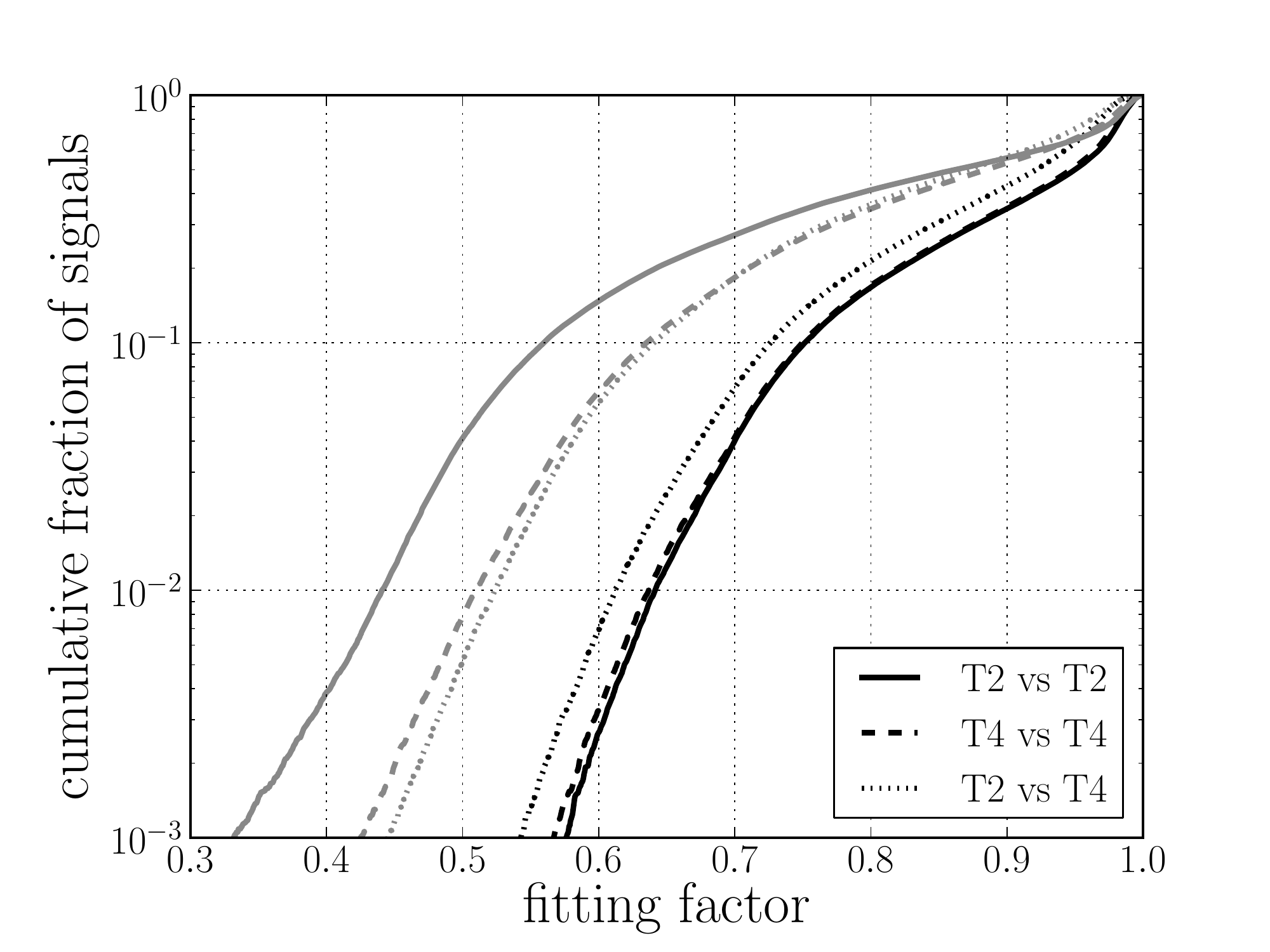}
\includegraphics[width=0.45\textwidth]
{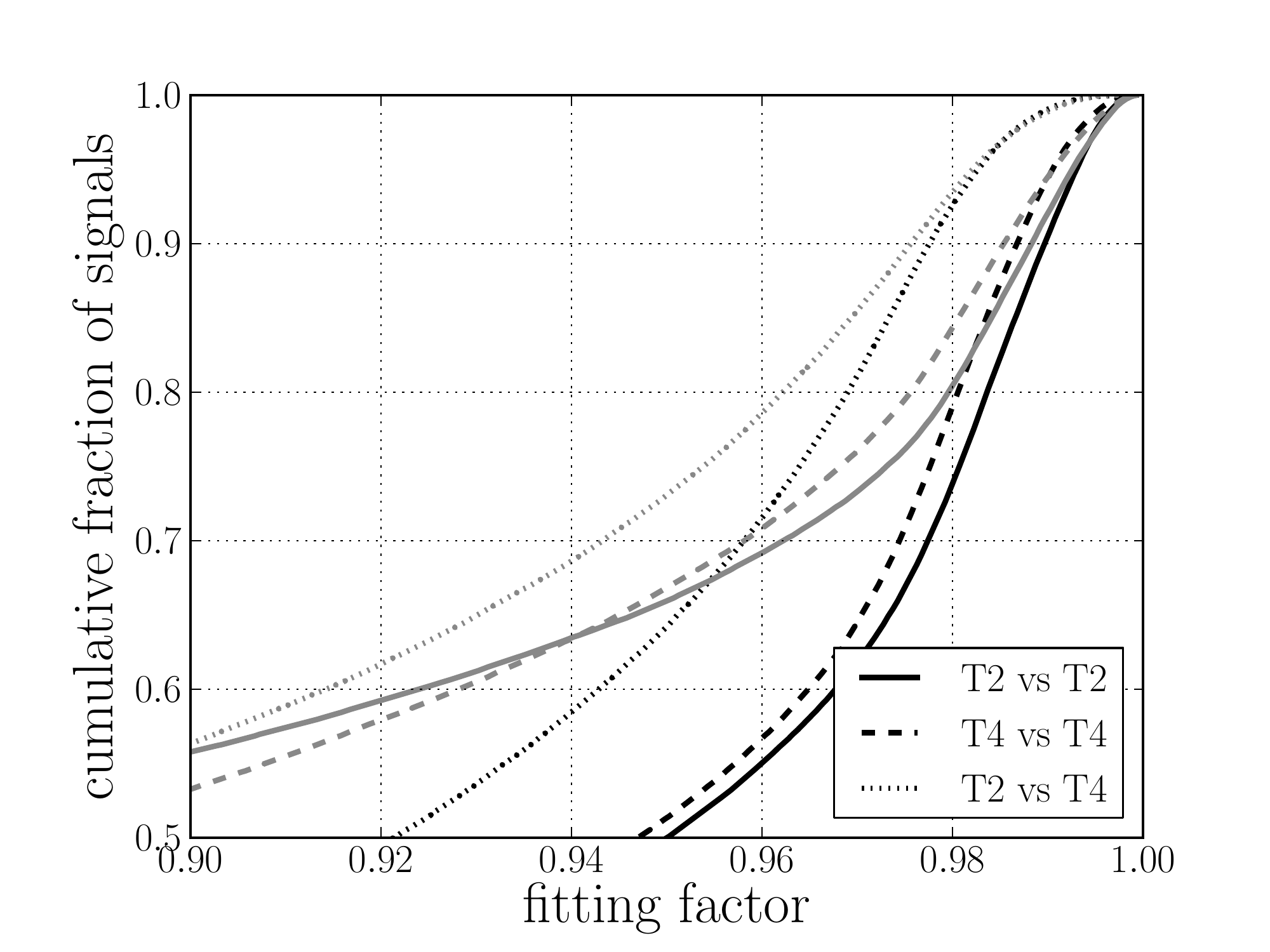}
\caption{\label{fig:aspineffectualness}
Fitting factor between a set of generic, precessing, NSBH signals and a
template bank of aligned-spin waveforms. Shown when both templates and signals
are generated using
the TaylorT2 approximant (black solid line) and the TaylorT4 approximant (black
dashed line).
Also shown when the templates are modeled using TaylorT2 and the signals are
modeled using TaylorT4 (black dotted line). For comparison the same results 
using a template bank of nonspinning waveforms are also plotted in gray. 
Plotted
over the full range of fitting factors (left) and zoomed in to show only
fitting factors greater than 0.9 (right). The
distribution that the NSBH signals are drawn from is described in Sec.
\ref{sec:nsbhpop}. The template bank construction is described in Sec.
\ref{sec:bank_construction}. Results obtained
using the zero-detuned, high-power advanced LIGO sensitivity curve with a 15 Hz
lower frequency cutoff.
}
\end{minipage}
\end{figure*}

The results of this simulation can be seen in 
Fig.~\ref{fig:aspineffectualness}. From this we can calculate the mean and 
median values of the fitting factor over the signal distribution that we used.
The mean fitting factor of the
signals is 0.82 (0.84) for the TaylorT2 (TaylorT4) approximant, while the median
fitting factor was 0.86 (0.88). In both cases the distributions have 
long tails, with some systems recovered with less than 30\% of their optimal
\ac{SNR}.
We also show results where we have modeled the templates using the TaylorT2
approximant and the signals using the TaylorT4 approximant. In this case
the mean fitting factor is 0.84 and the median is 0.87.
We notice that fewer signals are recovered with
high fitting factors ($> 0.95$) than in the other two cases, but we notice that
at lower values of fitting
factor the performance is very similar to the TaylorT4 vs TaylorT4 case.
The slight \emph{improvement}
of the TaylorT2 vs TaylorT4 case at lower fitting factors can be attributed to
the fact that the TaylorT2 bank is $\sim20\%$
larger than the TaylorT4 bank and therefore has more freedom to match
TaylorT4-modeled spinning signals. 

\begin{figure*}
    \centering
    \begin{minipage}[l]{2.0\columnwidth}
    \centering
\includegraphics[width=0.45\textwidth]
{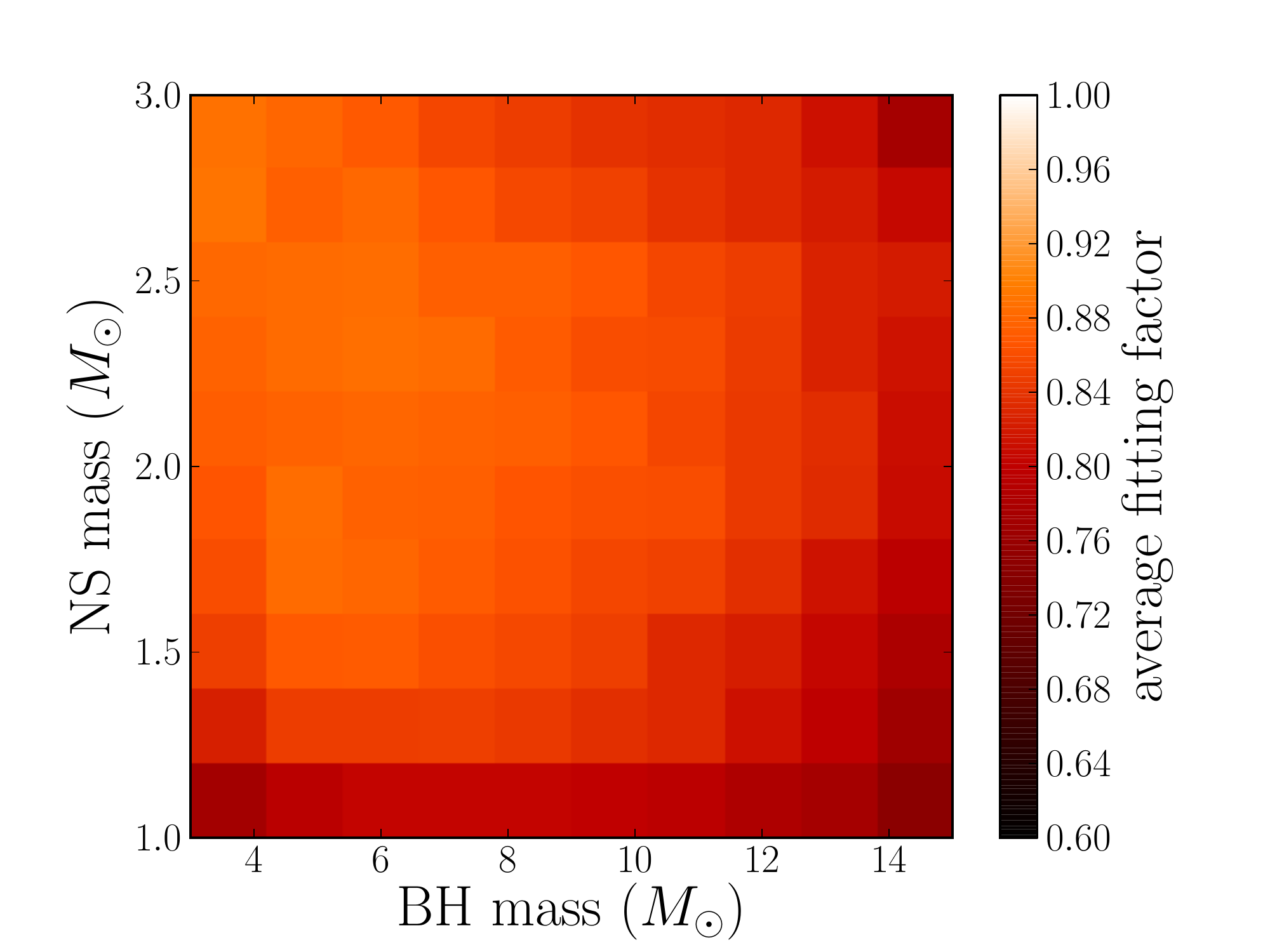}
\includegraphics[width=0.45\textwidth]
{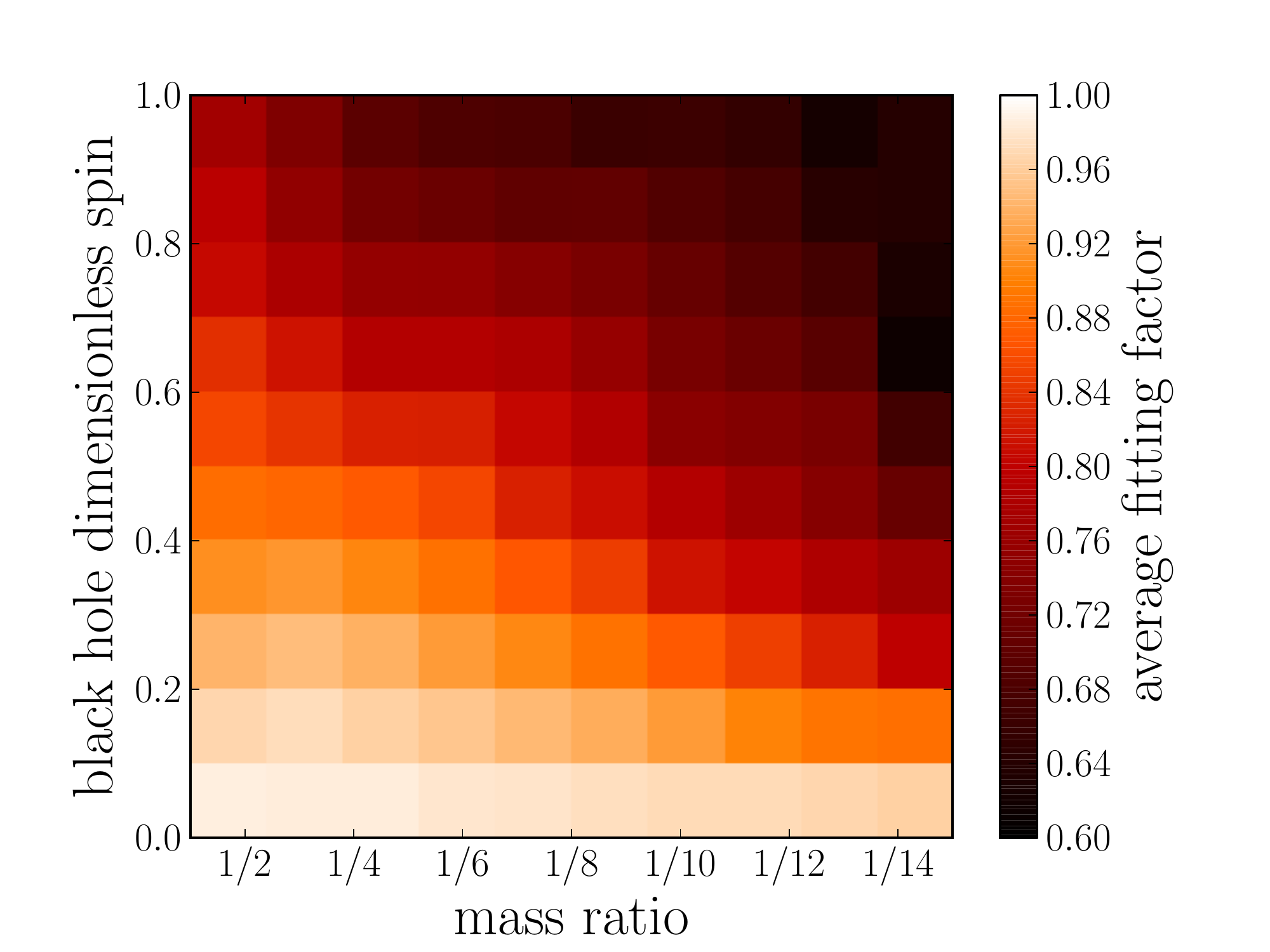}
\caption{\label{fig:nonspinavFF}
Average fitting factor between a set of generic, precessing, NSBH signals and a
template bank of nonspinning waveforms as a function of the component masses
(left) and as a function of the mass ratio and the black-hole dimensionless spin
magnitude (right). Both the signals and the template waveforms are modeled
using the TaylorT4 approximant. The distribution that the NSBH
signals are drawn from is described in Sec. \ref{sec:nsbhpop}.
Results obtained
using the zero-detuned, high-power advanced LIGO sensitivity curve with a 15 Hz
lower frequency cutoff.
}
\end{minipage}
\end{figure*}

\begin{figure*}
    \centering
    \begin{minipage}[l]{2.0\columnwidth}
    \centering
\includegraphics[width=0.45\textwidth]
{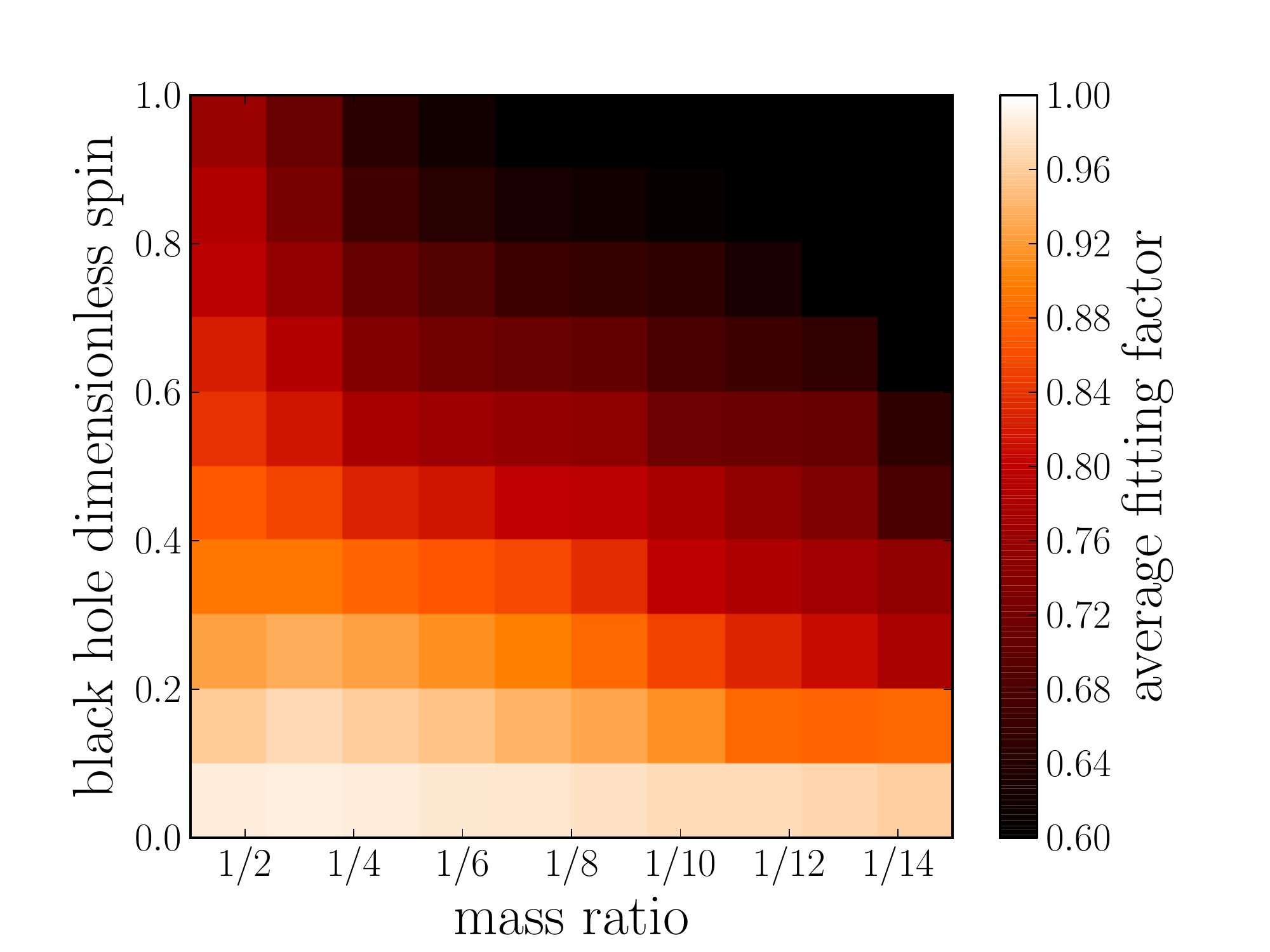}
\includegraphics[width=0.45\textwidth]
{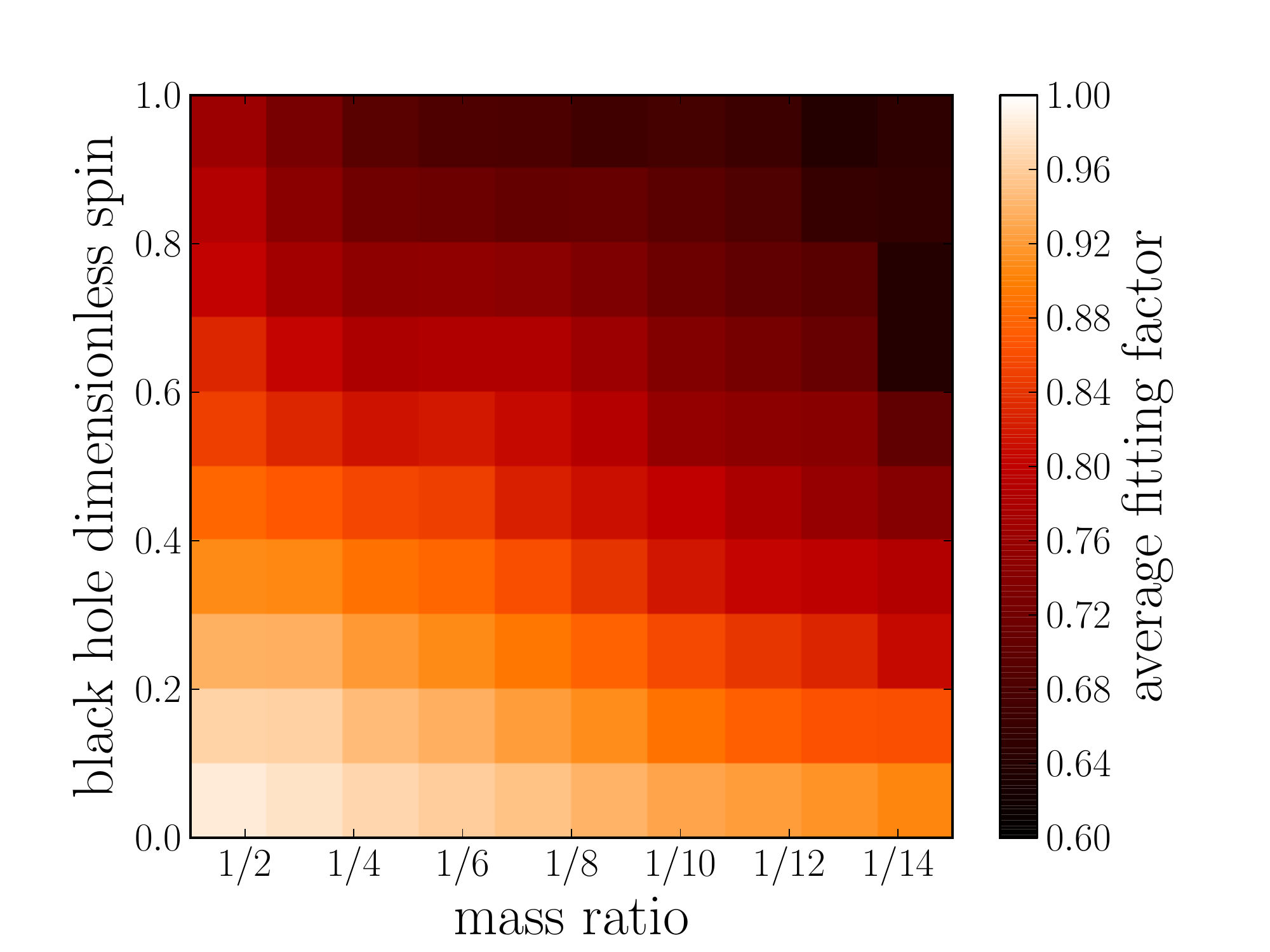}
\caption{\label{fig:nonspinavFFT2}
Average fitting factor between a set of generic, precessing, NSBH signals and a
template bank of nonspinning waveforms as a function of the mass ratio and the
black-hole dimensionless spin
magnitude (right). Shown when both the
template waveforms and signals are modeled with TaylorT2 (left) and when the
template waveforms are modeled with TaylorT2 and the signals are modeled with
TaylorT4 (right). The results in these plots are almost identical to each other 
and to the right panel of Fig.~\ref{fig:nonspinavFF}.
The distribution that the NSBH
signals are drawn from is described in Sec. \ref{sec:nsbhpop}.
Results obtained
using the zero-detuned, high-power advanced LIGO sensitivity curve with a 15 Hz
lower frequency cutoff.
}
\end{minipage}
\end{figure*}

In Fig.~\ref{fig:nonspinavFF}, we show the mean fitting factor as a function
of the intrinsic parameters of the system when both templates and signals
were modeled with the TaylorT4 approximant. For comparison, in
Fig.~\ref{fig:nonspinavFFT2} we show the mean fitting factor as a function of
the spin magnitude and mass ratio for the TaylorT2 vs TaylorT2 results and the
TaylorT2 vs TaylorT4 results. In both cases the results are similar to the
TaylorT4 vs TaylorT4 case, which indicates that the results are not suffering
from a significant bias due to the choice of waveform approximant. However, we
note that when using TaylorT2 as the signal model, the performance of the
nonspinning banks is worse for high spin, unequal mass systems than when
using TaylorT4 as the signal model.

\begin{figure*}
    \centering
    \begin{minipage}[l]{2.0\columnwidth}
    \centering
\includegraphics[width=0.45\textwidth]
{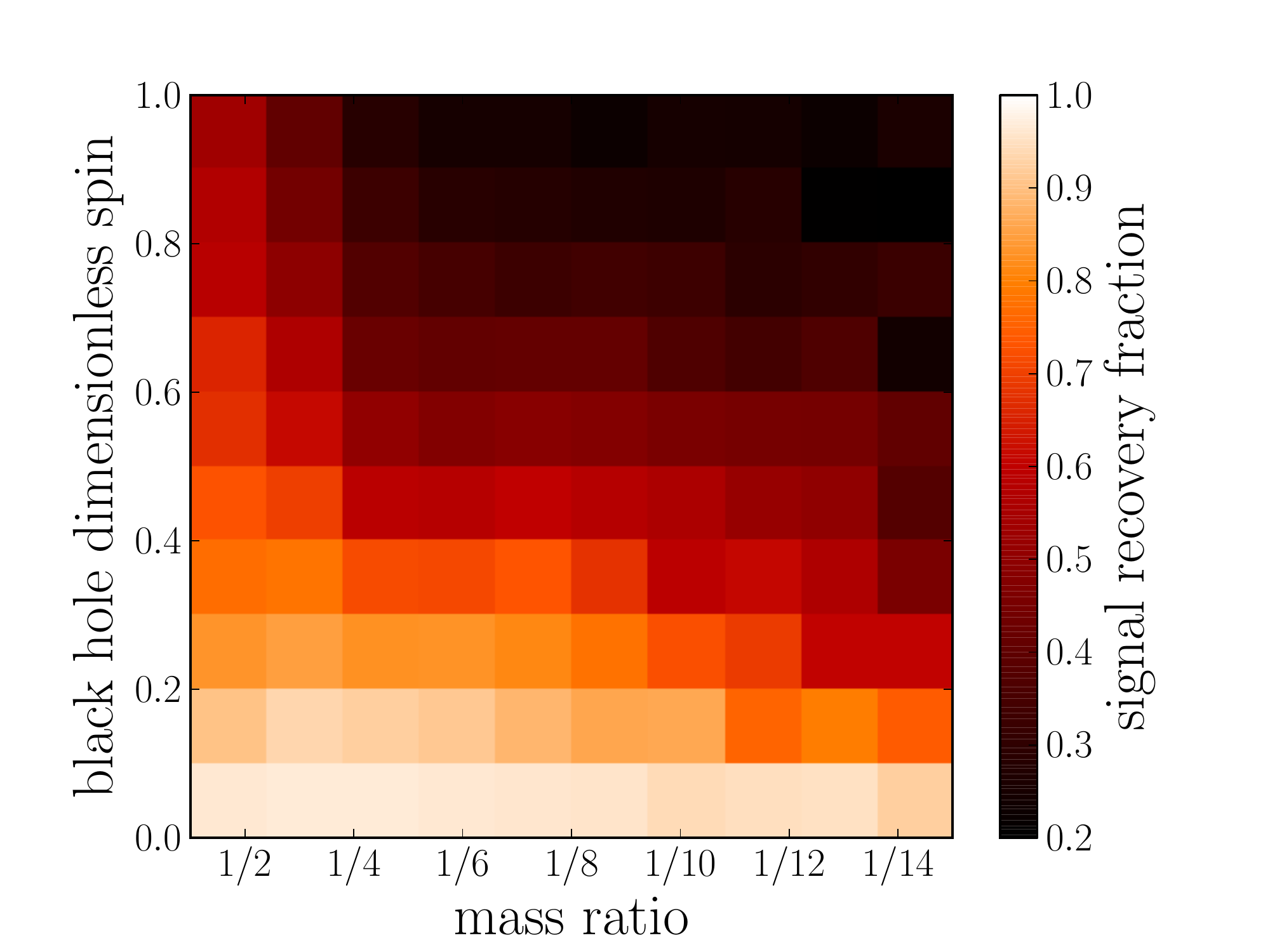}
\includegraphics[width=0.45\textwidth]
{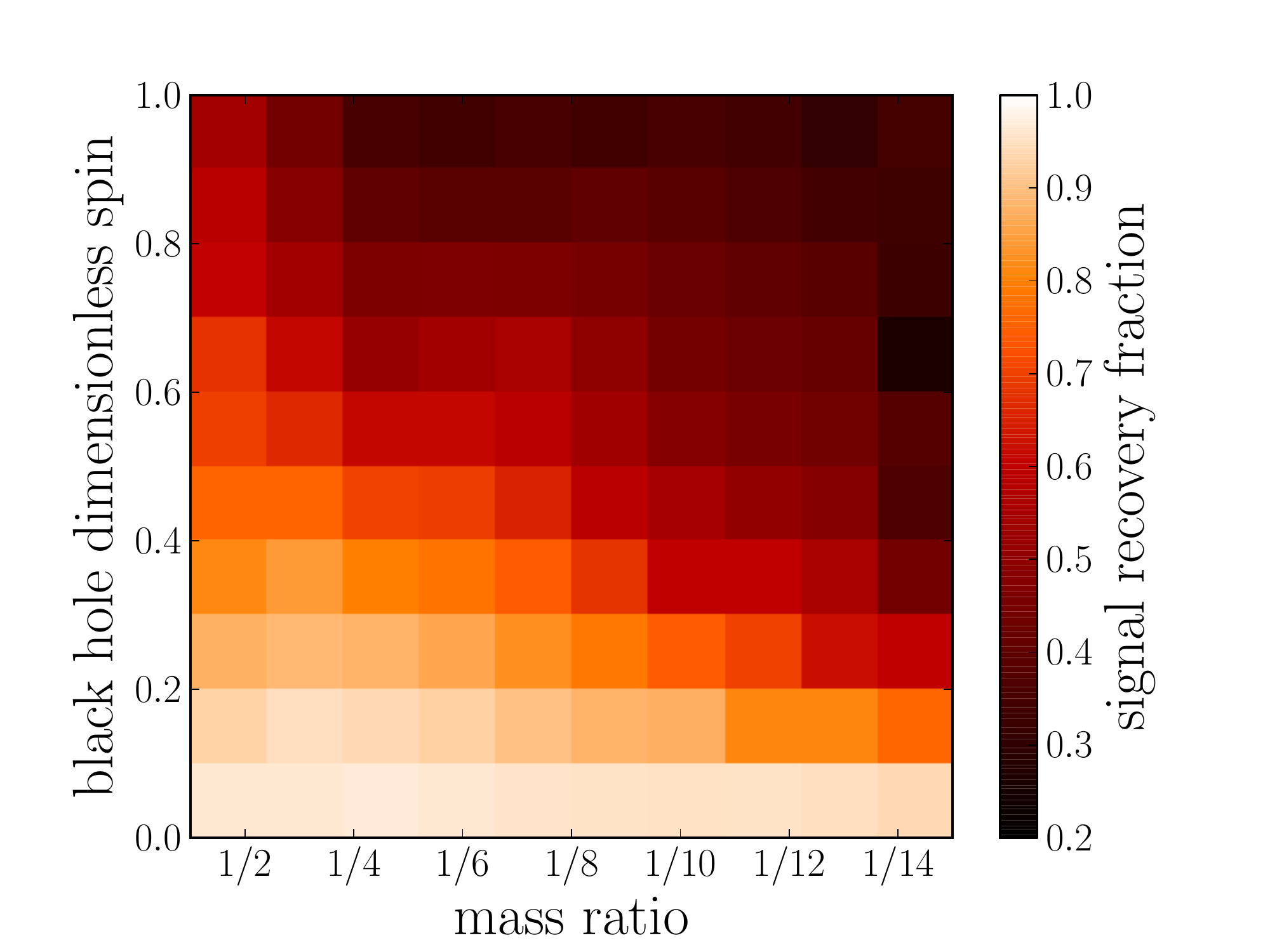}
\caption{\label{fig:nonspineffFF}
The signal recovery fraction obtained for a set of generic, precessing, NSBH
signals and a template bank of nonspinning waveforms as a function of the mass
ratio and the black-hole dimensionless spin. Shown when both the
template waveforms and the signals are modeled with TaylorT2 (left) and when
both the template waveforms and the signals are modeled with TaylorT4 (right).
The distribution of the signal recovery fraction over the mass space is very 
similar to the distribution of average fitting factors shown in 
Figs.~\ref{fig:nonspinavFF} and \ref{fig:nonspinavFFT2}.
The distribution that the NSBH
signals are drawn from is described in Sec. \ref{sec:nsbhpop}.
Results obtained
using the zero-detuned, high-power advanced LIGO sensitivity curve with a 15 Hz
lower frequency cutoff.
}
\end{minipage}
\end{figure*}

In Fig.~\ref{fig:nonspineffFF} we show
the signal recovery fraction as a function of the \ac{BH} spin magnitude and
the mass ratio. The signal recovery fraction is defined in Sec. 
\ref{sec:bank_testing}.
It is clear that using a nonspinning bank to search for \ac{NSBH} systems will
result in a considerable reduction
in the \ac{NSBH} detection rate. In addition, the ability to
detect systems with high spin, especially systems that also have unequal 
masses, is especially poor.  We note that these efficiencies would be
improved by using nonspinning templates outside of the chosen mass ranges, for
example \ac{BNS} or binary black-hole template waveforms, or even templates
with unphysical mass parameters \cite{Brown:2012qf,Baird:2012cu}.

\subsection{Performance of aligned-spin template banks when searching for 
generic NSBH signals}
\label{sec:aligned_spin}

With the template banks of aligned-spin systems described in Sec.
\ref{sec:bank_construction}, we are able to recover aligned-spin systems
modeled with either the TaylorT2 or TaylorT4 approximant with fitting factors
greater than 0.97 in $>99\%$ of cases, as shown in Sec.
\ref{sec:bank_validation}.
If we use these banks to search for precessing systems modeled with the same
approximants, any loss in signal power, beyond that lost due to the spacing of
the aligned-spin bank, is entirely due to precession.
We now assess the performance of these aligned-spin banks when searching for
generic, precessing \ac{NSBH} signals and identify regions of the
parameter space where precessional effects cause a significant loss in
detection rate.

\begin{figure*}
    \centering
    \begin{minipage}[l]{2.0\columnwidth}
    \centering
\includegraphics[width=0.45\textwidth]
{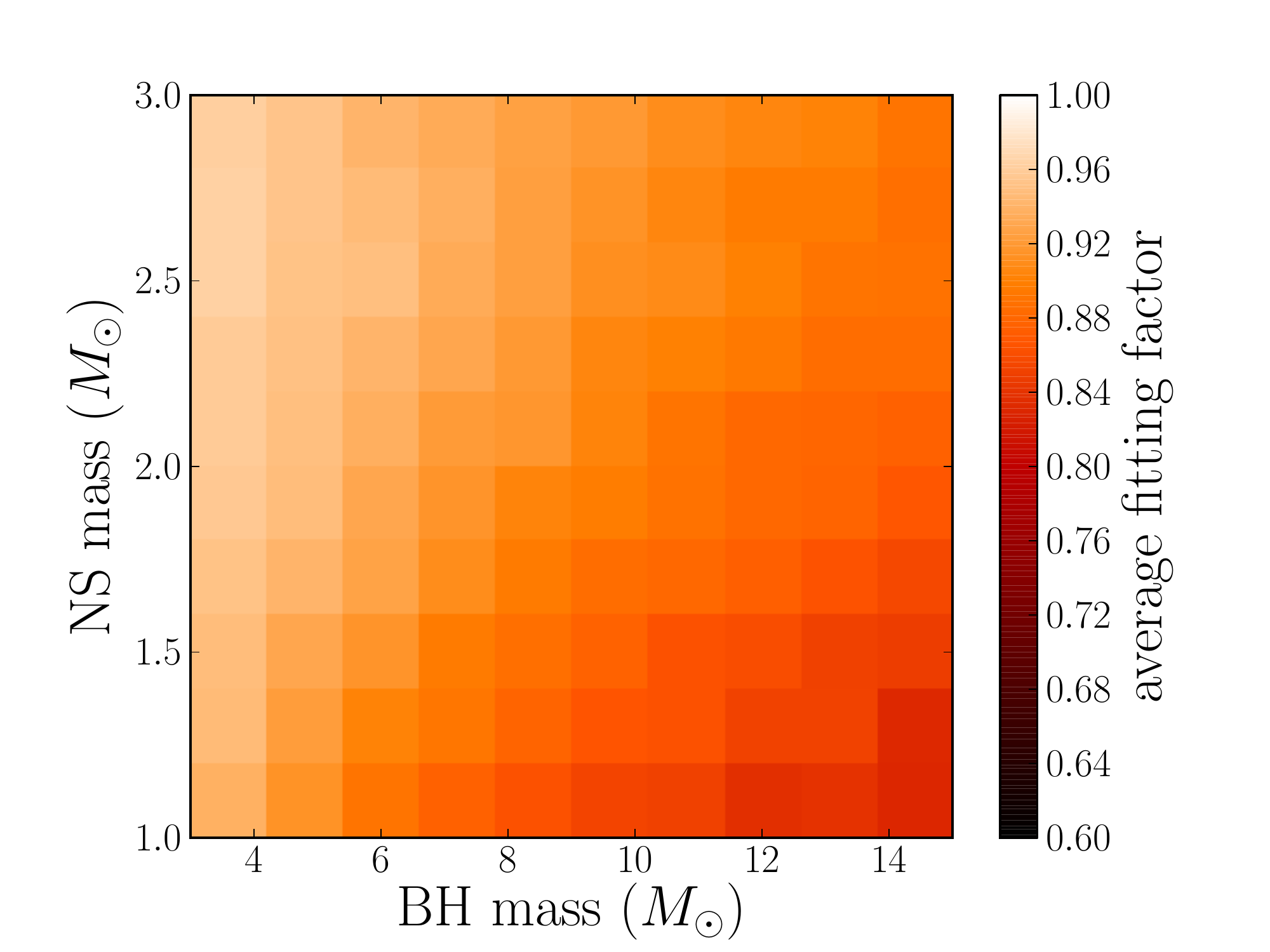}
\includegraphics[width=0.45\textwidth]
{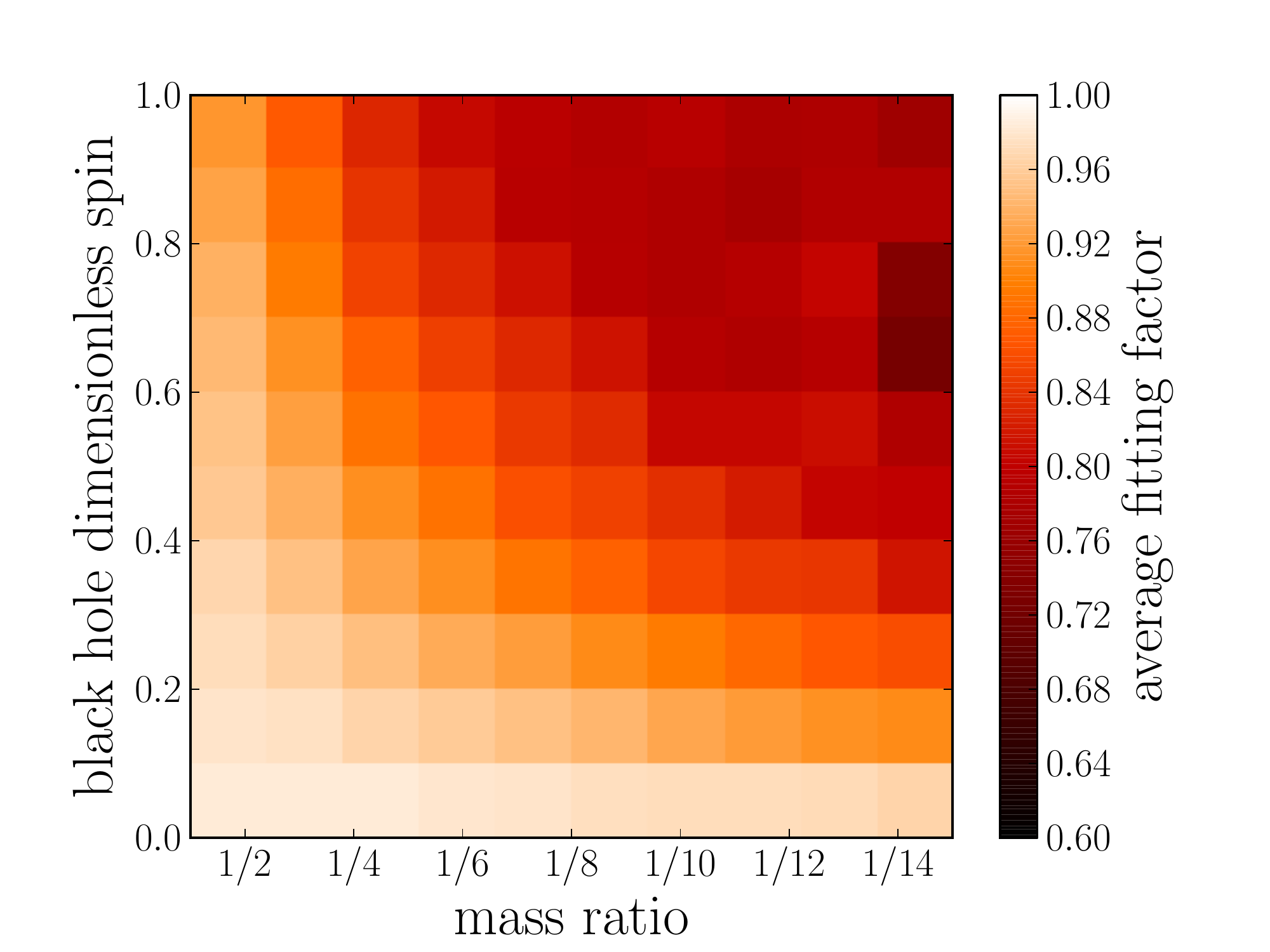}
\includegraphics[width=0.45\textwidth]
{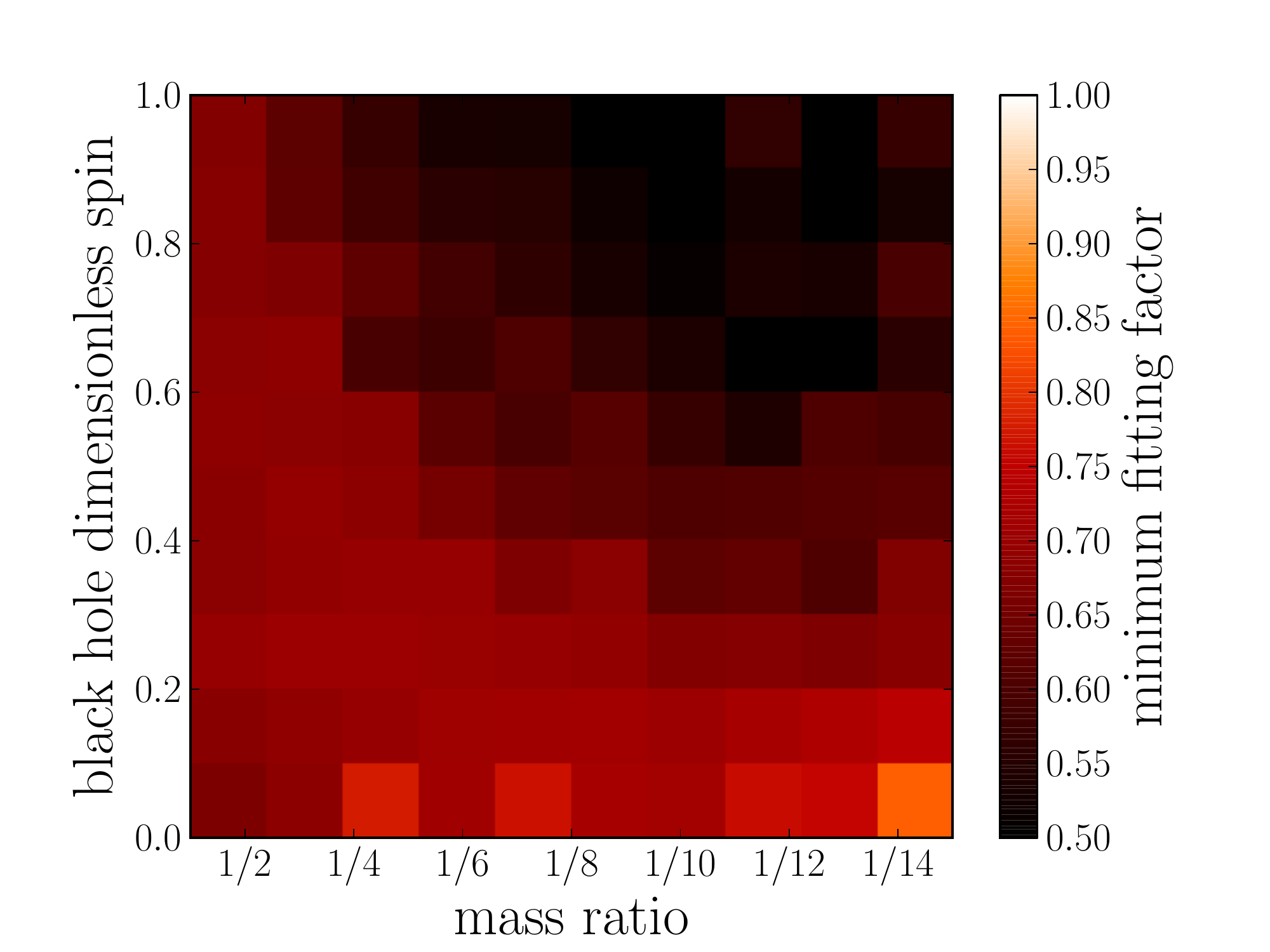}
\includegraphics[width=0.45\textwidth]
{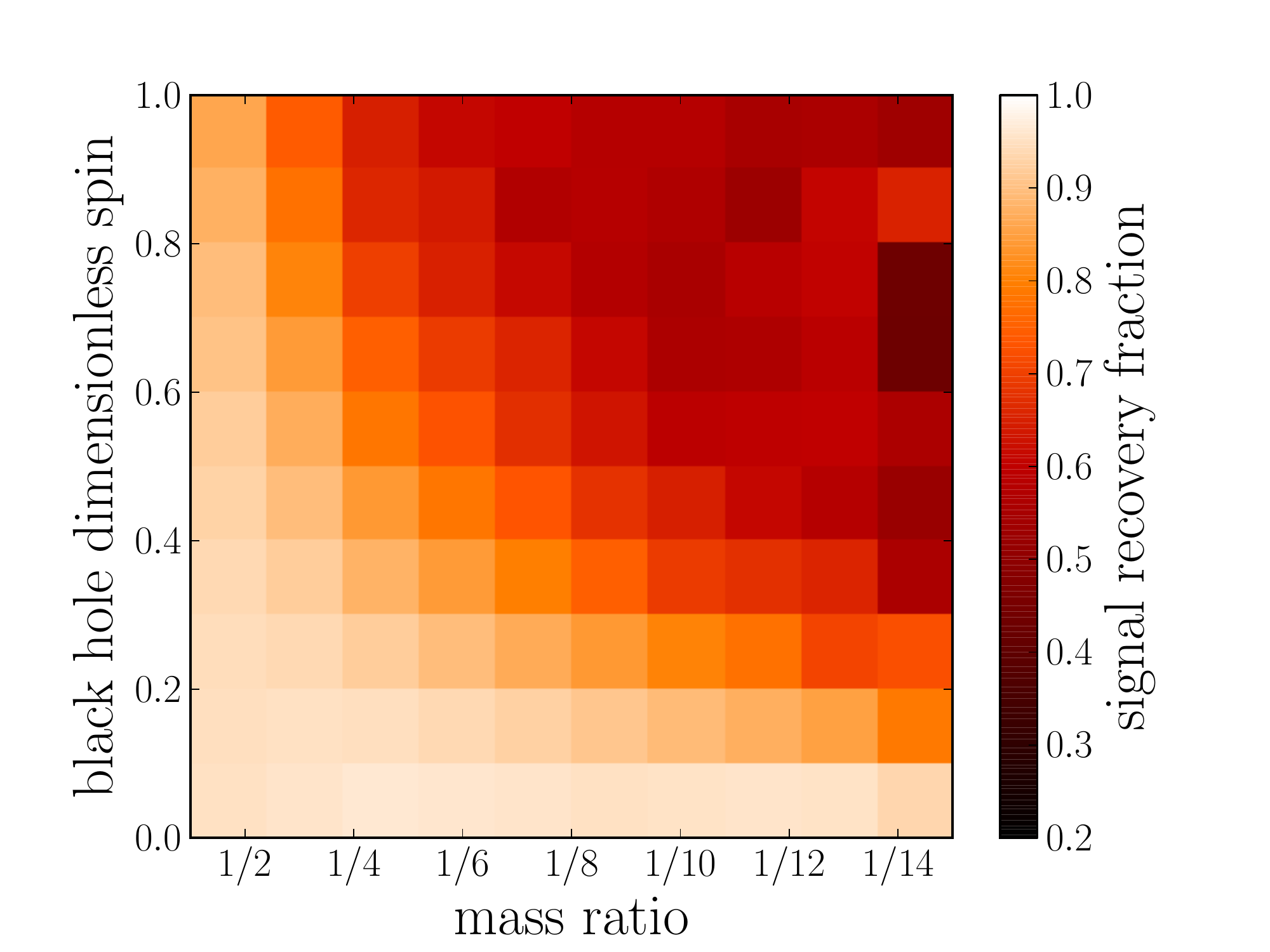}
\caption{\label{fig:aspinavFF}
Average fitting factor between a set of generic, precessing, NSBH
signals and a template bank of aligned-spin waveforms as a function of the
component masses (top left) and as a function of the
mass ratio and the black-hole dimensionless spin
magnitude (top right). Also plotted is the minimum fitting factor (bottom left) 
and the signal recovery fraction (bottom right) as a function of the
mass ratio and the black-hole dimensionless spin magnitude. Both signals and
template waveforms are modeled using the TaylorT4 approximant.
The distribution that the NSBH signals are drawn from
is described in Sec. \ref{sec:nsbhpop}. The template bank construction
is described in Sec. \ref{sec:bank_construction}. Results obtained
using the zero-detuned, high-power advanced LIGO sensitivity curve with a 15 Hz
lower frequency cutoff.
}
\end{minipage}
\end{figure*}

\begin{figure}
\includegraphics[width=0.45\textwidth]
{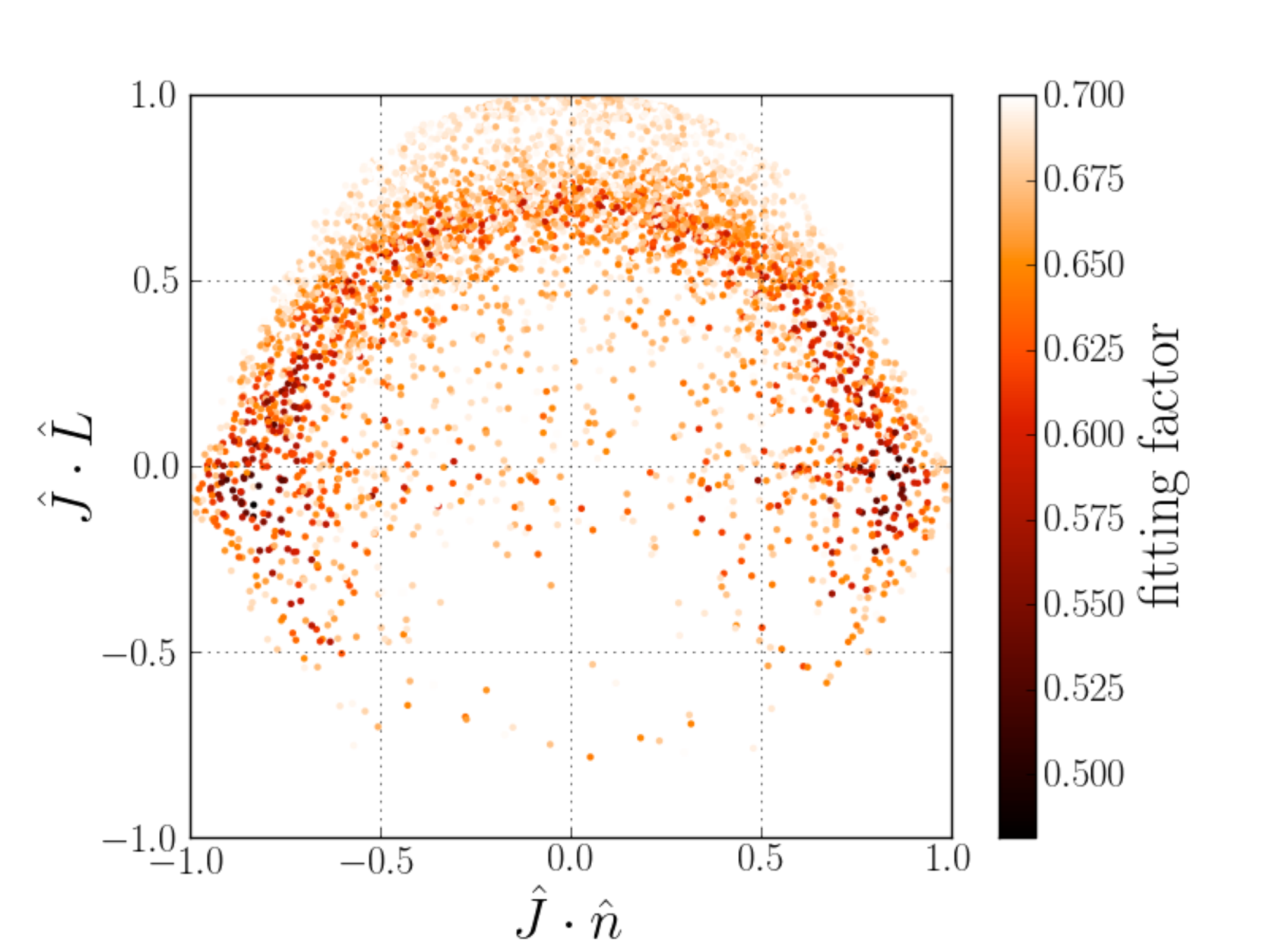}
\caption{\label{fig:arseofsauron}
The distribution of precessing NSBH signals that are recovered with fitting
factors $< 0.7$ when searching with an aligned-spin template bank. We use
$\hat{J}$ to denote the initial total angular momentum of the system, $\hat{n}$
denotes the line of sight towards the observer
and $\hat{L}$ denotes the orbital angular momentum when the gravitational-wave 
frequency is 60 Hz 
(at which point approximately half of the signal power has accumulated). 
Both signals and template waveforms are modeled using the TaylorT4 approximant.
The distribution that the NSBH signals are drawn from
is described in Sec. \ref{sec:nsbhpop}. The template bank construction
is described in Sec. \ref{sec:bank_construction}. Results obtained
using the zero-detuned, high-power advanced LIGO sensitivity curve with a 15 Hz
lower frequency cutoff.
}
\end{figure}

\begin{figure*}
    \centering 
    \begin{minipage}[l]{2.0\columnwidth}
    \centering
\includegraphics[width=0.45\textwidth]
{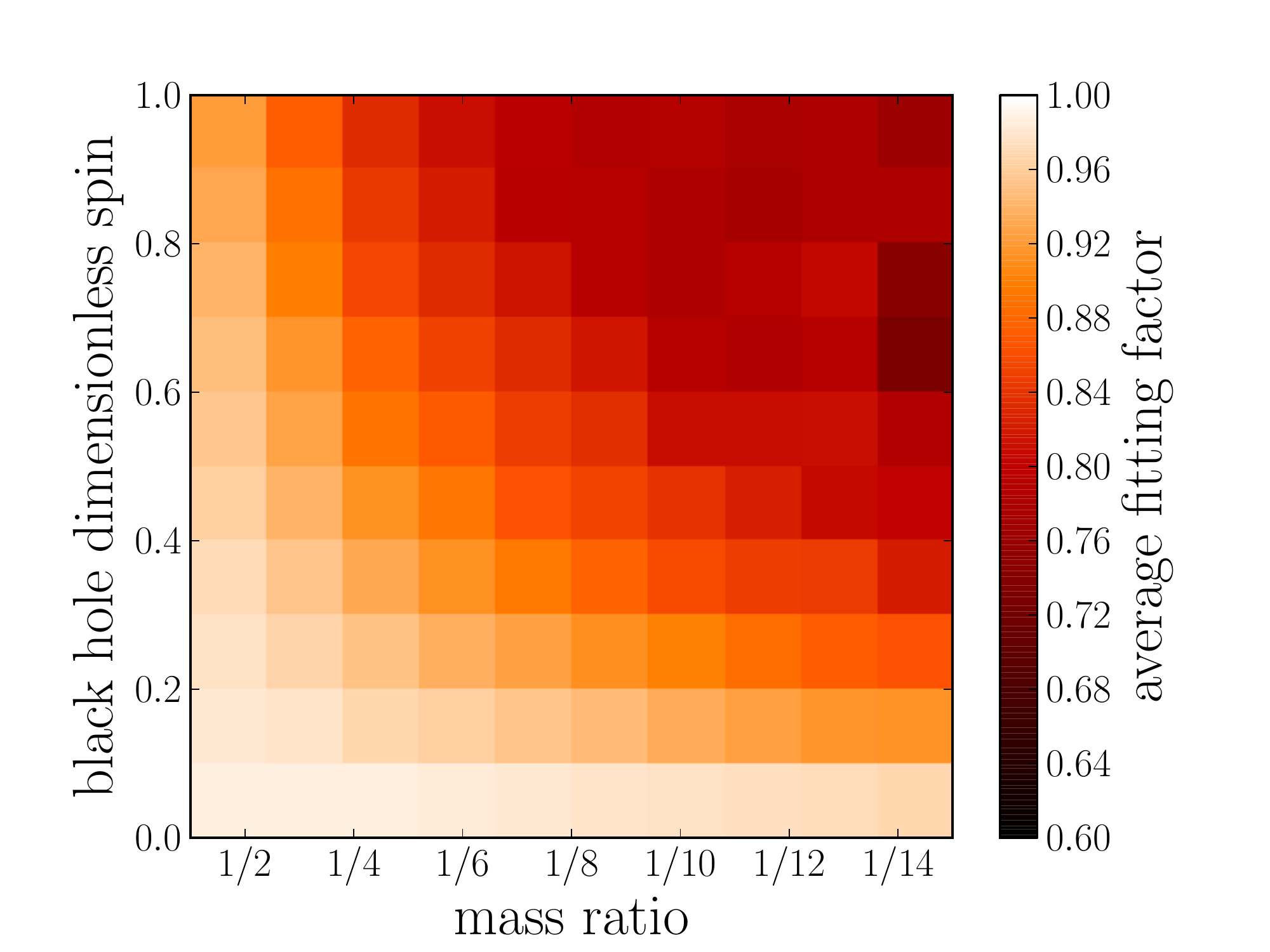}
\includegraphics[width=0.45\textwidth]
{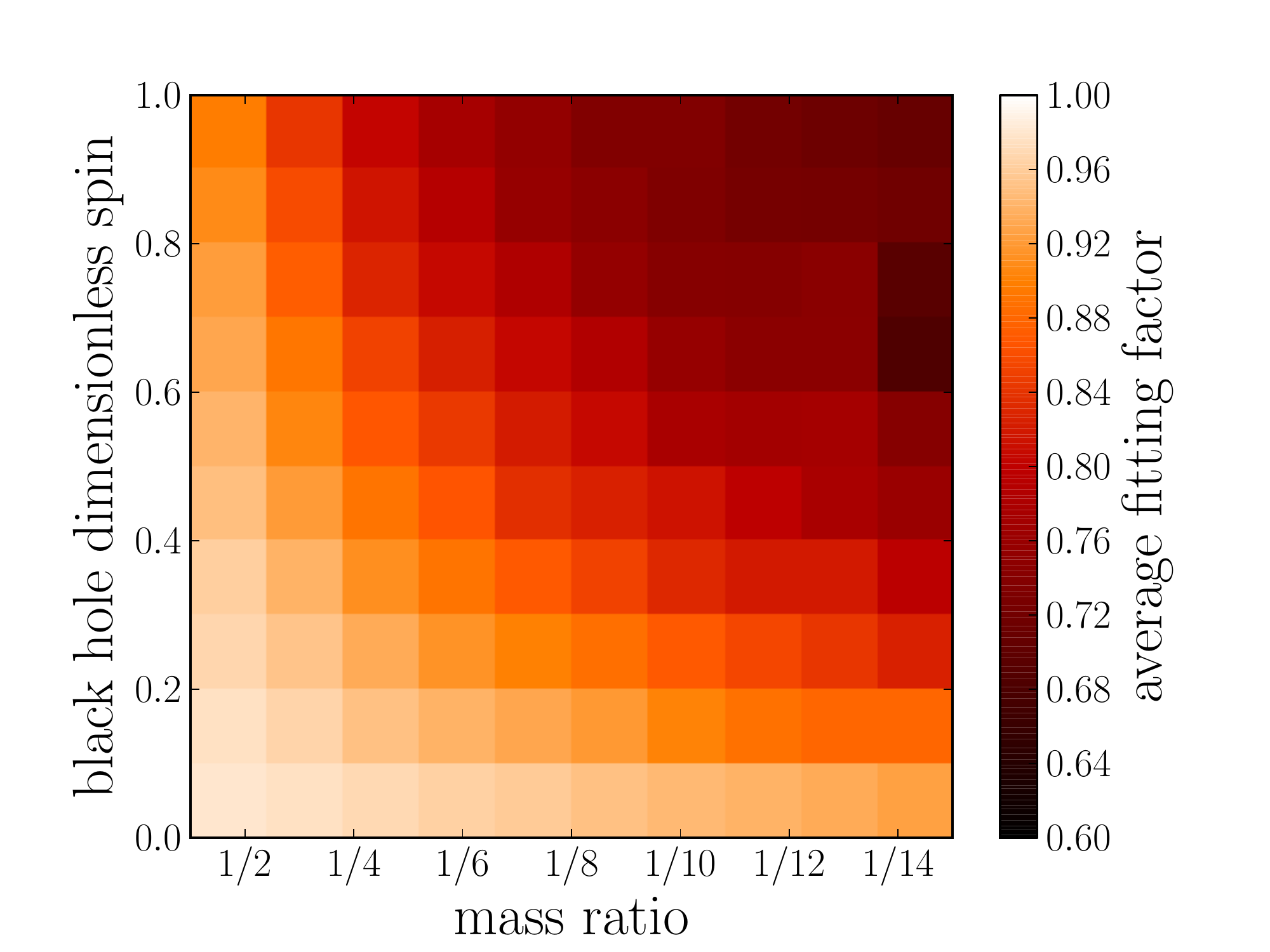}
\caption{\label{fig:aspinavFFT2}
Average fitting factor between a set of generic, precessing, NSBH
signals and a template bank of aligned-spin waveforms as a function of the
mass ratio and the black-hole dimensionless spin magnitude. Shown when both the
template waveforms and signals are modeled with TaylorT2 (left) and when the
template waveforms are modeled with TaylorT2 and the signals are modeled with
TaylorT4 (right). The results in these plots are almost identical to each other 
and to the top right panel of Fig.~\ref{fig:aspinavFF}.
The distribution that the NSBH
signals are drawn from is described in Sec. \ref{sec:nsbhpop}. The
template bank construction is described in Sec. \ref{sec:bank_construction}.
Results obtained
using the zero-detuned, high-power advanced LIGO sensitivity curve with a 15 Hz
lower frequency cutoff.
}
\end{minipage}
\end{figure*}

\begin{figure}
\includegraphics[width=0.45\textwidth]
{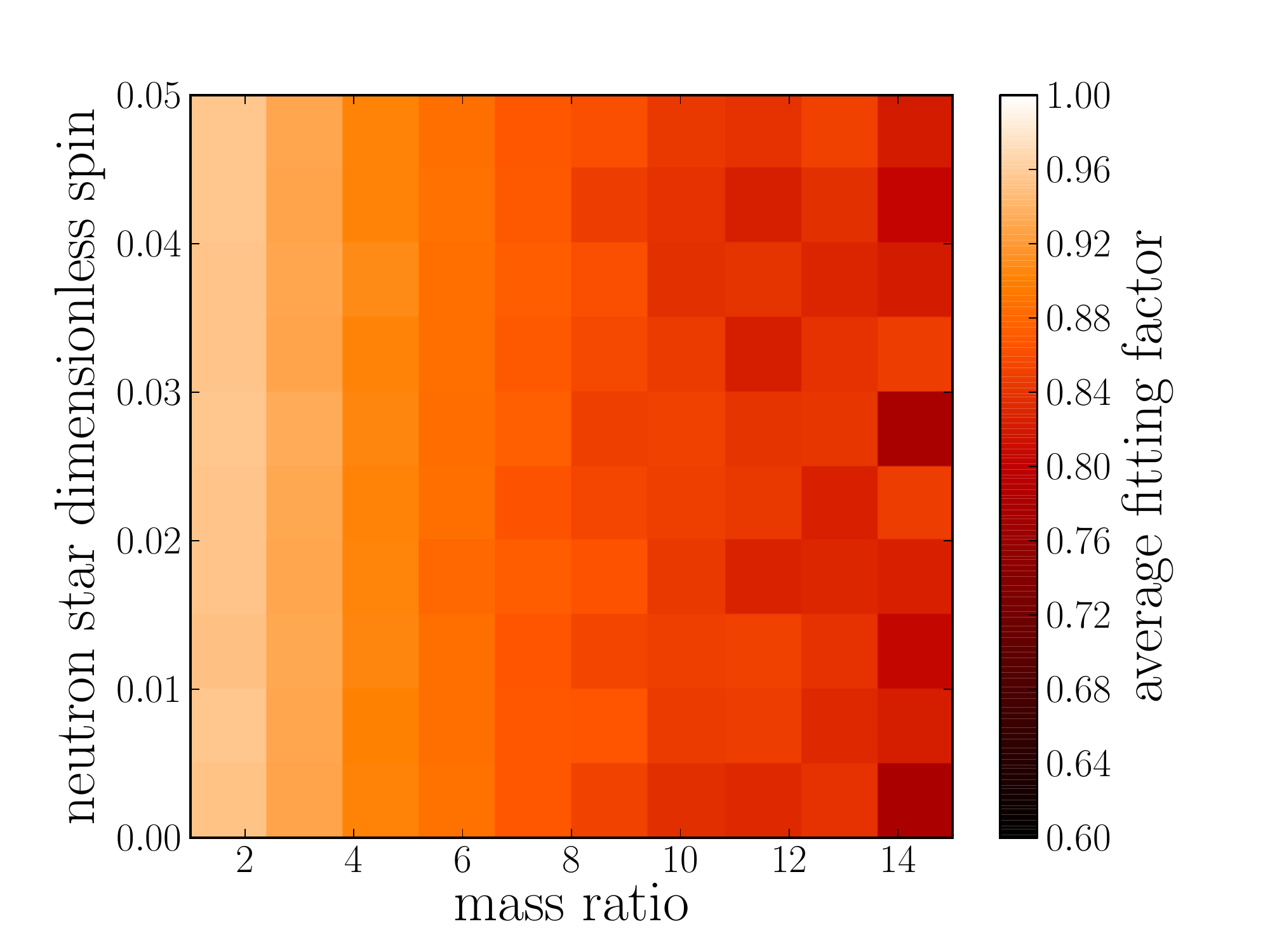}
\caption{\label{fig:nsspin}
Average fitting factor between a set of generic, precessing, NSBH
signals and a template bank of aligned-spin waveforms as a function of the
mass ratio and the neutron-star dimensionless spin
magnitude. Both signals and
template waveforms are modeled using the TaylorT4 approximant.
The distribution that the NSBH signals are drawn from
is described in Sec. \ref{sec:nsbhpop}. The template bank construction
is described in Sec. \ref{sec:bank_construction}. Results obtained
using the zero-detuned, high-power advanced LIGO sensitivity curve with a 15 Hz
lower frequency cutoff.
}
\end{figure}

\begin{figure}
\includegraphics[width=0.45\textwidth]
{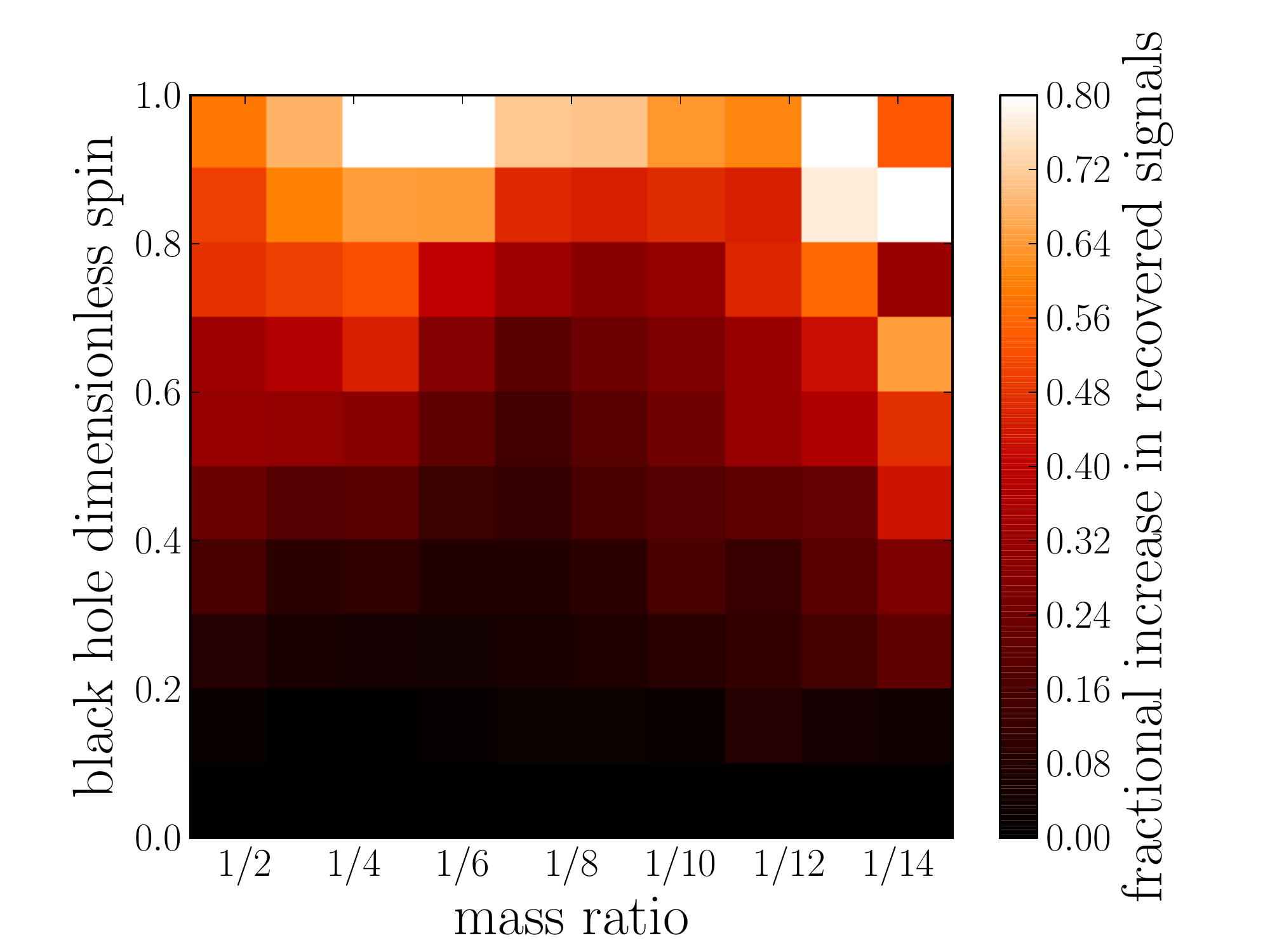}
\caption{\label{fig:aspinimpr}
The fractional increase in the number of recovered signals when
searching for generic, precessing, NSBH signals between
a template bank of aligned-spin waveforms and a template bank of nonspinning
waveforms. Both signals and template waveforms are modeled using the TaylorT4
approximant. The distribution that the NSBH
signals are drawn from is described in Sec. \ref{sec:nsbhpop}. The
template bank construction is described in Sec. \ref{sec:bank_construction}.
Results obtained
using the zero-detuned, high-power advanced LIGO sensitivity curve with a 15 Hz
lower frequency cutoff.
}
\end{figure}

Our signal population is a set of 100,000 precessing \ac{NSBH} signals. This 
distribution was
described in Sec. \ref{sec:nsbhpop}. For comparison this is the
\emph{same} set of signals as we used in Sec. \ref{sec:non_spinning}.
As before, we will assess fitting factors using both the TaylorT2 and TaylorT4
models to mitigate any bias arising from choice of waveform model.
When TaylorT2 is used as the signal model, we will use the bank of aligned-spin
systems that was placed using the TaylorF2 metric and a 1000 Hz
upper frequency cutoff and model the templates using the TaylorT2 approximant.
When TaylorT4 is used as the signal model, we will use the bank of aligned-spin
systems placed using the TaylorR2F4 metric and model the templates with
TaylorT4. The placement of these banks was described in Sec.
\ref{sec:bank_construction}.

The results of these simulations can be seen in 
Fig.~\ref{fig:aspineffectualness}, where we also compare with the
results obtained in Sec. \ref{sec:non_spinning} when using nonspinning
template banks. We can clearly see from Fig.~\ref{fig:aspineffectualness} that
the distribution of fitting factors for the
case when both signals and templates were modeled with TaylorT2 agrees well
with the case when both were modeled with TaylorT4. This indicates that
we have disentangled precessional effects from waveform-dependent effects and
our results are free of any bias due to the choice of waveform model.
The mismatches seen here, beyond that caused by the discreteness of the bank,
are due only to the effects of precession.
In both cases we observe a median fitting factor of
$\sim 0.95$ and a mean fitting factor of $\sim 0.91$. This is a clear
improvement
over the nonspinning results where the mean fitting factor was 0.82 (0.84) for
TaylorT2 (TaylorT4) and the median
fitting factor was 0.86 (0.88). 

In Fig.~\ref{fig:aspineffectualness} we also show results where the template
waveforms are modeled with TaylorT2 and the signals are modeled with TaylorT4.
In this case the performance is worse, with a median fitting factor of $\sim
0.92$ and a mean fitting factor of $\sim 0.88$.

In Fig.~\ref{fig:aspinavFF} we show the mean fitting factor as a function of
the intrinsic parameters
for our results with the TaylorT4 waveform. We also show the minimum fitting
factor and the signal recovery
fraction as a function of the \ac{BH} spin magnitude
and mass ratio for the same results.
The figure serves to highlight that there are certain
systems in certain regions of the parameter space where precessional effects
cause the \ac{NSBH} signals to have large mismatches with a bank of
aligned-spin templates. This is most prominent when $m_{BH} / m_{NS}$ and the
\ac{BH} spin magnitude are both large, i.e. where the black 
hole's angular momentum is particularly large relative to the orbital angular 
momentum. 
We explore this further in Fig.~\ref{fig:arseofsauron} where, following the
work of \cite{Brown:2012gs}, we show the distribution
of precessing systems recovered with fitting factors smaller than $0.7$. This
is plotted as a function of the angles between the total angular 
momentum, the orbital angular momentum and the line of sight to an observer. 
As predicted in \cite{Brown:2012gs}, there is clearly a correlation
between these angles and the systems recovered with the lowest fitting factors.
To demonstrate that these results are not specific to the TaylorT4 waveform, in
Fig.~\ref{fig:aspinavFFT2} we show the mean fitting factor as a function of
the \ac{BH} spin magnitude and mass ratio for our TaylorT2 vs TaylorT2 and
TaylorT2 vs TaylorT4 results. The TaylorT2 results are very similar to the
TaylorT4 results in Fig.~\ref{fig:aspinavFF}. This again demonstrates that the
choice of waveform is not affecting our statements regarding the effect
precession will have on searches for \ac{NSBH} signals using aligned-spin
template banks. When searching for TaylorT4 signals with TaylorT2 templates
we see lower fitting factors. The disagreement between these two waveform models
is a significant factor that will affect searches for \ac{NSBH} systems with
second-generation observatories. Computing higher order terms in the \acf{PN}
expansion of the center-of-mass energy and gravitational wave flux will help to
reduce this disagreement and produce waveforms that better match real
gravitational-wave signals.

To investigate whether the spin of the neutron star has any effect on these 
results, in Fig. \ref{fig:nsspin} we plot the average 
fitting factor as a function of the mass ratio and the \emph{neutron-star} 
dimensionless spin. There is not any noticeable correlation between the average 
fitting factor and the neutron star's spin. As a further test we generated a 
template bank of TaylorF2 waveforms, using the same parameters as the banks 
discussed in Sec. \ref{sec:bank_construction} and used in this section, 
except the neutron-star spin was only allowed to take a value of 0. We then 
evaluated the fitting factor between this bank and a set of TaylorF2 
\ac{NSBH} waveforms using the same distribution as described in Sec. 
\ref{sec:nsbhpop} except both spins were constrained to be aligned with the 
orbit and the neutron spin was either set to a value of 0.05 or -0.05. We found 
that 99.99\% of these signals had fitting factors larger than 0.96, which is the 
smallest fitting factor allowed by the placement algorithm. These results
indicate that as long as the \ac{NS} spin is $< 0.05$ then
it will be a negligible factor in searches for \ac{NSBH} binaries with 
\ac{aLIGO} and {AdV}.

We can also compare these results to the results we obtained using a
nonspinning template bank in Sec. \ref{sec:non_spinning}. In 
Fig.~\ref{fig:aspinimpr} we show the fractional increase in the number
of recovered signals between using nonspinning and aligned-spin template banks
for the TaylorT4 approximant. The fractional increase in the number
of recovered signals is calculated by taking the ratio of the signal recovery 
fraction when using a nonspinning bank and the signal recovery fraction when 
using an aligned-spin bank. This figure helps to
emphasize that a much greater fraction of systems with large spin would be
recovered when using an aligned-spin template bank. In Table 
\ref{tab:results_summary} we summarize the average signal
recovery fractions for the aligned-spin banks and
compare these numbers to the results obtained with nonspinning template banks. 
We remind the reader that we are comparing signal recovery at a 
fixed signal-to-noise ratio threshold. Signal recovery at a fixed false-alarm 
probability will depend on other factors, including the size of the parameter 
space covered by the template bank and the non-Gaussianity of the data. We 
discuss this further in the conclusion.

\begin{table*}
    \centering
    \begin{minipage}[l]{2.0\columnwidth}
    \centering
\begin{tabular}{p{1.6cm}|p{1.6cm}|p{1.85cm}|p{1.85cm}|p{1.85cm}|p{1.85cm}|p{
1.85cm}|p{1.85cm}}
\,\,\,\,\,\,Template  & \,\,\,\,\,\, Signal  & 
\multicolumn{2}{|p{3.7cm}|}{Signal recovery fraction
for non-spinning bank} & \multicolumn{2}{|p{3.7cm}|}
{Signal recovery fraction for aligned-spin bank} & 
\multicolumn{2}{|p{3.7cm}} {Fractional improvement
in signal recovery} \\ \cline{3-8}
 approximant & approximant & Average & $(10,1.4)M_{\odot}$ & Average
& $(10,1.4)M_{\odot}$ & Average & $(10,1.4)M_{\odot}$\\
\hline \hline
 TaylorT2 & TaylorT2 & 64\% & 63\% & 83\% & 74\% & 30\% & 17\% \\
 TaylorT4 & TaylorT4 & 69\% & 67\% & 82\% & 73\% & 19\% & 9\% \\
 TaylorT2 & TaylorT4 & 67\% & 64\% & 77\% & 67\% & 16\% & 5\% \\
\end{tabular}
\caption{\label{tab:results_summary}
The performance of our aligned-spin template banks when used to search for a
set of generic, precessing, NSBH signals using varying approximants for the
template and signal waveforms. We show both the mean signal recovery fraction 
over the full \ac{NSBH} signal population we consider and the signal recovery 
fraction for a \ac{NSBH} system with masses $(10\pm0.5,1.4\pm0.05)M_{\odot}$.
The distribution that
the NSBH signals are drawn from is described in Sec. \ref{sec:nsbhpop}. The
template bank construction is described in Sec. \ref{sec:bank_construction}.
Results obtained
using the zero-detuned, high-power advanced LIGO sensitivity curve with a 15 
Hz lower frequency cutoff and a 1000 Hz upper frequency cutoff.
}
\end{minipage}
\end{table*}

Finally, we compare our results with previous works. In \cite{Ajith:2012mn}
the authors presented an efficiency study when using a template bank of 
stochastically generated aligned-spin signals. We verified that when using the 
stochastic algorithm we used in this work, and using the same set of parameters 
as the study described in \cite{Ajith:2012mn}, we generated a bank with the 
same number of templates. We have therefore demonstrated that our template bank 
algorithm requires less templates to achieve the same level of coverage as the 
algorithm used in \cite{Ajith:2012mn}. In that work the effective fitting 
factor for a \ac{NSBH} system with masses given by $10 M_{\odot}$ ,
$1.4 M_{\odot}$ was estimated to be 0.95, which corresponds to a
signal recovery fraction of 86\%.
In contrast, our results show a lower signal recovery fraction 
for the same masses of $73\%-74\%$ when the same waveform model is used to 
model both the template and signal. It is not clear why this discrepancy 
occurs; however, it may be partially explained by the fact that 
the authors of \cite{Ajith:2012mn} used a lower frequency cutoff in their 
matched filters of 20 Hz, whereas we used 15 Hz, which is more appropriate for 
the 
predicted \ac{aLIGO} zero-detuned--high-power noise curve. 

In \cite{Brown:2012gs} the authors used a simplified model of precessing systems
to predict the distribution of fitting
factors for \ac{NSBH} systems. These results, shown in Fig. 11 of that work,
agree qualitatively with the results obtained here.
We also obtain quantitative agreement by comparing our simulations of 
generic precessing systems with TaylorT4 as the signal and template model with 
the values predicted by Eq. 46b of \cite{Brown:2012gs}. We find that 90\% of 
the fitting factors are within $0.03$ of the predicted values.
They also predicted the 
distribution of the signals that would be recovered with the lowest fitting 
factors as a function of the orientation of the black-hole spin and the 
orientation of the orbital plane with respect to the line of sight. We produce 
a similar distribution in Fig.~\ref{fig:arseofsauron}. 
A further exploration of the agreement of the fitting factors with 
this prediction will be carried out in a future work making use of these 
simulations.

\section{Conclusions}
\label{sec:conclusion}

In this work we have explored the effect that the angular momentum of the black
hole will have on searches for neutron-star black-hole binaries with
\ac{aLIGO}. The black hole's angular momentum will affect the phase evolution
of the emitted gravitational-wave signal, and, if the angular momentum is
misaligned with the orbital plane, will cause the system to precess. We have
found that if these effects are neglected in the filter waveforms used to
search for \ac{NSBH} binaries it will result in a loss in detection rate of
$31\%-36\%$ when searching for \ac{NSBH} systems with masses uniformly 
distributed in the range 
$(3-15,1-3)M_{\odot}$. When restricting the masses to 
$(9.5-10.5,1.35-1.45)M_{\odot}$ we find that the loss in detection rate is
$33\%-37\%$. The error in these measurements is due to uncertainty in 
the \ac{PN} waveform models used to simulate \ac{NSBH} gravitational-wave 
signals.  In a companion work we investigate how the uncertainty in waveform 
models used to simulate \ac{NSBH} waveforms will reduce detection 
efficiency~\cite{Nitz:2013mxa}. 

We have presented a new method to create a template bank of \ac{NSBH} filter
waveforms, where the black hole's angular momentum is
included but is restricted to be (anti)aligned with the orbit. These
waveforms will include the effect that the black hole's angular momentum has on
the phase evolution of the gravitational-wave signal, but will not include any
precessional effects. We have shown that this bank offers a
$16\%-30\%$ improvement in the detection rate of neutron-star--black-hole
mergers when compared to a nonspinning template bank when searching for 
\ac{NSBH} systems with masses in the range $(3-15,1-3)M_{\odot}$. However, when
searching for \ac{NSBH} systems with masses restricted to the range 
$(9.5-10.5,1.35-1.45)M_{\odot}$ we find the improvement is reduced to 
$5\%-17\%$.
Some systems are not recovered well with this new bank of filters. These systems
are ones where the black-hole spin is misaligned with the orbit and the waveform
is significantly modified due to precession of the orbital plane. This happens
most often when $m_{BH} / m_{NS}$ and the spin magnitude are both large. In
\cite{Brown:2012gs} the authors predict where in the parameter space to expect
\ac{NSBH} systems that will not be recovered well by nonprecessing template
banks. These predictions were
given in terms of the angles between the orbital plane, the black hole's angular
momentum and the line of sight to an observer. These predictions agree with the
results that we obtain in this work. In \cite{Ajith:2012mn} the authors claim
that an aligned-spin template bank will be effectual for detecting precessing
\ac{NSBH} systems. In this work, we find that with an aligned-spin template
bank $17\%-23\%$ of \ac{NSBH} systems will be missed compared to an ideal search
with
exactly matching filter waveforms. In reality this ideal search could never
be performed as it would require an infinite number of filter waveforms.
Template banks are usually constructed to allow for no more than a 3\% loss in
\ac{SNR}; therefore, we expect to lose up to $10\%$ of systems even if the
template bank fully covers the signal parameter space. We therefore conclude
that searches using precessing waveforms as templates could potentially
increase the detection rate of \ac{NSBH} signals, but not by more than $\sim
20\%$. Performing such a search would, however, remove an observational bias
against systems where precessional effects are most prevalent in the
gravitational-wave signal.

These figures are also affected by the 
parameter distribution chosen for the \ac{NSBH} systems. Here we chose a 
distribution that is uniform in mass, uniform in spin magnitudes, isotropic in 
spin orientations and isotropic in orientation parameters and sky location. We 
have, however, explored how the ability to detect precessing \ac{NSBH} signals 
varies as a function of the masses and spins as seen in Figs. 
\ref{fig:aspinavFF} and \ref{fig:arseofsauron}. 

When searching for \ac{NSBH} systems in \ac{aLIGO} one has to
consider the non-Gaussianity of the background noise, which we have not done in
this work. A non-Gaussian noise
artifact can produce \acp{SNR} that are considerably larger than those expected
from Gaussian noise fluctuations. To deal with this, numerous
consistency tests are used in the analyses to separate gravitational-wave
signals from instrumental noise artifacts \cite{Babak:2012zx}. It is possible
that the detection rate could be further reduced from the values we quote in
this work if some signals \emph{fail} these consistency tests and are
misclassified as non-Gaussian noise transients.
However, these signal consistency tests should only act to remove, or reduce the
significance of, events that already have low fitting factors and therefore do
not match well with the search templates. 
Another important consideration is that of the
number of templates used in the bank. To achieve higher fitting factors will
require more template waveforms, covering a larger signal space, which will 
allow more freedom in matching the
background noise and will mean that the \ac{SNR} of the loudest background
triggers will increase. Therefore signals will need slightly higher \acp{SNR} to
achieve the same false-alarm probability. However, a factor of 10 increase in
the number of \emph{independent} templates will only increase the expected
\ac{SNR} of the loudest background event by less than $5\%$, if 
Gaussian noise is assumed. Therefore, while we
are careful to note these considerations, we do not believe they will have a
large impact on the numbers we quote above and leave a detailed investigation of
such effects to future work.

In this work we have restricted ourselves to 
considering post-Newtonian, inspiral-only signal waveforms and consider only the 
case of two point particles. This was done as there is not currently any widely 
available waveform model that includes both the full evolution of a \ac{NSBH} 
coalescence \emph{and} includes precessional effects over the full parameter 
space that we consider. When such a model is available it may be that 
tidal forces and the merger component of the waveform may affect our 
conclusions. We believe that such effects will be limited as tidal effects are 
not expected to be important for detection of \ac{NSBH} systems with 
\ac{aLIGO}~\cite{Foucart:2013psa} 
and merger physics becomes increasingly important at higher 
masses~\cite{Brown:2012nn,Smith:2013mfa}, while we have restricted the 
black-hole mass to $< 15M_{\odot}$ in our simulations.
However it would be informative to repeat our simulations when a full \ac{NSBH} 
waveform model including tidal and merger physics is available.

\section*{Acknowledgements}
We thank Stefan Ballmer, Alessandra Buonanno, Eliu Huerta, Prayush Kumar, Richard O'Shaughnessy,
B.~S.~Sathyaprakash, Peter Saulson, Matt West and Karl Wette for useful 
discussions. We also thank Frank Ohme and the anonymous referee for providing 
thoughtful and insightful comments on this manuscript. This work is supported
by National Science Foundation awards PHY-0847611 (DAB, AHN), PHY-1205835
(AHN, IWH), PHY-0970074 (EO), PHY-0855589 (AL) and PHY11-25915 (DAB,IWH,EO,AL). 
DAB, IWH, AL, and EO thank the Kavli Institute for Theoretical Physics at
Santa Barbara University, supported in part by NSF grant PHY-0551164, for
hospitality during this work. DAB thanks the LIGO Laboratory Visitors Program, supported by NSF
cooperative agreement PHY-0757058, for hospitality.
DK and AL thank the Max
Planck Gesellschaft for support. DAB is supported by a Cottrell Scholar award
from the Research Corporation for Science Advancement.  Computations used in
this work were performed on the Syracuse University Gravitation and Relativity
cluster, which is supported by NSF awards PHY-1040231 and PHY-1104371.

\bibliography{references}

\begin{thebibliography}{92}%
\makeatletter
\providecommand \@ifxundefined [1]{%
 \@ifx{#1\undefined}
}%
\providecommand \@ifnum [1]{%
 \ifnum #1\expandafter \@firstoftwo
 \else \expandafter \@secondoftwo
 \fi
}%
\providecommand \@ifx [1]{%
 \ifx #1\expandafter \@firstoftwo
 \else \expandafter \@secondoftwo
 \fi
}%
\providecommand \natexlab [1]{#1}%
\providecommand \enquote  [1]{``#1''}%
\providecommand \bibnamefont  [1]{#1}%
\providecommand \bibfnamefont [1]{#1}%
\providecommand \citenamefont [1]{#1}%
\providecommand \href@noop [0]{\@secondoftwo}%
\providecommand \href [0]{\begingroup \@sanitize@url \@href}%
\providecommand \@href[1]{\@@startlink{#1}\@@href}%
\providecommand \@@href[1]{\endgroup#1\@@endlink}%
\providecommand \@sanitize@url [0]{\catcode `\\12\catcode `\$12\catcode
  `\&12\catcode `\#12\catcode `\^12\catcode `\_12\catcode `\%12\relax}%
\providecommand \@@startlink[1]{}%
\providecommand \@@endlink[0]{}%
\providecommand \url  [0]{\begingroup\@sanitize@url \@url }%
\providecommand \@url [1]{\endgroup\@href {#1}{\urlprefix }}%
\providecommand \urlprefix  [0]{URL }%
\providecommand \Eprint [0]{\href }%
\providecommand \doibase [0]{http://dx.doi.org/}%
\providecommand \selectlanguage [0]{\@gobble}%
\providecommand \bibinfo  [0]{\@secondoftwo}%
\providecommand \bibfield  [0]{\@secondoftwo}%
\providecommand \translation [1]{[#1]}%
\providecommand \BibitemOpen [0]{}%
\providecommand \bibitemStop [0]{}%
\providecommand \bibitemNoStop [0]{.\EOS\space}%
\providecommand \EOS [0]{\spacefactor3000\relax}%
\providecommand \BibitemShut  [1]{\csname bibitem#1\endcsname}%
\let\auto@bib@innerbib\@empty
\bibitem [{\citenamefont {Aasi}\ \emph {et~al.}(2013)\citenamefont {Aasi} \emph
  {et~al.}}]{Aasi:2013wya}%
  \BibitemOpen
  \bibfield  {author} {\bibinfo {author} {\bibfnamefont {J.}~\bibnamefont
  {Aasi}} \emph {et~al.} (\bibinfo {collaboration} {LIGO Scientific
  Collaboration, Virgo Collaboration}),\ }\href@noop {} {\  (\bibinfo {year}
  {2013})},\ \Eprint {http://arxiv.org/abs/1304.0670} {arXiv:1304.0670 [gr-qc]}
  \BibitemShut {NoStop}%
\bibitem [{\citenamefont {Harry}\ \emph {et~al.}(2010)\citenamefont {Harry}
  \emph {et~al.}}]{Harry:2010zz}%
  \BibitemOpen
  \bibfield  {author} {\bibinfo {author} {\bibfnamefont {G.~M.}\ \bibnamefont
  {Harry}} \emph {et~al.},\ }\href {\doibase 10.1088/0264-9381/27/8/084006}
  {\bibfield  {journal} {\bibinfo  {journal} {Class. Quant. Grav.}\ }\textbf
  {\bibinfo {volume} {27}},\ \bibinfo {pages} {084006} (\bibinfo {year}
  {2010})}\BibitemShut {NoStop}%
\bibitem [{\citenamefont {Acernese}\ \emph {et~al.}(2009)\citenamefont
  {Acernese} \emph {et~al.}}]{aVirgo}%
  \BibitemOpen
  \bibfield  {author} {\bibinfo {author} {\bibfnamefont {F.}~\bibnamefont
  {Acernese}} \emph {et~al.},\ }\href@noop {} {\  (\bibinfo {year} {2009})},\
  \bibinfo {note} {{Virgo Technical Report 0027A-09}}\BibitemShut {NoStop}%
\bibitem [{\citenamefont {Acernese}\ \emph {et~al.}()\citenamefont {Acernese}
  \emph {et~al.}}]{AdV2}%
  \BibitemOpen
  \bibfield  {author} {\bibinfo {author} {\bibfnamefont {F.}~\bibnamefont
  {Acernese}} \emph {et~al.},\ }\enquote {\bibinfo {title} {Plans for the
  upgrade of the gravitational wave detector virgo: Advanced virgo},}\ in\
  \href {\doibase 10.1142/9789814374552_0313} {\emph {\bibinfo {booktitle} {The
  Twelfth Marcel Grossmann Meeting}}},\ Chap.\ \bibinfo {chapter} {313}, pp.\
  \bibinfo {pages} {1738--1742}\BibitemShut {NoStop}%
\bibitem [{\citenamefont {Belczynski}\ \emph {et~al.}(2013)\citenamefont
  {Belczynski}, \citenamefont {Bulik}, \citenamefont {Mandel}, \citenamefont
  {Sathyaprakash}, \citenamefont {Zdziarski} \emph
  {et~al.}}]{Belczynski:2012jc}%
  \BibitemOpen
  \bibfield  {author} {\bibinfo {author} {\bibfnamefont {K.}~\bibnamefont
  {Belczynski}}, \bibinfo {author} {\bibfnamefont {T.}~\bibnamefont {Bulik}},
  \bibinfo {author} {\bibfnamefont {I.}~\bibnamefont {Mandel}}, \bibinfo
  {author} {\bibfnamefont {B.}~\bibnamefont {Sathyaprakash}}, \bibinfo {author}
  {\bibfnamefont {A.~A.}\ \bibnamefont {Zdziarski}},  \emph {et~al.},\ }\href
  {\doibase 10.1088/0004-637X/764/1/96} {\bibfield  {journal} {\bibinfo
  {journal} {Astrophys.J.}\ }\textbf {\bibinfo {volume} {764}},\ \bibinfo
  {pages} {96} (\bibinfo {year} {2013})},\ \Eprint
  {http://arxiv.org/abs/1209.2658} {arXiv:1209.2658 [astro-ph.HE]} \BibitemShut
  {NoStop}%
\bibitem [{\citenamefont {Abadie}\ \emph
  {et~al.}(2010{\natexlab{a}})\citenamefont {Abadie} \emph
  {et~al.}}]{Abadie:2010cf}%
  \BibitemOpen
  \bibfield  {author} {\bibinfo {author} {\bibfnamefont {J.}~\bibnamefont
  {Abadie}} \emph {et~al.} (\bibinfo {collaboration} {LIGO Scientific
  Collaboration, Virgo Collaboration}),\ }\href {\doibase
  10.1088/0264-9381/27/17/173001} {\bibfield  {journal} {\bibinfo  {journal}
  {Class.Quant.Grav.}\ }\textbf {\bibinfo {volume} {27}},\ \bibinfo {pages}
  {173001} (\bibinfo {year} {2010}{\natexlab{a}})},\ \Eprint
  {http://arxiv.org/abs/1003.2480} {arXiv:1003.2480 [astro-ph.HE]} \BibitemShut
  {NoStop}%
\bibitem [{\citenamefont {Cutler}\ \emph {et~al.}(1993)\citenamefont {Cutler},
  \citenamefont {Apostolatos}, \citenamefont {Bildsten}, \citenamefont {Finn},
  \citenamefont {Flanagan} \emph {et~al.}}]{Cutler:1992tc}%
  \BibitemOpen
  \bibfield  {author} {\bibinfo {author} {\bibfnamefont {C.}~\bibnamefont
  {Cutler}}, \bibinfo {author} {\bibfnamefont {T.~A.}\ \bibnamefont
  {Apostolatos}}, \bibinfo {author} {\bibfnamefont {L.}~\bibnamefont
  {Bildsten}}, \bibinfo {author} {\bibfnamefont {L.~S.}\ \bibnamefont {Finn}},
  \bibinfo {author} {\bibfnamefont {E.~E.}\ \bibnamefont {Flanagan}},  \emph
  {et~al.},\ }\href {\doibase 10.1103/PhysRevLett.70.2984} {\bibfield
  {journal} {\bibinfo  {journal} {Phys.Rev.Lett.}\ }\textbf {\bibinfo {volume}
  {70}},\ \bibinfo {pages} {2984} (\bibinfo {year} {1993})},\ \Eprint
  {http://arxiv.org/abs/astro-ph/9208005} {arXiv:astro-ph/9208005 [astro-ph]}
  \BibitemShut {NoStop}%
\bibitem [{\citenamefont {Apostolatos}\ \emph {et~al.}(1994)\citenamefont
  {Apostolatos}, \citenamefont {Cutler}, \citenamefont {Sussman},\ and\
  \citenamefont {Thorne}}]{Apostolatos:1994mx}%
  \BibitemOpen
  \bibfield  {author} {\bibinfo {author} {\bibfnamefont {T.~A.}\ \bibnamefont
  {Apostolatos}}, \bibinfo {author} {\bibfnamefont {C.}~\bibnamefont {Cutler}},
  \bibinfo {author} {\bibfnamefont {G.~J.}\ \bibnamefont {Sussman}}, \ and\
  \bibinfo {author} {\bibfnamefont {K.~S.}\ \bibnamefont {Thorne}},\ }\href
  {\doibase 10.1103/PhysRevD.49.6274} {\bibfield  {journal} {\bibinfo
  {journal} {Phys. Rev.}\ }\textbf {\bibinfo {volume} {D49}},\ \bibinfo {pages}
  {6274} (\bibinfo {year} {1994})}\BibitemShut {NoStop}%
\bibitem [{\citenamefont {Kidder}\ \emph {et~al.}(1993)\citenamefont {Kidder},
  \citenamefont {Will},\ and\ \citenamefont {Wiseman}}]{Kidder:1992fr}%
  \BibitemOpen
  \bibfield  {author} {\bibinfo {author} {\bibfnamefont {L.~E.}\ \bibnamefont
  {Kidder}}, \bibinfo {author} {\bibfnamefont {C.~M.}\ \bibnamefont {Will}}, \
  and\ \bibinfo {author} {\bibfnamefont {A.~G.}\ \bibnamefont {Wiseman}},\
  }\href {\doibase 10.1103/PhysRevD.47.R4183} {\bibfield  {journal} {\bibinfo
  {journal} {Phys.Rev.}\ }\textbf {\bibinfo {volume} {D47}},\ \bibinfo {pages}
  {4183} (\bibinfo {year} {1993})},\ \Eprint
  {http://arxiv.org/abs/gr-qc/9211025} {arXiv:gr-qc/9211025 [gr-qc]}
  \BibitemShut {NoStop}%
\bibitem [{\citenamefont {Kidder}(1995)}]{Kidder:1995zr}%
  \BibitemOpen
  \bibfield  {author} {\bibinfo {author} {\bibfnamefont {L.~E.}\ \bibnamefont
  {Kidder}},\ }\href {\doibase 10.1103/PhysRevD.52.821} {\bibfield  {journal}
  {\bibinfo  {journal} {Phys.Rev.}\ }\textbf {\bibinfo {volume} {D52}},\
  \bibinfo {pages} {821} (\bibinfo {year} {1995})},\ \Eprint
  {http://arxiv.org/abs/gr-qc/9506022} {arXiv:gr-qc/9506022 [gr-qc]}
  \BibitemShut {NoStop}%
\bibitem [{\citenamefont {Poisson}(1998)}]{Poisson:1997ha}%
  \BibitemOpen
  \bibfield  {author} {\bibinfo {author} {\bibfnamefont {E.}~\bibnamefont
  {Poisson}},\ }\href {\doibase 10.1103/PhysRevD.57.5287} {\bibfield  {journal}
  {\bibinfo  {journal} {Phys.Rev.}\ }\textbf {\bibinfo {volume} {D57}},\
  \bibinfo {pages} {5287} (\bibinfo {year} {1998})},\ \Eprint
  {http://arxiv.org/abs/gr-qc/9709032} {arXiv:gr-qc/9709032 [gr-qc]}
  \BibitemShut {NoStop}%
\bibitem [{\citenamefont {Mikoczi}\ \emph {et~al.}(2005)\citenamefont
  {Mikoczi}, \citenamefont {Vasuth},\ and\ \citenamefont
  {Gergely}}]{Mikoczi:2005dn}%
  \BibitemOpen
  \bibfield  {author} {\bibinfo {author} {\bibfnamefont {B.}~\bibnamefont
  {Mikoczi}}, \bibinfo {author} {\bibfnamefont {M.}~\bibnamefont {Vasuth}}, \
  and\ \bibinfo {author} {\bibfnamefont {L.~A.}\ \bibnamefont {Gergely}},\
  }\href {\doibase 10.1103/PhysRevD.71.124043} {\bibfield  {journal} {\bibinfo
  {journal} {Phys.Rev.}\ }\textbf {\bibinfo {volume} {D71}},\ \bibinfo {pages}
  {124043} (\bibinfo {year} {2005})},\ \Eprint
  {http://arxiv.org/abs/astro-ph/0504538} {arXiv:astro-ph/0504538 [astro-ph]}
  \BibitemShut {NoStop}%
\bibitem [{\citenamefont {Nitz}\ \emph {et~al.}(2013)\citenamefont {Nitz},
  \citenamefont {Lundgren}, \citenamefont {Brown}, \citenamefont {Ochsner},
  \citenamefont {Keppel} \emph {et~al.}}]{Nitz:2013mxa}%
  \BibitemOpen
  \bibfield  {author} {\bibinfo {author} {\bibfnamefont {A.~H.}\ \bibnamefont
  {Nitz}}, \bibinfo {author} {\bibfnamefont {A.}~\bibnamefont {Lundgren}},
  \bibinfo {author} {\bibfnamefont {D.~A.}\ \bibnamefont {Brown}}, \bibinfo
  {author} {\bibfnamefont {E.}~\bibnamefont {Ochsner}}, \bibinfo {author}
  {\bibfnamefont {D.}~\bibnamefont {Keppel}},  \emph {et~al.},\ }\href@noop {}
  {\  (\bibinfo {year} {2013})},\ \Eprint {http://arxiv.org/abs/1307.1757}
  {arXiv:1307.1757 [gr-qc]} \BibitemShut {NoStop}%
\bibitem [{\citenamefont {Wainstein}\ and\ \citenamefont
  {Zubakov}(1962)}]{Wainstein}%
  \BibitemOpen
  \bibfield  {author} {\bibinfo {author} {\bibfnamefont {L.~A.}\ \bibnamefont
  {Wainstein}}\ and\ \bibinfo {author} {\bibfnamefont {V.~D.}\ \bibnamefont
  {Zubakov}},\ }\href@noop {} {\emph {\bibinfo {title} {Extraction of Signals
  from Noise}}}\ (\bibinfo  {publisher} {Prentice-Hall},\ \bibinfo {address}
  {Englewood Cliffs},\ \bibinfo {year} {1962})\BibitemShut {NoStop}%
\bibitem [{\citenamefont {Wainstein}\ and\ \citenamefont
  {Zubakov}(1968)}]{Helstrom}%
  \BibitemOpen
  \bibfield  {author} {\bibinfo {author} {\bibfnamefont {L.~A.}\ \bibnamefont
  {Wainstein}}\ and\ \bibinfo {author} {\bibfnamefont {V.~D.}\ \bibnamefont
  {Zubakov}},\ }\href@noop {} {\emph {\bibinfo {title} {Statistical Theory of
  Signal Detection}}}\ (\bibinfo  {publisher} {Permagon},\ \bibinfo {address}
  {London},\ \bibinfo {year} {1968})\BibitemShut {NoStop}%
\bibitem [{\citenamefont {Allen}\ \emph {et~al.}(2012)\citenamefont {Allen},
  \citenamefont {Anderson}, \citenamefont {Brady}, \citenamefont {Brown},\ and\
  \citenamefont {Creighton}}]{Allen:2005fk}%
  \BibitemOpen
  \bibfield  {author} {\bibinfo {author} {\bibfnamefont {B.}~\bibnamefont
  {Allen}}, \bibinfo {author} {\bibfnamefont {W.~G.}\ \bibnamefont {Anderson}},
  \bibinfo {author} {\bibfnamefont {P.~R.}\ \bibnamefont {Brady}}, \bibinfo
  {author} {\bibfnamefont {D.~A.}\ \bibnamefont {Brown}}, \ and\ \bibinfo
  {author} {\bibfnamefont {J.~D.~E.}\ \bibnamefont {Creighton}},\ }\href
  {\doibase 10.1103/PhysRevD.85.122006} {\bibfield  {journal} {\bibinfo
  {journal} {Phys.Rev.}\ }\textbf {\bibinfo {volume} {D85}},\ \bibinfo {pages}
  {122006} (\bibinfo {year} {2012})},\ \Eprint
  {http://arxiv.org/abs/gr-qc/0509116} {arXiv:gr-qc/0509116 [gr-qc]}
  \BibitemShut {NoStop}%
\bibitem [{\citenamefont {Peters}\ and\ \citenamefont
  {Mathews}(1963)}]{Peters:1963ux}%
  \BibitemOpen
  \bibfield  {author} {\bibinfo {author} {\bibfnamefont {P.~C.}\ \bibnamefont
  {Peters}}\ and\ \bibinfo {author} {\bibfnamefont {J.}~\bibnamefont
  {Mathews}},\ }\href {\doibase 10.1103/PhysRev.131.435} {\bibfield  {journal}
  {\bibinfo  {journal} {Phys. Rev.}\ }\textbf {\bibinfo {volume} {131}},\
  \bibinfo {pages} {435} (\bibinfo {year} {1963})}\BibitemShut {NoStop}%
\bibitem [{\citenamefont {Thorne}(1987)}]{Th300}%
  \BibitemOpen
  \bibfield  {author} {\bibinfo {author} {\bibfnamefont {K.~S.}\ \bibnamefont
  {Thorne}},\ }in\ \href@noop {} {\emph {\bibinfo {booktitle} {Three hundred
  years of gravitation}}},\ \bibinfo {editor} {edited by\ \bibinfo {editor}
  {\bibfnamefont {S.}~\bibnamefont {Hawking}}\ and\ \bibinfo {editor}
  {\bibfnamefont {W.}~\bibnamefont {Israel}}}\ (\bibinfo  {publisher}
  {Cambridge University Press},\ \bibinfo {year} {1987})\ pp.\ \bibinfo {pages}
  {330--458}\BibitemShut {NoStop}%
\bibitem [{\citenamefont {Sathyaprakash}\ and\ \citenamefont
  {Dhurandhar}(1991)}]{Sathyaprakash:1991mt}%
  \BibitemOpen
  \bibfield  {author} {\bibinfo {author} {\bibfnamefont {B.~S.}\ \bibnamefont
  {Sathyaprakash}}\ and\ \bibinfo {author} {\bibfnamefont {S.~V.}\ \bibnamefont
  {Dhurandhar}},\ }\href {\doibase 10.1103/PhysRevD.44.3819} {\bibfield
  {journal} {\bibinfo  {journal} {Phys.Rev.}\ }\textbf {\bibinfo {volume}
  {D44}},\ \bibinfo {pages} {3819} (\bibinfo {year} {1991})}\BibitemShut
  {NoStop}%
\bibitem [{\citenamefont {Poisson}\ and\ \citenamefont
  {Will}(1995)}]{Poisson:1995ef}%
  \BibitemOpen
  \bibfield  {author} {\bibinfo {author} {\bibfnamefont {E.}~\bibnamefont
  {Poisson}}\ and\ \bibinfo {author} {\bibfnamefont {C.~M.}\ \bibnamefont
  {Will}},\ }\href {\doibase 10.1103/PhysRevD.52.848} {\bibfield  {journal}
  {\bibinfo  {journal} {Phys.Rev.}\ }\textbf {\bibinfo {volume} {D52}},\
  \bibinfo {pages} {848} (\bibinfo {year} {1995})},\ \Eprint
  {http://arxiv.org/abs/gr-qc/9502040} {arXiv:gr-qc/9502040 [gr-qc]}
  \BibitemShut {NoStop}%
\bibitem [{\citenamefont {Owen}(1996)}]{Owen:1995tm}%
  \BibitemOpen
  \bibfield  {author} {\bibinfo {author} {\bibfnamefont {B.~J.}\ \bibnamefont
  {Owen}},\ }\href {\doibase 10.1103/PhysRevD.53.6749} {\bibfield  {journal}
  {\bibinfo  {journal} {Phys.Rev.}\ }\textbf {\bibinfo {volume} {D53}},\
  \bibinfo {pages} {6749} (\bibinfo {year} {1996})},\ \Eprint
  {http://arxiv.org/abs/gr-qc/9511032} {arXiv:gr-qc/9511032 [gr-qc]}
  \BibitemShut {NoStop}%
\bibitem [{\citenamefont {Owen}\ and\ \citenamefont
  {Sathyaprakash}(1999)}]{Owen:1998dk}%
  \BibitemOpen
  \bibfield  {author} {\bibinfo {author} {\bibfnamefont {B.~J.}\ \bibnamefont
  {Owen}}\ and\ \bibinfo {author} {\bibfnamefont {B.~S.}\ \bibnamefont
  {Sathyaprakash}},\ }\href {\doibase 10.1103/PhysRevD.60.022002} {\bibfield
  {journal} {\bibinfo  {journal} {Phys.Rev.}\ }\textbf {\bibinfo {volume}
  {D60}},\ \bibinfo {pages} {022002} (\bibinfo {year} {1999})},\ \Eprint
  {http://arxiv.org/abs/gr-qc/9808076} {arXiv:gr-qc/9808076 [gr-qc]}
  \BibitemShut {NoStop}%
\bibitem [{\citenamefont {Babak}\ \emph {et~al.}(2006)\citenamefont {Babak},
  \citenamefont {Balasubramanian}, \citenamefont {Churches}, \citenamefont
  {Cokelaer},\ and\ \citenamefont {Sathyaprakash}}]{Babak:2006ty}%
  \BibitemOpen
  \bibfield  {author} {\bibinfo {author} {\bibfnamefont {S.}~\bibnamefont
  {Babak}}, \bibinfo {author} {\bibfnamefont {R.}~\bibnamefont
  {Balasubramanian}}, \bibinfo {author} {\bibfnamefont {D.}~\bibnamefont
  {Churches}}, \bibinfo {author} {\bibfnamefont {T.}~\bibnamefont {Cokelaer}},
  \ and\ \bibinfo {author} {\bibfnamefont {B.}~\bibnamefont {Sathyaprakash}},\
  }\href {\doibase 10.1088/0264-9381/23/18/002} {\bibfield  {journal} {\bibinfo
   {journal} {Class.Quant.Grav.}\ }\textbf {\bibinfo {volume} {23}},\ \bibinfo
  {pages} {5477} (\bibinfo {year} {2006})},\ \Eprint
  {http://arxiv.org/abs/gr-qc/0604037} {arXiv:gr-qc/0604037 [gr-qc]}
  \BibitemShut {NoStop}%
\bibitem [{\citenamefont {Balasubramanian}\ \emph {et~al.}(1996)\citenamefont
  {Balasubramanian}, \citenamefont {Sathyaprakash},\ and\ \citenamefont
  {Dhurandhar}}]{Balasubramanian:1995bm}%
  \BibitemOpen
  \bibfield  {author} {\bibinfo {author} {\bibfnamefont {R.}~\bibnamefont
  {Balasubramanian}}, \bibinfo {author} {\bibfnamefont {B.~S.}\ \bibnamefont
  {Sathyaprakash}}, \ and\ \bibinfo {author} {\bibfnamefont {S.~V.}\
  \bibnamefont {Dhurandhar}},\ }\href {\doibase 10.1103/PhysRevD.54.1860.2,
  10.1103/PhysRevD.53.3033} {\bibfield  {journal} {\bibinfo  {journal}
  {Phys.Rev.}\ }\textbf {\bibinfo {volume} {D53}},\ \bibinfo {pages} {3033}
  (\bibinfo {year} {1996})},\ \Eprint {http://arxiv.org/abs/gr-qc/9508011}
  {arXiv:gr-qc/9508011 [gr-qc]} \BibitemShut {NoStop}%
\bibitem [{\citenamefont {Cokelaer}(2007)}]{Cokelaer:2007kx}%
  \BibitemOpen
  \bibfield  {author} {\bibinfo {author} {\bibfnamefont {T.}~\bibnamefont
  {Cokelaer}},\ }\href {\doibase 10.1103/PhysRevD.76.102004} {\bibfield
  {journal} {\bibinfo  {journal} {Phys.Rev.}\ }\textbf {\bibinfo {volume}
  {D76}},\ \bibinfo {pages} {102004} (\bibinfo {year} {2007})},\ \Eprint
  {http://arxiv.org/abs/0706.4437} {arXiv:0706.4437 [gr-qc]} \BibitemShut
  {NoStop}%
\bibitem [{\citenamefont {Babak}\ \emph {et~al.}(2013)\citenamefont {Babak},
  \citenamefont {Biswas}, \citenamefont {Brady}, \citenamefont {Brown},
  \citenamefont {Cannon} \emph {et~al.}}]{Babak:2012zx}%
  \BibitemOpen
  \bibfield  {author} {\bibinfo {author} {\bibfnamefont {S.}~\bibnamefont
  {Babak}}, \bibinfo {author} {\bibfnamefont {R.}~\bibnamefont {Biswas}},
  \bibinfo {author} {\bibfnamefont {P.}~\bibnamefont {Brady}}, \bibinfo
  {author} {\bibfnamefont {D.}~\bibnamefont {Brown}}, \bibinfo {author}
  {\bibfnamefont {K.}~\bibnamefont {Cannon}},  \emph {et~al.},\ }\href
  {\doibase 10.1103/PhysRevD.87.024033} {\bibfield  {journal} {\bibinfo
  {journal} {Phys.Rev.}\ }\textbf {\bibinfo {volume} {D87}},\ \bibinfo {pages}
  {024033} (\bibinfo {year} {2013})},\ \Eprint {http://arxiv.org/abs/1208.3491}
  {arXiv:1208.3491 [gr-qc]} \BibitemShut {NoStop}%
\bibitem [{\citenamefont {Abbott}\ \emph
  {et~al.}(2009{\natexlab{a}})\citenamefont {Abbott} \emph
  {et~al.}}]{Abbott:2009tt}%
  \BibitemOpen
  \bibfield  {author} {\bibinfo {author} {\bibfnamefont {B.~P.}\ \bibnamefont
  {Abbott}} \emph {et~al.} (\bibinfo {collaboration} {LIGO Scientific
  Collaboration}),\ }\href {\doibase 10.1103/PhysRevD.79.122001} {\bibfield
  {journal} {\bibinfo  {journal} {Phys. Rev.}\ }\textbf {\bibinfo {volume}
  {D79}},\ \bibinfo {pages} {122001} (\bibinfo {year} {2009}{\natexlab{a}})},\
  \Eprint {http://arxiv.org/abs/0901.0302} {arXiv:0901.0302 [gr-qc]}
  \BibitemShut {NoStop}%
\bibitem [{\citenamefont {Abbott}\ \emph
  {et~al.}(2009{\natexlab{b}})\citenamefont {Abbott} \emph
  {et~al.}}]{Abbott:2009qj}%
  \BibitemOpen
  \bibfield  {author} {\bibinfo {author} {\bibfnamefont {B.~P.}\ \bibnamefont
  {Abbott}} \emph {et~al.} (\bibinfo {collaboration} {LIGO Scientific
  Collaboration}),\ }\href {\doibase 10.1103/PhysRevD.80.047101} {\bibfield
  {journal} {\bibinfo  {journal} {Phys. Rev.}\ }\textbf {\bibinfo {volume}
  {D80}},\ \bibinfo {pages} {047101} (\bibinfo {year} {2009}{\natexlab{b}})},\
  \Eprint {http://arxiv.org/abs/0905.3710} {arXiv:0905.3710 [gr-qc]}
  \BibitemShut {NoStop}%
\bibitem [{\citenamefont {Abadie}\ \emph
  {et~al.}(2010{\natexlab{b}})\citenamefont {Abadie} \emph
  {et~al.}}]{Abadie:2010yba}%
  \BibitemOpen
  \bibfield  {author} {\bibinfo {author} {\bibfnamefont {J.}~\bibnamefont
  {Abadie}} \emph {et~al.} (\bibinfo {collaboration} {LIGO and Virgo Scientific
  Collaborations}),\ }\href {\doibase 10.1103/PhysRevD.82.102001} {\bibfield
  {journal} {\bibinfo  {journal} {Phys. Rev.}\ }\textbf {\bibinfo {volume}
  {D82}},\ \bibinfo {pages} {102001} (\bibinfo {year} {2010}{\natexlab{b}})},\
  \Eprint {http://arxiv.org/abs/1005.4655} {arXiv:1005.4655 [gr-qc]}
  \BibitemShut {NoStop}%
\bibitem [{\citenamefont {Abadie}\ \emph {et~al.}(2012)\citenamefont {Abadie}
  \emph {et~al.}}]{Abadie:2011nz}%
  \BibitemOpen
  \bibfield  {author} {\bibinfo {author} {\bibfnamefont {J.}~\bibnamefont
  {Abadie}} \emph {et~al.},\ }\href {\doibase 10.1103/PhysRevD.85.082002}
  {\bibfield  {journal} {\bibinfo  {journal} {Phys.Rev.}\ }\textbf {\bibinfo
  {volume} {D85}},\ \bibinfo {pages} {082002} (\bibinfo {year} {2012})},\
  \Eprint {http://arxiv.org/abs/1111.7314} {arXiv:1111.7314 [gr-qc]}
  \BibitemShut {NoStop}%
\bibitem [{\citenamefont {Babak}(2008)}]{Babak:2008rb}%
  \BibitemOpen
  \bibfield  {author} {\bibinfo {author} {\bibfnamefont {S.}~\bibnamefont
  {Babak}},\ }\href {\doibase 10.1088/0264-9381/25/19/195011} {\bibfield
  {journal} {\bibinfo  {journal} {Class. Quant. Grav.}\ }\textbf {\bibinfo
  {volume} {25}},\ \bibinfo {pages} {195011} (\bibinfo {year} {2008})},\
  \Eprint {http://arxiv.org/abs/0801.4070} {arXiv:0801.4070 [gr-qc]}
  \BibitemShut {NoStop}%
\bibitem [{\citenamefont {Harry}\ \emph {et~al.}(2009)\citenamefont {Harry},
  \citenamefont {Allen},\ and\ \citenamefont {Sathyaprakash}}]{Harry:2009ea}%
  \BibitemOpen
  \bibfield  {author} {\bibinfo {author} {\bibfnamefont {I.~W.}\ \bibnamefont
  {Harry}}, \bibinfo {author} {\bibfnamefont {B.}~\bibnamefont {Allen}}, \ and\
  \bibinfo {author} {\bibfnamefont {B.~S.}\ \bibnamefont {Sathyaprakash}},\
  }\href {\doibase 10.1103/PhysRevD.80.104014} {\bibfield  {journal} {\bibinfo
  {journal} {Phys.Rev.}\ }\textbf {\bibinfo {volume} {D80}},\ \bibinfo {pages}
  {104014} (\bibinfo {year} {2009})},\ \Eprint {http://arxiv.org/abs/0908.2090}
  {arXiv:0908.2090 [gr-qc]} \BibitemShut {NoStop}%
\bibitem [{\citenamefont {Manca}\ and\ \citenamefont
  {Vallisneri}(2010)}]{Manca:2009xw}%
  \BibitemOpen
  \bibfield  {author} {\bibinfo {author} {\bibfnamefont {G.~M.}\ \bibnamefont
  {Manca}}\ and\ \bibinfo {author} {\bibfnamefont {M.}~\bibnamefont
  {Vallisneri}},\ }\href {\doibase 10.1103/PhysRevD.81.024004} {\bibfield
  {journal} {\bibinfo  {journal} {Phys.Rev.}\ }\textbf {\bibinfo {volume}
  {D81}},\ \bibinfo {pages} {024004} (\bibinfo {year} {2010})},\ \Eprint
  {http://arxiv.org/abs/0909.0563} {arXiv:0909.0563 [gr-qc]} \BibitemShut
  {NoStop}%
\bibitem [{\citenamefont {Ajith}\ \emph {et~al.}(2012)\citenamefont {Ajith},
  \citenamefont {Fotopoulos}, \citenamefont {Privitera}, \citenamefont
  {Neunzert},\ and\ \citenamefont {Weinstein}}]{Ajith:2012mn}%
  \BibitemOpen
  \bibfield  {author} {\bibinfo {author} {\bibfnamefont {P.}~\bibnamefont
  {Ajith}}, \bibinfo {author} {\bibfnamefont {N.}~\bibnamefont {Fotopoulos}},
  \bibinfo {author} {\bibfnamefont {S.}~\bibnamefont {Privitera}}, \bibinfo
  {author} {\bibfnamefont {A.}~\bibnamefont {Neunzert}}, \ and\ \bibinfo
  {author} {\bibfnamefont {A.}~\bibnamefont {Weinstein}},\ }\href@noop {} {\
  (\bibinfo {year} {2012})},\ \Eprint {http://arxiv.org/abs/1210.6666}
  {arXiv:1210.6666 [gr-qc]} \BibitemShut {NoStop}%
\bibitem [{\citenamefont {Brown}\ \emph
  {et~al.}(2012{\natexlab{a}})\citenamefont {Brown}, \citenamefont {Harry},
  \citenamefont {Lundgren},\ and\ \citenamefont {Nitz}}]{Brown:2012qf}%
  \BibitemOpen
  \bibfield  {author} {\bibinfo {author} {\bibfnamefont {D.~A.}\ \bibnamefont
  {Brown}}, \bibinfo {author} {\bibfnamefont {I.}~\bibnamefont {Harry}},
  \bibinfo {author} {\bibfnamefont {A.}~\bibnamefont {Lundgren}}, \ and\
  \bibinfo {author} {\bibfnamefont {A.~H.}\ \bibnamefont {Nitz}},\ }\href
  {\doibase 10.1103/PhysRevD.86.084017} {\bibfield  {journal} {\bibinfo
  {journal} {Phys.Rev.}\ }\textbf {\bibinfo {volume} {D86}},\ \bibinfo {pages}
  {084017} (\bibinfo {year} {2012}{\natexlab{a}})},\ \Eprint
  {http://arxiv.org/abs/1207.6406} {arXiv:1207.6406 [gr-qc]} \BibitemShut
  {NoStop}%
\bibitem [{\citenamefont {Apostolatos}(1996)}]{Apostolatos:1996rf}%
  \BibitemOpen
  \bibfield  {author} {\bibinfo {author} {\bibfnamefont {T.~A.}\ \bibnamefont
  {Apostolatos}},\ }\href {\doibase 10.1103/PhysRevD.54.2421} {\bibfield
  {journal} {\bibinfo  {journal} {Phys.Rev.}\ }\textbf {\bibinfo {volume}
  {D54}},\ \bibinfo {pages} {2421} (\bibinfo {year} {1996})}\BibitemShut
  {NoStop}%
\bibitem [{\citenamefont {Buonanno}\ \emph {et~al.}(2003)\citenamefont
  {Buonanno}, \citenamefont {Chen},\ and\ \citenamefont
  {Vallisneri}}]{Buonanno:2002fy}%
  \BibitemOpen
  \bibfield  {author} {\bibinfo {author} {\bibfnamefont {A.}~\bibnamefont
  {Buonanno}}, \bibinfo {author} {\bibfnamefont {Y.-b.}\ \bibnamefont {Chen}},
  \ and\ \bibinfo {author} {\bibfnamefont {M.}~\bibnamefont {Vallisneri}},\
  }\href {\doibase 10.1103/PhysRevD.67.104025, 10.1103/PhysRevD.74.029904}
  {\bibfield  {journal} {\bibinfo  {journal} {Phys.Rev.}\ }\textbf {\bibinfo
  {volume} {D67}},\ \bibinfo {pages} {104025} (\bibinfo {year} {2003})},\
  \Eprint {http://arxiv.org/abs/gr-qc/0211087} {arXiv:gr-qc/0211087 [gr-qc]}
  \BibitemShut {NoStop}%
\bibitem [{\citenamefont {Grandclement}\ \emph {et~al.}(2003)\citenamefont
  {Grandclement}, \citenamefont {Kalogera},\ and\ \citenamefont
  {Vecchio}}]{Grandclement:2002dv}%
  \BibitemOpen
  \bibfield  {author} {\bibinfo {author} {\bibfnamefont {P.}~\bibnamefont
  {Grandclement}}, \bibinfo {author} {\bibfnamefont {V.}~\bibnamefont
  {Kalogera}}, \ and\ \bibinfo {author} {\bibfnamefont {A.}~\bibnamefont
  {Vecchio}},\ }\href {\doibase 10.1103/PhysRevD.67.042003} {\bibfield
  {journal} {\bibinfo  {journal} {Phys. Rev.}\ }\textbf {\bibinfo {volume}
  {D67}},\ \bibinfo {pages} {042003} (\bibinfo {year} {2003})},\ \Eprint
  {http://arxiv.org/abs/gr-qc/0207062} {arXiv:gr-qc/0207062} \BibitemShut
  {NoStop}%
\bibitem [{\citenamefont {Grandclement}\ and\ \citenamefont
  {Kalogera}(2003)}]{Grandclement:2002vx}%
  \BibitemOpen
  \bibfield  {author} {\bibinfo {author} {\bibfnamefont {P.}~\bibnamefont
  {Grandclement}}\ and\ \bibinfo {author} {\bibfnamefont {V.}~\bibnamefont
  {Kalogera}},\ }\href {\doibase 10.1103/PhysRevD.67.082002} {\bibfield
  {journal} {\bibinfo  {journal} {Phys.Rev.}\ }\textbf {\bibinfo {volume}
  {D67}},\ \bibinfo {pages} {082002} (\bibinfo {year} {2003})},\ \Eprint
  {http://arxiv.org/abs/gr-qc/0211075} {arXiv:gr-qc/0211075 [gr-qc]}
  \BibitemShut {NoStop}%
\bibitem [{\citenamefont {Grandclement}\ \emph {et~al.}(2004)\citenamefont
  {Grandclement}, \citenamefont {Ihm}, \citenamefont {Kalogera},\ and\
  \citenamefont {Belczynski}}]{Grandclement:2003ck}%
  \BibitemOpen
  \bibfield  {author} {\bibinfo {author} {\bibfnamefont {P.}~\bibnamefont
  {Grandclement}}, \bibinfo {author} {\bibfnamefont {M.}~\bibnamefont {Ihm}},
  \bibinfo {author} {\bibfnamefont {V.}~\bibnamefont {Kalogera}}, \ and\
  \bibinfo {author} {\bibfnamefont {K.}~\bibnamefont {Belczynski}},\ }\href
  {\doibase 10.1103/PhysRevD.69.102002} {\bibfield  {journal} {\bibinfo
  {journal} {Phys.Rev.}\ }\textbf {\bibinfo {volume} {D69}},\ \bibinfo {pages}
  {102002} (\bibinfo {year} {2004})},\ \Eprint
  {http://arxiv.org/abs/gr-qc/0312084} {arXiv:gr-qc/0312084 [gr-qc]}
  \BibitemShut {NoStop}%
\bibitem [{\citenamefont {Pan}\ \emph {et~al.}(2004)\citenamefont {Pan},
  \citenamefont {Buonanno}, \citenamefont {Chen},\ and\ \citenamefont
  {Vallisneri}}]{Pan:2003qt}%
  \BibitemOpen
  \bibfield  {author} {\bibinfo {author} {\bibfnamefont {Y.}~\bibnamefont
  {Pan}}, \bibinfo {author} {\bibfnamefont {A.}~\bibnamefont {Buonanno}},
  \bibinfo {author} {\bibfnamefont {Y.-b.}\ \bibnamefont {Chen}}, \ and\
  \bibinfo {author} {\bibfnamefont {M.}~\bibnamefont {Vallisneri}},\ }\href
  {\doibase 10.1103/PhysRevD.69.104017, 10.1103/PhysRevD.74.029905} {\bibfield
  {journal} {\bibinfo  {journal} {Phys.Rev.}\ }\textbf {\bibinfo {volume}
  {D69}},\ \bibinfo {pages} {104017} (\bibinfo {year} {2004})},\ \Eprint
  {http://arxiv.org/abs/gr-qc/0310034} {arXiv:gr-qc/0310034 [gr-qc]}
  \BibitemShut {NoStop}%
\bibitem [{\citenamefont {Buonanno}\ \emph {et~al.}(2004)\citenamefont
  {Buonanno}, \citenamefont {Chen}, \citenamefont {Pan},\ and\ \citenamefont
  {Vallisneri}}]{Buonanno:2004yd}%
  \BibitemOpen
  \bibfield  {author} {\bibinfo {author} {\bibfnamefont {A.}~\bibnamefont
  {Buonanno}}, \bibinfo {author} {\bibfnamefont {Y.-b.}\ \bibnamefont {Chen}},
  \bibinfo {author} {\bibfnamefont {Y.}~\bibnamefont {Pan}}, \ and\ \bibinfo
  {author} {\bibfnamefont {M.}~\bibnamefont {Vallisneri}},\ }\href {\doibase
  10.1103/PhysRevD.74.029902, 10.1103/PhysRevD.70.104003} {\bibfield  {journal}
  {\bibinfo  {journal} {Phys.Rev.}\ }\textbf {\bibinfo {volume} {D70}},\
  \bibinfo {pages} {104003} (\bibinfo {year} {2004})},\ \Eprint
  {http://arxiv.org/abs/gr-qc/0405090} {arXiv:gr-qc/0405090 [gr-qc]}
  \BibitemShut {NoStop}%
\bibitem [{\citenamefont {Buonanno}\ \emph {et~al.}(2005)\citenamefont
  {Buonanno}, \citenamefont {Chen}, \citenamefont {Pan}, \citenamefont
  {Tagoshi},\ and\ \citenamefont {Vallisneri}}]{Buonanno:2005pt}%
  \BibitemOpen
  \bibfield  {author} {\bibinfo {author} {\bibfnamefont {A.}~\bibnamefont
  {Buonanno}}, \bibinfo {author} {\bibfnamefont {Y.}~\bibnamefont {Chen}},
  \bibinfo {author} {\bibfnamefont {Y.}~\bibnamefont {Pan}}, \bibinfo {author}
  {\bibfnamefont {H.}~\bibnamefont {Tagoshi}}, \ and\ \bibinfo {author}
  {\bibfnamefont {M.}~\bibnamefont {Vallisneri}},\ }\href {\doibase
  10.1103/PhysRevD.72.084027} {\bibfield  {journal} {\bibinfo  {journal}
  {Phys.Rev.}\ }\textbf {\bibinfo {volume} {D72}},\ \bibinfo {pages} {084027}
  (\bibinfo {year} {2005})},\ \Eprint {http://arxiv.org/abs/gr-qc/0508064}
  {arXiv:gr-qc/0508064 [gr-qc]} \BibitemShut {NoStop}%
\bibitem [{\citenamefont {Abbott}\ \emph {et~al.}(2008)\citenamefont {Abbott}
  \emph {et~al.}}]{Abbott:2007ai}%
  \BibitemOpen
  \bibfield  {author} {\bibinfo {author} {\bibfnamefont {B.}~\bibnamefont
  {Abbott}} \emph {et~al.} (\bibinfo {collaboration} {LIGO Scientific
  Collaboration}),\ }\href {\doibase 10.1103/PhysRevD.78.042002} {\bibfield
  {journal} {\bibinfo  {journal} {Phys. Rev.}\ }\textbf {\bibinfo {volume}
  {D78}},\ \bibinfo {pages} {042002} (\bibinfo {year} {2008})},\ \Eprint
  {http://arxiv.org/abs/0712.2050} {arXiv:0712.2050 [gr-qc]} \BibitemShut
  {NoStop}%
\bibitem [{\citenamefont {Van~Den~Broeck}\ \emph {et~al.}(2009)\citenamefont
  {Van~Den~Broeck}, \citenamefont {Brown}, \citenamefont {Cokelaer},
  \citenamefont {Harry}, \citenamefont {Jones} \emph
  {et~al.}}]{VanDenBroeck:2009gd}%
  \BibitemOpen
  \bibfield  {author} {\bibinfo {author} {\bibfnamefont {C.}~\bibnamefont
  {Van~Den~Broeck}}, \bibinfo {author} {\bibfnamefont {D.~A.}\ \bibnamefont
  {Brown}}, \bibinfo {author} {\bibfnamefont {T.}~\bibnamefont {Cokelaer}},
  \bibinfo {author} {\bibfnamefont {I.}~\bibnamefont {Harry}}, \bibinfo
  {author} {\bibfnamefont {G.}~\bibnamefont {Jones}},  \emph {et~al.},\ }\href
  {\doibase 10.1103/PhysRevD.80.024009} {\bibfield  {journal} {\bibinfo
  {journal} {Phys.Rev.}\ }\textbf {\bibinfo {volume} {D80}},\ \bibinfo {pages}
  {024009} (\bibinfo {year} {2009})},\ \Eprint {http://arxiv.org/abs/0904.1715}
  {arXiv:0904.1715 [gr-qc]} \BibitemShut {NoStop}%
\bibitem [{\citenamefont {Fazi}(2009)}]{Fazi:2009}%
  \BibitemOpen
  \bibfield  {author} {\bibinfo {author} {\bibfnamefont {D.}~\bibnamefont
  {Fazi}},\ }\emph {\bibinfo {title} {Development of a physical-template search
  for gravitational waves from spinning compact-object binaries with LIGO}},\
  \href@noop {} {Ph.D. thesis},\ \bibinfo  {school} {Universit\`{a} di Bologna}
  (\bibinfo {year} {2009})\BibitemShut {NoStop}%
\bibitem [{\citenamefont {Harry}\ and\ \citenamefont
  {Fairhurst}(2011)}]{Harry:2011qh}%
  \BibitemOpen
  \bibfield  {author} {\bibinfo {author} {\bibfnamefont {I.}~\bibnamefont
  {Harry}}\ and\ \bibinfo {author} {\bibfnamefont {S.}~\bibnamefont
  {Fairhurst}},\ }\href {\doibase 10.1088/0264-9381/28/13/134008} {\bibfield
  {journal} {\bibinfo  {journal} {Class.Quant.Grav.}\ }\textbf {\bibinfo
  {volume} {28}},\ \bibinfo {pages} {134008} (\bibinfo {year} {2011})},\
  \Eprint {http://arxiv.org/abs/1101.1459} {arXiv:1101.1459 [gr-qc]}
  \BibitemShut {NoStop}%
\bibitem [{\citenamefont {Brown}\ \emph
  {et~al.}(2012{\natexlab{b}})\citenamefont {Brown}, \citenamefont {Lundgren},\
  and\ \citenamefont {O'Shaughnessy}}]{Brown:2012gs}%
  \BibitemOpen
  \bibfield  {author} {\bibinfo {author} {\bibfnamefont {D.~A.}\ \bibnamefont
  {Brown}}, \bibinfo {author} {\bibfnamefont {A.}~\bibnamefont {Lundgren}}, \
  and\ \bibinfo {author} {\bibfnamefont {R.}~\bibnamefont {O'Shaughnessy}},\
  }\href {\doibase 10.1103/PhysRevD.86.064020} {\bibfield  {journal} {\bibinfo
  {journal} {Phys.Rev.}\ }\textbf {\bibinfo {volume} {D86}},\ \bibinfo {pages}
  {064020} (\bibinfo {year} {2012}{\natexlab{b}})},\ \Eprint
  {http://arxiv.org/abs/1203.6060} {arXiv:1203.6060 [gr-qc]} \BibitemShut
  {NoStop}%
\bibitem [{\citenamefont {Lundgren}\ and\ \citenamefont
  {O'Shaughnessy}(2013)}]{Lundgren:2013jla}%
  \BibitemOpen
  \bibfield  {author} {\bibinfo {author} {\bibfnamefont {A.}~\bibnamefont
  {Lundgren}}\ and\ \bibinfo {author} {\bibfnamefont {R.}~\bibnamefont
  {O'Shaughnessy}},\ }\href@noop {} {\  (\bibinfo {year} {2013})},\ \Eprint
  {http://arxiv.org/abs/1304.3332} {arXiv:1304.3332 [gr-qc]} \BibitemShut
  {NoStop}%
\bibitem [{\citenamefont {Duez}(2010)}]{Duez:2009yz}%
  \BibitemOpen
  \bibfield  {author} {\bibinfo {author} {\bibfnamefont {M.~D.}\ \bibnamefont
  {Duez}},\ }\href {\doibase 10.1088/0264-9381/27/11/114002} {\bibfield
  {journal} {\bibinfo  {journal} {Class.Quant.Grav.}\ }\textbf {\bibinfo
  {volume} {27}},\ \bibinfo {pages} {114002} (\bibinfo {year} {2010})},\
  \Eprint {http://arxiv.org/abs/0912.3529} {arXiv:0912.3529 [astro-ph.HE]}
  \BibitemShut {NoStop}%
\bibitem [{\citenamefont {Shibata}\ and\ \citenamefont
  {Taniguchi}(2011)}]{Shibata:2011jka}%
  \BibitemOpen
  \bibfield  {author} {\bibinfo {author} {\bibfnamefont {M.}~\bibnamefont
  {Shibata}}\ and\ \bibinfo {author} {\bibfnamefont {K.}~\bibnamefont
  {Taniguchi}},\ }\href {\doibase 10.12942/lrr-2011-6} {\bibfield  {journal}
  {\bibinfo  {journal} {Living Rev.Rel.}\ }\textbf {\bibinfo {volume} {14}},\
  \bibinfo {pages} {6} (\bibinfo {year} {2011})}\BibitemShut {NoStop}%
\bibitem [{\citenamefont {Pannarale}(2012)}]{Pannarale:2012ux}%
  \BibitemOpen
  \bibfield  {author} {\bibinfo {author} {\bibfnamefont {F.}~\bibnamefont
  {Pannarale}},\ }\href@noop {} {\  (\bibinfo {year} {2012})},\ \Eprint
  {http://arxiv.org/abs/1208.5869} {arXiv:1208.5869 [gr-qc]} \BibitemShut
  {NoStop}%
\bibitem [{\citenamefont {Lackey}\ \emph {et~al.}(2013)\citenamefont {Lackey},
  \citenamefont {Kyutoku}, \citenamefont {Shibata}, \citenamefont {Brady},\
  and\ \citenamefont {Friedman}}]{Lackey:2013axa}%
  \BibitemOpen
  \bibfield  {author} {\bibinfo {author} {\bibfnamefont {B.~D.}\ \bibnamefont
  {Lackey}}, \bibinfo {author} {\bibfnamefont {K.}~\bibnamefont {Kyutoku}},
  \bibinfo {author} {\bibfnamefont {M.}~\bibnamefont {Shibata}}, \bibinfo
  {author} {\bibfnamefont {P.~R.}\ \bibnamefont {Brady}}, \ and\ \bibinfo
  {author} {\bibfnamefont {J.~L.}\ \bibnamefont {Friedman}},\ }\href@noop {} {\
   (\bibinfo {year} {2013})},\ \Eprint {http://arxiv.org/abs/1303.6298}
  {arXiv:1303.6298 [gr-qc]} \BibitemShut {NoStop}%
\bibitem [{\citenamefont {Foucart}\ \emph {et~al.}(2013)\citenamefont
  {Foucart}, \citenamefont {Buchman}, \citenamefont {Duez}, \citenamefont
  {Grudich}, \citenamefont {Kidder} \emph {et~al.}}]{Foucart:2013psa}%
  \BibitemOpen
  \bibfield  {author} {\bibinfo {author} {\bibfnamefont {F.}~\bibnamefont
  {Foucart}}, \bibinfo {author} {\bibfnamefont {L.}~\bibnamefont {Buchman}},
  \bibinfo {author} {\bibfnamefont {M.~D.}\ \bibnamefont {Duez}}, \bibinfo
  {author} {\bibfnamefont {M.}~\bibnamefont {Grudich}}, \bibinfo {author}
  {\bibfnamefont {L.~E.}\ \bibnamefont {Kidder}},  \emph {et~al.},\ }\href@noop
  {} {\  (\bibinfo {year} {2013})},\ \Eprint {http://arxiv.org/abs/1307.7685}
  {arXiv:1307.7685 [gr-qc]} \BibitemShut {NoStop}%
\bibitem [{\citenamefont {Brown}\ \emph
  {et~al.}(2012{\natexlab{c}})\citenamefont {Brown}, \citenamefont {Kumar},\
  and\ \citenamefont {Nitz}}]{Brown:2012nn}%
  \BibitemOpen
  \bibfield  {author} {\bibinfo {author} {\bibfnamefont {D.~A.}\ \bibnamefont
  {Brown}}, \bibinfo {author} {\bibfnamefont {P.}~\bibnamefont {Kumar}}, \ and\
  \bibinfo {author} {\bibfnamefont {A.~H.}\ \bibnamefont {Nitz}},\ }\href@noop
  {} {\  (\bibinfo {year} {2012}{\natexlab{c}})},\ \Eprint
  {http://arxiv.org/abs/1211.6184} {arXiv:1211.6184 [gr-qc]} \BibitemShut
  {NoStop}%
\bibitem [{\citenamefont {Smith}\ \emph {et~al.}(2013)\citenamefont {Smith},
  \citenamefont {Mandel},\ and\ \citenamefont {Vecchio}}]{Smith:2013mfa}%
  \BibitemOpen
  \bibfield  {author} {\bibinfo {author} {\bibfnamefont {R.~J.~E.}\
  \bibnamefont {Smith}}, \bibinfo {author} {\bibfnamefont {I.}~\bibnamefont
  {Mandel}}, \ and\ \bibinfo {author} {\bibfnamefont {A.}~\bibnamefont
  {Vecchio}},\ }\href {\doibase 10.1103/PhysRevD.88.044010} {\bibfield
  {journal} {\bibinfo  {journal} {Phys.Rev.}\ }\textbf {\bibinfo {volume}
  {D88}},\ \bibinfo {pages} {044010} (\bibinfo {year} {2013})},\ \Eprint
  {http://arxiv.org/abs/1302.6049} {arXiv:1302.6049 [astro-ph.HE]} \BibitemShut
  {NoStop}%
\bibitem [{\citenamefont {Ozel}\ \emph {et~al.}(2010)\citenamefont {Ozel},
  \citenamefont {Psaltis}, \citenamefont {Narayan},\ and\ \citenamefont
  {McClintock}}]{Ozel:2010su}%
  \BibitemOpen
  \bibfield  {author} {\bibinfo {author} {\bibfnamefont {F.}~\bibnamefont
  {Ozel}}, \bibinfo {author} {\bibfnamefont {D.}~\bibnamefont {Psaltis}},
  \bibinfo {author} {\bibfnamefont {R.}~\bibnamefont {Narayan}}, \ and\
  \bibinfo {author} {\bibfnamefont {J.~E.}\ \bibnamefont {McClintock}},\ }\href
  {\doibase 10.1088/0004-637X/725/2/1918} {\bibfield  {journal} {\bibinfo
  {journal} {Astrophys.J.}\ }\textbf {\bibinfo {volume} {725}},\ \bibinfo
  {pages} {1918} (\bibinfo {year} {2010})},\ \Eprint
  {http://arxiv.org/abs/1006.2834} {arXiv:1006.2834 [astro-ph.GA]} \BibitemShut
  {NoStop}%
\bibitem [{\citenamefont {Farr}\ \emph {et~al.}(2011)\citenamefont {Farr},
  \citenamefont {Sravan}, \citenamefont {Cantrell}, \citenamefont {Kreidberg},
  \citenamefont {Bailyn} \emph {et~al.}}]{Farr:2010tu}%
  \BibitemOpen
  \bibfield  {author} {\bibinfo {author} {\bibfnamefont {W.~M.}\ \bibnamefont
  {Farr}}, \bibinfo {author} {\bibfnamefont {N.}~\bibnamefont {Sravan}},
  \bibinfo {author} {\bibfnamefont {A.}~\bibnamefont {Cantrell}}, \bibinfo
  {author} {\bibfnamefont {L.}~\bibnamefont {Kreidberg}}, \bibinfo {author}
  {\bibfnamefont {C.~D.}\ \bibnamefont {Bailyn}},  \emph {et~al.},\ }\href
  {\doibase 10.1088/0004-637X/741/2/103} {\bibfield  {journal} {\bibinfo
  {journal} {Astrophys.J.}\ }\textbf {\bibinfo {volume} {741}},\ \bibinfo
  {pages} {103} (\bibinfo {year} {2011})},\ \Eprint
  {http://arxiv.org/abs/1011.1459} {arXiv:1011.1459 [astro-ph.GA]} \BibitemShut
  {NoStop}%
\bibitem [{\citenamefont {Kreidberg}\ \emph {et~al.}(2012)\citenamefont
  {Kreidberg}, \citenamefont {Bailyn}, \citenamefont {Farr},\ and\
  \citenamefont {Kalogera}}]{Kreidberg:2012ud}%
  \BibitemOpen
  \bibfield  {author} {\bibinfo {author} {\bibfnamefont {L.}~\bibnamefont
  {Kreidberg}}, \bibinfo {author} {\bibfnamefont {C.~D.}\ \bibnamefont
  {Bailyn}}, \bibinfo {author} {\bibfnamefont {W.~M.}\ \bibnamefont {Farr}}, \
  and\ \bibinfo {author} {\bibfnamefont {V.}~\bibnamefont {Kalogera}},\ }\href
  {\doibase 10.1088/0004-637X/757/1/36} {\bibfield  {journal} {\bibinfo
  {journal} {Astrophys.J.}\ }\textbf {\bibinfo {volume} {757}},\ \bibinfo
  {pages} {36} (\bibinfo {year} {2012})},\ \Eprint
  {http://arxiv.org/abs/1205.1805} {arXiv:1205.1805 [astro-ph.HE]} \BibitemShut
  {NoStop}%
\bibitem [{\citenamefont {Prestwich}\ \emph {et~al.}(2007)\citenamefont
  {Prestwich}, \citenamefont {Kilgard}, \citenamefont {Crowther}, \citenamefont
  {Carpano}, \citenamefont {Pollock} \emph {et~al.}}]{Prestwich:2007mj}%
  \BibitemOpen
  \bibfield  {author} {\bibinfo {author} {\bibfnamefont {A.}~\bibnamefont
  {Prestwich}}, \bibinfo {author} {\bibfnamefont {R.}~\bibnamefont {Kilgard}},
  \bibinfo {author} {\bibfnamefont {P.}~\bibnamefont {Crowther}}, \bibinfo
  {author} {\bibfnamefont {S.}~\bibnamefont {Carpano}}, \bibinfo {author}
  {\bibfnamefont {A.}~\bibnamefont {Pollock}},  \emph {et~al.},\ }\href
  {\doibase 10.1086/523755} {\bibfield  {journal} {\bibinfo  {journal}
  {Astrophys.J.}\ }\textbf {\bibinfo {volume} {669}},\ \bibinfo {pages} {L21}
  (\bibinfo {year} {2007})},\ \Eprint {http://arxiv.org/abs/0709.2892}
  {arXiv:0709.2892 [astro-ph]} \BibitemShut {NoStop}%
\bibitem [{\citenamefont {Silverman}\ and\ \citenamefont
  {Filippenko}(2008)}]{Silverman:2008ss}%
  \BibitemOpen
  \bibfield  {author} {\bibinfo {author} {\bibfnamefont {J.~M.}\ \bibnamefont
  {Silverman}}\ and\ \bibinfo {author} {\bibfnamefont {A.~V.}\ \bibnamefont
  {Filippenko}},\ }\href {\doibase 10.1086/588096} {\bibfield  {journal}
  {\bibinfo  {journal} {Astrophys.J.}\ }\textbf {\bibinfo {volume} {678}},\
  \bibinfo {pages} {L17} (\bibinfo {year} {2008})},\ \Eprint
  {http://arxiv.org/abs/0802.2716} {arXiv:0802.2716 [astro-ph]} \BibitemShut
  {NoStop}%
\bibitem [{\citenamefont {Miller}\ \emph {et~al.}(2009)\citenamefont {Miller},
  \citenamefont {Reynolds}, \citenamefont {Fabian}, \citenamefont {Miniutti},\
  and\ \citenamefont {Gallo}}]{Miller:2009cw}%
  \BibitemOpen
  \bibfield  {author} {\bibinfo {author} {\bibfnamefont {J.}~\bibnamefont
  {Miller}}, \bibinfo {author} {\bibfnamefont {C.}~\bibnamefont {Reynolds}},
  \bibinfo {author} {\bibfnamefont {A.}~\bibnamefont {Fabian}}, \bibinfo
  {author} {\bibfnamefont {G.}~\bibnamefont {Miniutti}}, \ and\ \bibinfo
  {author} {\bibfnamefont {L.}~\bibnamefont {Gallo}},\ }\href {\doibase
  10.1088/0004-637X/697/1/900} {\bibfield  {journal} {\bibinfo  {journal}
  {Astrophys.J.}\ }\textbf {\bibinfo {volume} {697}},\ \bibinfo {pages} {900}
  (\bibinfo {year} {2009})},\ \Eprint {http://arxiv.org/abs/0902.2840}
  {arXiv:0902.2840 [astro-ph.HE]} \BibitemShut {NoStop}%
\bibitem [{\citenamefont {Ozel}\ \emph {et~al.}(2012)\citenamefont {Ozel},
  \citenamefont {Psaltis}, \citenamefont {Narayan},\ and\ \citenamefont
  {Villarreal}}]{Ozel:2012ax}%
  \BibitemOpen
  \bibfield  {author} {\bibinfo {author} {\bibfnamefont {F.}~\bibnamefont
  {Ozel}}, \bibinfo {author} {\bibfnamefont {D.}~\bibnamefont {Psaltis}},
  \bibinfo {author} {\bibfnamefont {R.}~\bibnamefont {Narayan}}, \ and\
  \bibinfo {author} {\bibfnamefont {A.~S.}\ \bibnamefont {Villarreal}},\ }\href
  {\doibase 10.1088/0004-637X/757/1/55} {\bibfield  {journal} {\bibinfo
  {journal} {Astrophys.J.}\ }\textbf {\bibinfo {volume} {757}},\ \bibinfo
  {pages} {55} (\bibinfo {year} {2012})},\ \Eprint
  {http://arxiv.org/abs/1201.1006} {arXiv:1201.1006 [astro-ph.HE]} \BibitemShut
  {NoStop}%
\bibitem [{\citenamefont {Kiziltan}\ \emph {et~al.}(2010)\citenamefont
  {Kiziltan}, \citenamefont {Kottas},\ and\ \citenamefont
  {Thorsett}}]{Kiziltan:2010ct}%
  \BibitemOpen
  \bibfield  {author} {\bibinfo {author} {\bibfnamefont {B.}~\bibnamefont
  {Kiziltan}}, \bibinfo {author} {\bibfnamefont {A.}~\bibnamefont {Kottas}}, \
  and\ \bibinfo {author} {\bibfnamefont {S.~E.}\ \bibnamefont {Thorsett}},\
  }\href@noop {} {\  (\bibinfo {year} {2010})},\ \Eprint
  {http://arxiv.org/abs/1011.4291} {arXiv:1011.4291 [astro-ph.GA]} \BibitemShut
  {NoStop}%
\bibitem [{\citenamefont {Freire}\ \emph {et~al.}(2007)\citenamefont {Freire},
  \citenamefont {Ransom}, \citenamefont {Begin}, \citenamefont {Stairs},
  \citenamefont {Hessels} \emph {et~al.}}]{Freire:2007jd}%
  \BibitemOpen
  \bibfield  {author} {\bibinfo {author} {\bibfnamefont {P.~C.~C.}\
  \bibnamefont {Freire}}, \bibinfo {author} {\bibfnamefont {S.~M.}\
  \bibnamefont {Ransom}}, \bibinfo {author} {\bibfnamefont {S.}~\bibnamefont
  {Begin}}, \bibinfo {author} {\bibfnamefont {I.~H.}\ \bibnamefont {Stairs}},
  \bibinfo {author} {\bibfnamefont {J.~W.}\ \bibnamefont {Hessels}},  \emph
  {et~al.},\ }\href@noop {} {\bibfield  {journal} {\bibinfo  {journal}
  {Astrophys.J.}\ } (\bibinfo {year} {2007})},\ \Eprint
  {http://arxiv.org/abs/0711.0925} {arXiv:0711.0925 [astro-ph]} \BibitemShut
  {NoStop}%
\bibitem [{\citenamefont {Rhoades}\ and\ \citenamefont
  {Ruffini}(1974)}]{Rhoades:1974fn}%
  \BibitemOpen
  \bibfield  {author} {\bibinfo {author} {\bibfnamefont {J.}~\bibnamefont
  {Rhoades}, \bibfnamefont {Clifford~E.}}\ and\ \bibinfo {author}
  {\bibfnamefont {R.}~\bibnamefont {Ruffini}},\ }\href@noop {} {\bibfield
  {journal} {\bibinfo  {journal} {Phys.Rev.Lett.}\ }\textbf {\bibinfo {volume}
  {32}},\ \bibinfo {pages} {324} (\bibinfo {year} {1974})}\BibitemShut
  {NoStop}%
\bibitem [{\citenamefont {Lo}\ and\ \citenamefont {Lin}(2011)}]{Lo:2010bj}%
  \BibitemOpen
  \bibfield  {author} {\bibinfo {author} {\bibfnamefont {K.-W.}\ \bibnamefont
  {Lo}}\ and\ \bibinfo {author} {\bibfnamefont {L.-M.}\ \bibnamefont {Lin}},\
  }\href {\doibase 10.1088/0004-637X/728/1/12} {\bibfield  {journal} {\bibinfo
  {journal} {Astrophys.J.}\ }\textbf {\bibinfo {volume} {728}},\ \bibinfo
  {pages} {12} (\bibinfo {year} {2011})},\ \Eprint
  {http://arxiv.org/abs/1011.3563} {arXiv:1011.3563 [astro-ph.HE]} \BibitemShut
  {NoStop}%
\bibitem [{\citenamefont {Lorimer}(2008)}]{Lorimer:2008se}%
  \BibitemOpen
  \bibfield  {author} {\bibinfo {author} {\bibfnamefont {D.}~\bibnamefont
  {Lorimer}},\ }\href@noop {} {\bibfield  {journal} {\bibinfo  {journal}
  {Living Rev.Rel.}\ }\textbf {\bibinfo {volume} {11}},\ \bibinfo {pages} {8}
  (\bibinfo {year} {2008})},\ \Eprint {http://arxiv.org/abs/0811.0762}
  {arXiv:0811.0762 [astro-ph]} \BibitemShut {NoStop}%
\bibitem [{\citenamefont {Mandel}\ and\ \citenamefont
  {O'Shaughnessy}(2010)}]{Mandel:2009nx}%
  \BibitemOpen
  \bibfield  {author} {\bibinfo {author} {\bibfnamefont {I.}~\bibnamefont
  {Mandel}}\ and\ \bibinfo {author} {\bibfnamefont {R.}~\bibnamefont
  {O'Shaughnessy}},\ }\href {\doibase 10.1088/0264-9381/27/11/114007}
  {\bibfield  {journal} {\bibinfo  {journal} {Class.Quant.Grav.}\ }\textbf
  {\bibinfo {volume} {27}},\ \bibinfo {pages} {114007} (\bibinfo {year}
  {2010})},\ \Eprint {http://arxiv.org/abs/0912.1074} {arXiv:0912.1074
  [astro-ph.HE]} \BibitemShut {NoStop}%
\bibitem [{\citenamefont {Bildsten}\ \emph {et~al.}(1997)\citenamefont
  {Bildsten}, \citenamefont {Chakrabarty}, \citenamefont {Chiu}, \citenamefont
  {Finger}, \citenamefont {Koh} \emph {et~al.}}]{Bildsten:1997vw}%
  \BibitemOpen
  \bibfield  {author} {\bibinfo {author} {\bibfnamefont {L.}~\bibnamefont
  {Bildsten}}, \bibinfo {author} {\bibfnamefont {D.}~\bibnamefont
  {Chakrabarty}}, \bibinfo {author} {\bibfnamefont {J.}~\bibnamefont {Chiu}},
  \bibinfo {author} {\bibfnamefont {M.~H.}\ \bibnamefont {Finger}}, \bibinfo
  {author} {\bibfnamefont {D.~T.}\ \bibnamefont {Koh}},  \emph {et~al.},\
  }\href@noop {} {\bibfield  {journal} {\bibinfo  {journal} {Astrophys J.
  Supp.}\ }\textbf {\bibinfo {volume} {113}},\ \bibinfo {pages} {367} (\bibinfo
  {year} {1997})},\ \Eprint {http://arxiv.org/abs/astro-ph/9707125}
  {arXiv:astro-ph/9707125 [astro-ph]} \BibitemShut {NoStop}%
\bibitem [{\citenamefont {Chakrabarty}(2008)}]{Chakrabarty:2008gz}%
  \BibitemOpen
  \bibfield  {author} {\bibinfo {author} {\bibfnamefont {D.}~\bibnamefont
  {Chakrabarty}},\ }\href@noop {} {\bibfield  {journal} {\bibinfo  {journal}
  {AIP Conference Series}\ }\bibinfo {series} {American Institute of Physics
  Conference Series},\ \textbf {\bibinfo {volume} {1068}},\ \bibinfo {pages}
  {67} (\bibinfo {year} {2008})},\ \Eprint {http://arxiv.org/abs/0809.4031}
  {arXiv:0809.4031 [astro-ph]} \BibitemShut {NoStop}%
\bibitem [{\citenamefont {Burgay}\ \emph {et~al.}(2003)\citenamefont {Burgay}
  \emph {et~al.}}]{Burgay:2003jj}%
  \BibitemOpen
  \bibfield  {author} {\bibinfo {author} {\bibfnamefont {M.}~\bibnamefont
  {Burgay}} \emph {et~al.},\ }\href {\doibase 10.1038/nature02124} {\bibfield
  {journal} {\bibinfo  {journal} {Nature}\ }\textbf {\bibinfo {volume} {426}},\
  \bibinfo {pages} {531} (\bibinfo {year} {2003})}\BibitemShut {NoStop}%
\bibitem [{\citenamefont {Damour}\ \emph {et~al.}(2001)\citenamefont {Damour},
  \citenamefont {Iyer},\ and\ \citenamefont {Sathyaprakash}}]{Damour:2000zb}%
  \BibitemOpen
  \bibfield  {author} {\bibinfo {author} {\bibfnamefont {T.}~\bibnamefont
  {Damour}}, \bibinfo {author} {\bibfnamefont {B.~R.}\ \bibnamefont {Iyer}}, \
  and\ \bibinfo {author} {\bibfnamefont {B.~S.}\ \bibnamefont
  {Sathyaprakash}},\ }\href {\doibase 10.1103/PhysRevD.63.044023} {\bibfield
  {journal} {\bibinfo  {journal} {Phys. Rev.}\ }\textbf {\bibinfo {volume}
  {D63}},\ \bibinfo {pages} {044023} (\bibinfo {year} {2001})},\ \Eprint
  {http://arxiv.org/abs/gr-qc/0010009} {arXiv:gr-qc/0010009} \BibitemShut
  {NoStop}%
\bibitem [{\citenamefont {Peters}(1964)}]{Peters:1964zz}%
  \BibitemOpen
  \bibfield  {author} {\bibinfo {author} {\bibfnamefont {P.}~\bibnamefont
  {Peters}},\ }\href {\doibase 10.1103/PhysRev.136.B1224} {\bibfield  {journal}
  {\bibinfo  {journal} {Phys.Rev.}\ }\textbf {\bibinfo {volume} {136}},\
  \bibinfo {pages} {B1224} (\bibinfo {year} {1964})}\BibitemShut {NoStop}%
\bibitem [{\citenamefont {Arun}\ \emph {et~al.}(2009)\citenamefont {Arun},
  \citenamefont {Buonanno}, \citenamefont {Faye},\ and\ \citenamefont
  {Ochsner}}]{Arun:2008kb}%
  \BibitemOpen
  \bibfield  {author} {\bibinfo {author} {\bibfnamefont {K.~G.}\ \bibnamefont
  {Arun}}, \bibinfo {author} {\bibfnamefont {A.}~\bibnamefont {Buonanno}},
  \bibinfo {author} {\bibfnamefont {G.}~\bibnamefont {Faye}}, \ and\ \bibinfo
  {author} {\bibfnamefont {E.}~\bibnamefont {Ochsner}},\ }\href {\doibase
  10.1103/PhysRevD.79.104023} {\bibfield  {journal} {\bibinfo  {journal} {Phys.
  Rev.}\ }\textbf {\bibinfo {volume} {D79}},\ \bibinfo {pages} {104023}
  (\bibinfo {year} {2009})},\ \Eprint {http://arxiv.org/abs/0810.5336}
  {arXiv:0810.5336 [gr-qc]} \BibitemShut {NoStop}%
\bibitem [{\citenamefont {Buonanno}\ \emph {et~al.}(2009)\citenamefont
  {Buonanno}, \citenamefont {Iyer}, \citenamefont {Ochsner}, \citenamefont
  {Pan},\ and\ \citenamefont {Sathyaprakash}}]{Buonanno:2009zt}%
  \BibitemOpen
  \bibfield  {author} {\bibinfo {author} {\bibfnamefont {A.}~\bibnamefont
  {Buonanno}}, \bibinfo {author} {\bibfnamefont {B.~R.}\ \bibnamefont {Iyer}},
  \bibinfo {author} {\bibfnamefont {E.}~\bibnamefont {Ochsner}}, \bibinfo
  {author} {\bibfnamefont {Y.}~\bibnamefont {Pan}}, \ and\ \bibinfo {author}
  {\bibfnamefont {B.~S.}\ \bibnamefont {Sathyaprakash}},\ }\href {\doibase
  10.1103/PhysRevD.80.084043} {\bibfield  {journal} {\bibinfo  {journal} {Phys.
  Rev.}\ }\textbf {\bibinfo {volume} {D80}},\ \bibinfo {pages} {084043}
  (\bibinfo {year} {2009})},\ \Eprint {http://arxiv.org/abs/0907.0700}
  {arXiv:0907.0700 [gr-qc]} \BibitemShut {NoStop}%
\bibitem [{\citenamefont {Wiseman}(1993)}]{Wiseman:1993aj}%
  \BibitemOpen
  \bibfield  {author} {\bibinfo {author} {\bibfnamefont {A.~G.}\ \bibnamefont
  {Wiseman}},\ }\href {\doibase 10.1103/PhysRevD.48.4757} {\bibfield  {journal}
  {\bibinfo  {journal} {Phys.Rev.}\ }\textbf {\bibinfo {volume} {D48}},\
  \bibinfo {pages} {4757} (\bibinfo {year} {1993})}\BibitemShut {NoStop}%
\bibitem [{\citenamefont {Blanchet}\ \emph
  {et~al.}(1995{\natexlab{a}})\citenamefont {Blanchet}, \citenamefont
  {Damour},\ and\ \citenamefont {Iyer}}]{Blanchet:1995fg}%
  \BibitemOpen
  \bibfield  {author} {\bibinfo {author} {\bibfnamefont {L.}~\bibnamefont
  {Blanchet}}, \bibinfo {author} {\bibfnamefont {T.}~\bibnamefont {Damour}}, \
  and\ \bibinfo {author} {\bibfnamefont {B.~R.}\ \bibnamefont {Iyer}},\ }\href
  {\doibase 10.1103/PhysRevD.51.5360, 10.1103/PhysRevD.54.1860} {\bibfield
  {journal} {\bibinfo  {journal} {Phys.Rev.}\ }\textbf {\bibinfo {volume}
  {D51}},\ \bibinfo {pages} {5360} (\bibinfo {year} {1995}{\natexlab{a}})},\
  \Eprint {http://arxiv.org/abs/gr-qc/9501029} {arXiv:gr-qc/9501029 [gr-qc]}
  \BibitemShut {NoStop}%
\bibitem [{\citenamefont {Blanchet}\ \emph
  {et~al.}(1995{\natexlab{b}})\citenamefont {Blanchet}, \citenamefont {Damour},
  \citenamefont {Iyer}, \citenamefont {Will},\ and\ \citenamefont
  {Wiseman}}]{Blanchet:1995ez}%
  \BibitemOpen
  \bibfield  {author} {\bibinfo {author} {\bibfnamefont {L.}~\bibnamefont
  {Blanchet}}, \bibinfo {author} {\bibfnamefont {T.}~\bibnamefont {Damour}},
  \bibinfo {author} {\bibfnamefont {B.~R.}\ \bibnamefont {Iyer}}, \bibinfo
  {author} {\bibfnamefont {C.~M.}\ \bibnamefont {Will}}, \ and\ \bibinfo
  {author} {\bibfnamefont {A.~G.}\ \bibnamefont {Wiseman}},\ }\href {\doibase
  10.1103/PhysRevLett.74.3515} {\bibfield  {journal} {\bibinfo  {journal}
  {Phys.Rev.Lett.}\ }\textbf {\bibinfo {volume} {74}},\ \bibinfo {pages} {3515}
  (\bibinfo {year} {1995}{\natexlab{b}})},\ \Eprint
  {http://arxiv.org/abs/gr-qc/9501027} {arXiv:gr-qc/9501027 [gr-qc]}
  \BibitemShut {NoStop}%
\bibitem [{\citenamefont {Blanchet}\ \emph {et~al.}(1996)\citenamefont
  {Blanchet}, \citenamefont {Iyer}, \citenamefont {Will},\ and\ \citenamefont
  {Wiseman}}]{Blanchet:1996pi}%
  \BibitemOpen
  \bibfield  {author} {\bibinfo {author} {\bibfnamefont {L.}~\bibnamefont
  {Blanchet}}, \bibinfo {author} {\bibfnamefont {B.~R.}\ \bibnamefont {Iyer}},
  \bibinfo {author} {\bibfnamefont {C.~M.}\ \bibnamefont {Will}}, \ and\
  \bibinfo {author} {\bibfnamefont {A.~G.}\ \bibnamefont {Wiseman}},\ }\href
  {\doibase 10.1088/0264-9381/13/4/002} {\bibfield  {journal} {\bibinfo
  {journal} {Class.Quant.Grav.}\ }\textbf {\bibinfo {volume} {13}},\ \bibinfo
  {pages} {575} (\bibinfo {year} {1996})},\ \Eprint
  {http://arxiv.org/abs/gr-qc/9602024} {arXiv:gr-qc/9602024 [gr-qc]}
  \BibitemShut {NoStop}%
\bibitem [{\citenamefont {Blanchet}\ \emph {et~al.}(2002)\citenamefont
  {Blanchet}, \citenamefont {Faye}, \citenamefont {Iyer},\ and\ \citenamefont
  {Joguet}}]{Blanchet:2001ax}%
  \BibitemOpen
  \bibfield  {author} {\bibinfo {author} {\bibfnamefont {L.}~\bibnamefont
  {Blanchet}}, \bibinfo {author} {\bibfnamefont {G.}~\bibnamefont {Faye}},
  \bibinfo {author} {\bibfnamefont {B.~R.}\ \bibnamefont {Iyer}}, \ and\
  \bibinfo {author} {\bibfnamefont {B.}~\bibnamefont {Joguet}},\ }\href
  {\doibase 10.1103/PhysRevD.71.129902, 10.1103/PhysRevD.65.061501} {\bibfield
  {journal} {\bibinfo  {journal} {Phys.Rev.}\ }\textbf {\bibinfo {volume}
  {D65}},\ \bibinfo {pages} {061501} (\bibinfo {year} {2002})},\ \Eprint
  {http://arxiv.org/abs/gr-qc/0105099} {arXiv:gr-qc/0105099 [gr-qc]}
  \BibitemShut {NoStop}%
\bibitem [{\citenamefont {Blanchet}\ \emph {et~al.}(2004)\citenamefont
  {Blanchet}, \citenamefont {Damour}, \citenamefont {Esposito-Farese},\ and\
  \citenamefont {Iyer}}]{Blanchet:2004ek}%
  \BibitemOpen
  \bibfield  {author} {\bibinfo {author} {\bibfnamefont {L.}~\bibnamefont
  {Blanchet}}, \bibinfo {author} {\bibfnamefont {T.}~\bibnamefont {Damour}},
  \bibinfo {author} {\bibfnamefont {G.}~\bibnamefont {Esposito-Farese}}, \ and\
  \bibinfo {author} {\bibfnamefont {B.~R.}\ \bibnamefont {Iyer}},\ }\href
  {\doibase 10.1103/PhysRevLett.93.091101} {\bibfield  {journal} {\bibinfo
  {journal} {Phys.Rev.Lett.}\ }\textbf {\bibinfo {volume} {93}},\ \bibinfo
  {pages} {091101} (\bibinfo {year} {2004})},\ \Eprint
  {http://arxiv.org/abs/gr-qc/0406012} {arXiv:gr-qc/0406012 [gr-qc]}
  \BibitemShut {NoStop}%
\bibitem [{\citenamefont {Blanchet}\ \emph {et~al.}(2012)\citenamefont
  {Blanchet}, \citenamefont {Buonanno},\ and\ \citenamefont
  {Faye}}]{Blanchet:2012sm}%
  \BibitemOpen
  \bibfield  {author} {\bibinfo {author} {\bibfnamefont {L.}~\bibnamefont
  {Blanchet}}, \bibinfo {author} {\bibfnamefont {A.}~\bibnamefont {Buonanno}},
  \ and\ \bibinfo {author} {\bibfnamefont {G.}~\bibnamefont {Faye}},\
  }\href@noop {} {\  (\bibinfo {year} {2012})},\ \Eprint
  {http://arxiv.org/abs/1210.0764} {arXiv:1210.0764 [gr-qc]} \BibitemShut
  {NoStop}%
\bibitem [{\citenamefont {Bohe}\ \emph {et~al.}(2013)\citenamefont {Bohe},
  \citenamefont {Marsat},\ and\ \citenamefont {Blanchet}}]{Bohe:2013cla}%
  \BibitemOpen
  \bibfield  {author} {\bibinfo {author} {\bibfnamefont {A.}~\bibnamefont
  {Bohe}}, \bibinfo {author} {\bibfnamefont {S.}~\bibnamefont {Marsat}}, \ and\
  \bibinfo {author} {\bibfnamefont {L.}~\bibnamefont {Blanchet}},\ }\href@noop
  {} {\  (\bibinfo {year} {2013})},\ \Eprint {http://arxiv.org/abs/1303.7412}
  {arXiv:1303.7412 [gr-qc]} \BibitemShut {NoStop}%
\bibitem [{\citenamefont {Cutler}\ and\ \citenamefont
  {Flanagan}(1994)}]{Cutler:1994ys}%
  \BibitemOpen
  \bibfield  {author} {\bibinfo {author} {\bibfnamefont {C.}~\bibnamefont
  {Cutler}}\ and\ \bibinfo {author} {\bibfnamefont {E.~E.}\ \bibnamefont
  {Flanagan}},\ }\href {\doibase 10.1103/PhysRevD.49.2658} {\bibfield
  {journal} {\bibinfo  {journal} {Phys.Rev.}\ }\textbf {\bibinfo {volume}
  {D49}},\ \bibinfo {pages} {2658} (\bibinfo {year} {1994})},\ \Eprint
  {http://arxiv.org/abs/gr-qc/9402014} {arXiv:gr-qc/9402014 [gr-qc]}
  \BibitemShut {NoStop}%
\bibitem [{\citenamefont {Droz}\ \emph {et~al.}(1999)\citenamefont {Droz},
  \citenamefont {Knapp}, \citenamefont {Poisson},\ and\ \citenamefont
  {Owen}}]{Droz:1999qx}%
  \BibitemOpen
  \bibfield  {author} {\bibinfo {author} {\bibfnamefont {S.}~\bibnamefont
  {Droz}}, \bibinfo {author} {\bibfnamefont {D.~J.}\ \bibnamefont {Knapp}},
  \bibinfo {author} {\bibfnamefont {E.}~\bibnamefont {Poisson}}, \ and\
  \bibinfo {author} {\bibfnamefont {B.~J.}\ \bibnamefont {Owen}},\ }\href
  {\doibase 10.1103/PhysRevD.59.124016} {\bibfield  {journal} {\bibinfo
  {journal} {Phys.Rev.}\ }\textbf {\bibinfo {volume} {D59}},\ \bibinfo {pages}
  {124016} (\bibinfo {year} {1999})},\ \Eprint
  {http://arxiv.org/abs/gr-qc/9901076} {arXiv:gr-qc/9901076 [gr-qc]}
  \BibitemShut {NoStop}%
\bibitem [{\citenamefont {Blanchet}(2006)}]{Blanchet:2006zz}%
  \BibitemOpen
  \bibfield  {author} {\bibinfo {author} {\bibfnamefont {L.}~\bibnamefont
  {Blanchet}},\ }\href@noop {} {\bibfield  {journal} {\bibinfo  {journal}
  {Living Rev.Rel.}\ }\textbf {\bibinfo {volume} {9}} (\bibinfo {year}
  {2006})}\BibitemShut {NoStop}%
\bibitem [{\citenamefont {Apostolatos}(1995)}]{Apostolatos:1995pj}%
  \BibitemOpen
  \bibfield  {author} {\bibinfo {author} {\bibfnamefont {T.~A.}\ \bibnamefont
  {Apostolatos}},\ }\href {\doibase 10.1103/PhysRevD.52.605} {\bibfield
  {journal} {\bibinfo  {journal} {Phys.Rev.}\ }\textbf {\bibinfo {volume}
  {D52}},\ \bibinfo {pages} {605} (\bibinfo {year} {1995})}\BibitemShut
  {NoStop}%
\bibitem [{\citenamefont {Ohme}\ \emph {et~al.}(2013)\citenamefont {Ohme},
  \citenamefont {Nielsen}, \citenamefont {Keppel},\ and\ \citenamefont
  {Lundgren}}]{Ohme:2013nsa}%
  \BibitemOpen
  \bibfield  {author} {\bibinfo {author} {\bibfnamefont {F.}~\bibnamefont
  {Ohme}}, \bibinfo {author} {\bibfnamefont {A.~B.}\ \bibnamefont {Nielsen}},
  \bibinfo {author} {\bibfnamefont {D.}~\bibnamefont {Keppel}}, \ and\ \bibinfo
  {author} {\bibfnamefont {A.}~\bibnamefont {Lundgren}},\ }\href@noop {} {\
  (\bibinfo {year} {2013})},\ \Eprint {http://arxiv.org/abs/1304.7017}
  {arXiv:1304.7017 [gr-qc]} \BibitemShut {NoStop}%
\bibitem [{\citenamefont {Conway}\ and\ \citenamefont
  {Sloane}(1993)}]{Conway:1993}%
  \BibitemOpen
  \bibfield  {author} {\bibinfo {author} {\bibfnamefont {J.}~\bibnamefont
  {Conway}}\ and\ \bibinfo {author} {\bibfnamefont {N.}~\bibnamefont
  {Sloane}},\ }\href@noop {} {\emph {\bibinfo {title} {Sphere Packings,
  Lattices and Groups}}},\ \bibinfo {edition} {2nd}\ ed.\ (\bibinfo
  {publisher} {Springer-Verlag, New York},\ \bibinfo {year} {1993})\BibitemShut
  {NoStop}%
\bibitem [{\citenamefont {Baird}\ \emph {et~al.}(2013)\citenamefont {Baird},
  \citenamefont {Fairhurst}, \citenamefont {Hannam},\ and\ \citenamefont
  {Murphy}}]{Baird:2012cu}%
  \BibitemOpen
  \bibfield  {author} {\bibinfo {author} {\bibfnamefont {E.}~\bibnamefont
  {Baird}}, \bibinfo {author} {\bibfnamefont {S.}~\bibnamefont {Fairhurst}},
  \bibinfo {author} {\bibfnamefont {M.}~\bibnamefont {Hannam}}, \ and\ \bibinfo
  {author} {\bibfnamefont {P.}~\bibnamefont {Murphy}},\ }\href {\doibase
  10.1103/PhysRevD.87.024035} {\bibfield  {journal} {\bibinfo  {journal}
  {Phys.Rev.}\ }\textbf {\bibinfo {volume} {D87}},\ \bibinfo {pages} {024035}
  (\bibinfo {year} {2013})},\ \Eprint {http://arxiv.org/abs/1211.0546}
  {arXiv:1211.0546 [gr-qc]} \BibitemShut {NoStop}%
\bibitem [{\citenamefont {Keppel}(2013)}]{Keppel:2013yia}%
  \BibitemOpen
  \bibfield  {author} {\bibinfo {author} {\bibfnamefont {D.}~\bibnamefont
  {Keppel}},\ }\href@noop {} {\  (\bibinfo {year} {2013})},\ \Eprint
  {http://arxiv.org/abs/1303.2005} {arXiv:1303.2005 [physics.data-an]}
  \BibitemShut {NoStop}%
\end{thebibliography}%

\end{document}